\documentclass[12pt,a4paper]{article}
\pdfoutput=1
\usepackage{jheppub}

\usepackage{epstopdf}
\usepackage{graphicx}
\usepackage{ulem}
\usepackage{epsfig}
\usepackage{dcolumn}
\usepackage{bm}
\usepackage{amssymb}
\usepackage{amsmath,bm}
\usepackage{amsfonts}
\usepackage{slashed}
\usepackage[mathscr]{euscript}
\usepackage{epsfig}
\usepackage{verbatim}
\usepackage[english]{babel}
\usepackage{color}
\usepackage{subfigure}

\newdimen\tableauside\tableauside=1.0ex
\newdimen\tableaurule\tableaurule=0.4pt
\newdimen\tableaustep
\def\phantomhrule#1{\hbox{\vbox to0pt{\hrule height\tableaurule width#1\vss}}}
\def\phantomvrule#1{\vbox{\hbox to0pt{\vrule width\tableaurule height#1\hss}}}
\def\sqr{\vbox{%
                \phantomhrule\tableaustep
                \hbox{\phantomvrule\tableaustep\kern\tableaustep\phantomvrule\tableaustep}%
                \hbox{\vbox{\phantomhrule\tableauside}\kern-\tableaurule}}}
\def\squares#1{\hbox{\count0=#1\noindent\loop\sqr
                \advance\count0 by-1 \ifnum\count0>0\repeat}}
\def\tableau#1{\vcenter{\offinterlineskip
                \tableaustep=\tableauside\advance\tableaustep by-\tableaurule
                \kern\normallineskip\hbox
                {\kern\normallineskip\vbox
                        {\gettableau#1 0 }%
                        \kern\normallineskip\kern\tableaurule}%
                \kern\normallineskip\kern\tableaurule}}
\def\gettableau#1 {\ifnum#1=0\let\next=\null\else
        \squares{#1}\let\next=\gettableau\fi\next}

\newcommand{\be}{ \begin{equation}}
\newcommand{\ee}{\end{equation}}
\newcommand{\bea}{\begin{eqnarray}}
\newcommand{\eea}{\end{eqnarray}}

\def\ZZZ{{\hskip-3pt\hbox{ Z\kern-1.6mm Z}}}
\def\zzz{{\hskip-3pt\hbox{ z\kern-1mm z}}}

\def\bal#1\eal{\begin{align}#1\end{align}}

\def\one{{\hbox{ 1\kern-.8mm l}}}
\def\zero{{\hbox{ 0\kern-1.5mm 0}}}

\def\e{\,{\rm e}}

\preprint{YITP-18-40}

\title{Notes on the Causal Structure in a tensor network
}

\author{Arpan Bhattacharyya${}^{d}$, Long Cheng$^{b}$, Ling-Yan Hung$^{a,b,c}$, Sirui Ning$^{e}$, Zhi Yang$^{b}$}

\affiliation{$^a$ State Key Laboratory of Surface Physics and Department of Physics, Fudan University,\\
        \hspace*{0.3cm}220 Handan Road, 200433 Shanghai, P. R. China}

\affiliation{$^b$ Department of Physics and Center for Field Theory and Particle Physics,\\ Fudan University,  \hspace*{0.3cm}220 Handan Road, 200433 Shanghai, P. R. China}

\affiliation{$^c$ Collaborative Innovation Center of Advanced  Microstructures,
Nanjing University,\\        \hspace*{0.3cm}Nanjing, 210093, P. R. China.}

\affiliation{$^d$ Yukawa Institute for Theoretical Physics (YITP), Kyoto University, \\        \hspace*{0.3cm}Kitashirakawa Oiwakecho, Sakyo-ku, Kyoto 606-8502, Japan}

\affiliation{$e$ School of the Gifted Young, University of Science and Technology of China, \\ \hspace{0.3cm}Hefei, 230026  Anhui, P.R.China}

\emailAdd{bhattacharyya.arpan@yahoo.com, lcheng@fudan.edu.cn, elektron.janethung@gmail.com, cmicheal@mail.ustc.edu.cn, zyang14@fudan.edu.cn}

\abstract{
In this paper we attempt to understand Lorentzian tensor networks, as a preparation for constructing tensor networks that can describe more exotic backgrounds such as black holes.  To define notions of reference frames and switching of reference frames on a tensor network, we will borrow ideas from the algebraic quantum field theory literature.  With these definitions, we construct simple examples of Lorentzian tensor networks and solve the spectrum for a choice of ``inertial frame''  based on Gaussian models of fermions and integrable models.  In particular, the tensor network can be viewed as a periodically driven Floquet system, that by-pass the ``doubling problem'' and gives rise to fermions with exactly linear dispersion relations. We will find that a boost operator connecting different inertial frames, and notions of ``Rindler observers'' can be defined, and that important physics in Lorentz invariant QFT, such as the Unruh effect, can be captured by such skeleton of spacetime. We find interesting subtleties when the same approach is directly applied to bosons -- the operator algebra contains commutators that take the wrong sign -- resembling bosons behind horizons.
}

\begin{document}

\maketitle

\makeatletter
\g@addto@macro\bfseries{\boldmath}
\makeatother

\section{Introduction}

Interesting aspects of the AdS/CFT, such as reconstructing black hole physics from CFT data, are crucial for understanding quantum gravity. There has been increasing amount of evidence showing that the AdS/CFT can be understood in terms of a tensor network since the possibility was first pointed out in \cite{Swingle:2009bg}. There are interesting toy construction of tensor networks attempting to capture the physics particularly of black holes. For example, it is noted that a black hole should probably behave like a region of particularly high bond dimension in a tensor network \cite{Pastawski:2015qua,Hayden:2016cfa}. These constructions, however, are (mostly) based on static spacetimes, and the tensor network considered essentially describes the Euclidean geometry of some time slice.  The complete description of black holes, particularly if we were to address questions such as unitarity, is a time-dependent question. Interesting physics is associated with the black hole horizon, and it would thus require notions of  null surfaces, and therefore also notions of spacelike and timelike separations in a tensor network construction. Therefore, to gain an understanding of gravitational dynamics via tensor networks requires us at ground zero to define how space-like/time-like separations are described in a tensor network in the first place.

There is some recent progress based on a random tensor network that also attempts to describe a covariant version of the AdS/CFT \cite{Qi:2018shh}. The current paper has a relatively modest goal, taking a step to understand what it means to represent space-time on a tensor network. In this paper we would like to explore how one should recover crucial features of a Lorentzian space-time and understand the causal structures and their implications for different space-time observers. To that end, we also need to set up the problem and define notions such as observers on a tensor network.

Our paper is divided into two parts. First,  we would like to put together the language that has been developed in the tensor network/error correcting code/bulk-reconstruction literature  with notions developed in algebraic QFT.  In section 2, we review just enough basics so that in section 3, we define notions of causality, Cauchy surfaces and frames of references in a tensor network.

Then in section 4, using these definitions, we construct explicit models. We start with a quadratic fermionic model, and demonstrate, at least in some limits very explicitly, that notions such as Lorentz transformation and the Unruh effect can be described to very good approximation in the tensor network. In fact, the tensor network evolution is effectively turning the system into a periodically driven Floquet system that could give rise to a linear dispersion relation free of the ``doubling problem''. (For a review of the problem, see, for example \cite{lattice_book}.) These fermions can have exactly linear dispersion relations with emergent Lorentz invariance that becomes evident in the computation of the correlation functions and anti-commutation relations.  We will also see that light cones depend on the explicit choice of the tensors furbishing the tensor network.

This hopefully lays out some basic features of  tensor networks describing Lorentzian space-time, and serves as preparation as we move on to a covariant construction of more general interesting spacetimes based on the tensor networks.

\section{Axioms of AQFT}

In this section we give a very brief summary of the basic axioms defining an AQFT, listing all the ingredients that are going to have a natural realization in a tensor network.
Our discussion is heavily based on \cite{Fredenhagen:2014lda} which gives a relatively gentle review of the subject. We also find \cite{Wald:1995yp} a concise and physical exposition of the subject.
The Kaag-Hastler axioms are motivated by incorporating locality and causality into an operator algebra that in turn defines a QFT.

The ingredients involved are therefore spacetime manifold $\mathcal{M}$ on the one hand, and some operator algebra $\mathcal{A}$ on the other.
There are various conditions imposed on $\mathcal{M}$. Of course traditionally, the discussion is based on smooth manifolds. We will list all the ingredients and discuss which can make direct contact with the tensor network.
First, we need some constraints on the set of spacetimes concerned.
\begin{enumerate}
\item $\mathcal{M}$ as a topological space is Hausdorff, connected and paracompact.
\item $\mathcal{M}$ has a pseudo-Riemannian metric $g$. This defines a causal structure. For a smooth curve $\gamma(t)$ in $\mathcal{M}$, it can either be
space-like if $g(\dot\gamma, \dot\gamma) <0$; causal (time-like or null like) if $g(\dot\gamma, \dot\gamma) \ge 0$
\item $\mathcal{M}$ is globally hyperbolic such that it does not contain closed causal curves and for any two points $x$ and $y$, $J_+(x)\cap J_-(y) $ is compact. Here $J_+(x)$ denotes the collection of points that are in the "future" of $x$ and  that can be connected to $x$ by a future directed causal curve. $J_-(x)$ corresponds to the causal past of $x$.
\item Having a causal structure, it also means that we can define an order relation $x\prec y$ i.e. $x$ precedes $y$ if there exists a future directing causal curve connecting $x$ to $y$. This relation can be generalized to subsets in $\mathcal{M}$. Consider two sets $O_1$ and $O_2$.  They are space-like separated if they cannot be connected by a casual curve -- if for all $x\in \bar{O_1}$, $J_{\pm}(x)$ has empty intersection with $O_2$.
\item There exist foliations by Cauchy surfaces, (locally) diffeomorphic to $\Sigma \times R$
\item There exist ``admissible embeddings" $\chi: \mathcal{M}\to \mathcal{N}$ for globally hyperbolic spacetimes $\mathcal{M}$, $\mathcal{N}$, such that the map preserves the metric, orientations and causal
structure.
\item Collection $B$ of bounded subsets of $\mathcal{M}$ forms a directed set. There exists a reflexive and transitive binary relation : for a pair $O_1,O_2\in B, \,\exists O : O_1\subseteq O $ and $O_2 \subseteq O$ .
\end{enumerate}

These spacetimes form a ``category'' \emph{LOC} -- category of locally hyperbolic, oriented and time-oriented spacetimes.

With such a space-time as a starting point, the AQFT is a rule of assigning an operator algebra to hyperbolic spacetimes.  Mathematically, an AQFT is a functor that maps
between two categories. \emph{LOC} on  one hand, and \emph{OBS} --  the category of unital C* algebra corresponding to the algebra of operators of physical observables-- on the other.

In QFT, it is taken that observables form a $C^*$-algebra.  For completeness, we provide the definition of $C^*$- algebra below.
A $C^*$-algebra $A$ comes with a norm $||x||$ and a map * that takes $A\to A$.
For $x \in A$, where $A$
\begin{eqnarray}
&&x^{**} = (x^*)^* = x, \\
&&(x+y)^* = x^* + y^*, \\
&&(\lambda x)^* = \bar\lambda x^*\\
&&||x^*\,x|| = ||x|| ||x^*||.
\end{eqnarray}
The ``C'' stood for ``(norm)-closed''. The norm provides the algebra a topological structure.
Let us make contact with quantum mechanics.
Consider $H$ to be a complex Hilbert space with inner product denoted $\langle \cdot, \cdot\rangle$. The collection of bounded linear operators on $H$, denoted by $B(H)$, is a $C^*$-algebra. The linear structure is clear. The product is by composition of operators. The * operation is the adjoint; for any operator $a$ on H, its adjoint is defined by the equation $\langle a^*\zeta,\eta \rangle = \langle \zeta,a\eta \rangle$, for all $\zeta$ and $\eta$ in $H$. Finally, the norm is given by $||a|| = \textrm{sup}\{||a \zeta|| \vert \zeta \in H, |\zeta| \le 1\}$,
for any $a$ in $B(H)$.

The axioms constraining the AQFT functor  $\mathcal{U}$ are as follows:
$\mathcal{U}$ assigns to each bounded subset $O \in B $  a $C^*$-algebra $\mathcal{U}(O)$. The algebra of spacetime $\mathcal{M}$ is defined as the {\it inductive} limit
\be
\mathcal{U}(\mathcal{M}) = \overline{\cup_O\,\, \mathcal{U}(O)}.
\ee
(This can be understood as a generalization of direct sum. ) Since this is a map from a directed set $B$ to another set, it forms a {\it net} of $C^*$-algebras.
For any subset $\mathcal{N}\subset \mathcal{M}$,  $\mathcal{U}(\mathcal{N})$ is generated by $\mathcal{U}(O)$ where $O \subset \mathcal{N}$,
\begin{enumerate}
\item Isotony -- For $O\subset \tilde{O} , \,\, \mathcal{U}(O) \subset \mathcal{U}(\tilde{O})$. This expresses the fact that the operator algebra abides by the notion of $\subset$ in spacetime $\mathcal{M}$.
\item Locality (Einstein causality). When $O_1$ and $O_2$ are space-like separated as defined above, $[A,B]=0,$ for all $A \in \mathcal{U}(O_1)$ and $B\in \mathcal{U}(O_2)$.
\item Time slice axiom.  The solvability of the initial value problem is translated into a requirement of the operator algebra. The algebra $\mathcal{U}(N)$ is isomorphic to $\mathcal{U}(\mathcal{M})$ for any causally convex neighbourhood of a Cauchy surface $\Sigma$.  Note that a causally convex neighbourhood $N$ is one in which no causal curve inside $\mathcal{M}$ meets $N$ in a disconnected set.
\item In a generally curved spacetime, it is more suitable to invoke local charts and they can be understood as the admissible maps $\chi$ discussed above. For each admissible embedding $\chi : N \to M$, there is an injective homomorphism $\alpha_\chi : \mathcal{U}(N) \to \mathcal{U}(M)$. If  $\chi_1: M\to N$ and $\chi_2: N \to L$ then we have
\be
\alpha_{\chi_2 \circ \chi_2} = \alpha_{\chi_2}\circ \alpha_{\chi_1}
\ee
i.e. altogether, the assignment of algebra via $\mathcal{U}$ has to be compatible with the structures of embeddings leading to maps between algebras. That makes an AQFT a covariant functor between {\bf LOC} and {\bf OBS}.
In terms of these embedding maps, the {\bf Einstein Causality constraint} can be phrased as follows: if $\chi_1(M_1) \subset M$ and $\chi_2(M_2) \subset M$ are causally disconnected, then
\be
[\alpha_{\chi_1}(\mathcal{U}(M_1)), \alpha_{\chi_2}(\mathcal{U}(M_2))] = 0
\ee
Similarly, the {\bf time-slice axiom} can be restated if we introduce an embedding map $\chi: N\to M$, where $N$ is the causally convex neighbourhood of a Cauchy surface, we have
$\alpha_{\chi}$ an {\it isomorphism}.
\item  The above isomorphism describes general covariance. If we only have global isometries, then the statement is instead restricted to the existence of an isomorphism $\alpha_L$ that maps
$\alpha_L: \mathcal{U}(O) \to \mathcal{U}(LO)$, where $LO$ is the region $O$ transformed via some isometry $L$ (such as Poincare symmetry in Minkowski spacetime).
\item Time evolution between any two Cauchy surfaces: given two Cauchy surfaces  $\Sigma_{1,2}$, since the neighbourhood $N_i $ of each can be associated with an embedding map $\chi_i: N_i \to M$ such that $\alpha_{\chi_i}$ is an algebra isomorphism from $\mathcal{U}(N_i) \to \mathcal{U}(M)$. One can obtain an algebra isomorphism between $\alpha_{1\to 2}: \mathcal{U}(N_1) \to \mathcal{U}(N_2)$, where
\be
\alpha_{1\to2} = \alpha^{-1}_{\chi_2} \circ \alpha_{\chi_1}
\ee
\end{enumerate}

\section {Constructing a causal spacetime using tensor network}
We would like to build a causal spacetime using a tensor network by associating building components of a tensor network to ingredients in an AQFT discussed above.

A tensor network is a collection of tensors contracted with each other, with each tensor represented as a vertex, and shared contracted indices between tensors represented by an edge connecting these vertices.
Therefore in this simplest version, the tensor network associates a graph to a collection of tensors.
\subsection{Topological space}

A graph $F$ can be viewed as a topological space naturally endowed with 0 simplices (vertices) and 1-simplices (edges). This topological space is the arena  that plays the role of the spacetime manifold $M$. Clearly there are more structures to spacetimes than such a skeleton, although not everything has an immediate analogue in the tensor network.

Notwithstanding,  a graph is a set of vertices and edges, and as such, the notion of $O_i \subset O_j$, where $O_i$ and $O_j$ are subgraphs of $F$ is well defined. Therefore, this is also a directed set.

\subsection{Hilbert space and local operator algebra}
Each edge is associated to a contracted index of a tensor. Therefore each edge can be naturally associated with a Hilbert space. (Perhaps practically finite dimensional.) Operators acting on this Hilbert space would thus form a C* algebra, as discussed in the previous section. Associating an operator algebra to links do not constitute $\mathcal{U}(O)$. These operators are related to each other. Such relations will be laid out below.

\subsection{Causal structure: timelike vs spacelike separation}

The graph does not generically possess any causal structure. The associated tensor network without any causal structure might well be associated to a Euclidean version of spacetime.
To describe Lorentzian signature, the graph needs to acquire a causal structure.

One way of incorporating causal structure in the tensor network is based on local unitaries.

Consider the special case in which each tensor $T$ has an even number of legs, each with the same bond dimension. We can split the indices into two equal groups $g_{1,2}$. If $T^{\beta_j \in g_2}_{\alpha_i \in g_1}$ is a unitary from $\alpha_i$ to $\beta_j$, then we would consider the vertices connected via these edges to be causally connected. This thus assigns a causal ordering in the graph.

Such orderings can only be made consistent with each other, however, as follows. Consider two indices $\alpha_{1} \in g_1$ and $\beta_1\in g_2$, where we have assigned the ordering  $g_1 \prec g_2$. If there exists another split into two groups $g_3$ and $g_4$ such that $\alpha_1$ and $\beta_1$ belong to $g_3$, then at least one index in $g_1$ must now belong to $g_4$ and at least another one in $g_2$ now belongs to $g_3$. In such a case,  requiring either $g_3 \prec g_4$ or vice versa would be inconsistent with $g_1 \prec g_2$. Therefore a consistent causal ordering assignment can be made only if there is a unique splitting $g_1,g_2$ such that $T$ is unitary. In which case, edges within the same group can be considered space-like separated.  For each set of unitary evolution we are defining a set of observers, or a frame.

For $T$ being {\it perfect tensors} (a 2n-index tensor $T_{a1a2...a2n}$ is a perfect tensor if, for any bipartition of its indices into a set $A$ and complementary set $A^c$ with $|A|\leq|A^c|$, T is proportional to an isometric tensor from $A$ to $A^c$.\cite{Pastawski:2015qua}) for example, then every pair of indices is time-like separated, and yet there is no ordering agreed by all frames. In other words, there is no consistent assignment of a causal precedence in this case.

Such an assignment is local. The above assignment makes it natural to include arrows in the edges to denote causal precedence. We can put in-going arrows among edges in $g_1$ and out-going arrows in edges in $g_2$. As a unitary matrix, the number of arrows is conserved across each vertex.

For a global assignment of causal structure, one needs to pay special attention to how edges are contracted with each other. With the arrow assignment, a global causal structure would follow when these directed tensors are assembled together, where each  out-going edge proceeds to become an in-going edge in the other vertex it connects to.

Global assignment of a causal structure thus requires that the graph $F$ is orientable. The requirement of the absence of closed time-like curves becomes the requirement that the graph is a {\it directed acyclic graph}, which is a finite directed graph with no directed cycles.
The causal structure is borne out by the building block being local unitaries.  This is similar to the consideration in  the causal-set approach to quantum gravity. (This is a huge subject in its own right which is impossible to review here. We refer interested readers to one of the original papers \cite{Bombelli:1987aa} and \cite{Sorkin:2003bx}  and references therein for more recent discussions.) In the current paper,  the emergent causality in a tensor network is ultimately a measurement problem of commutators based on Einstein locality. They are not pre-determined at the level of the graph even though the structure of the graph could preclude various causal structures.

This point of view will be emphasized again in section \ref{Einstein_local} below.
In the context of the tensor networks it is not clear whether  graphs that are transitively closed/complete -- graphs where there exists an edge connecting any two causally related points-- play any special role.

To summarise, it appears that the conservation of in-going and out-going legs on a directed acyclic graph matched with local tensors with a unique ``unitary direction'' has replaced the notion of global hyperbolicity.

\subsubsection{Operator pushing and local unitaries}

The local unitaries $T$ defines isomorphisms between operator algebras. Operators from in-going legs are related to operators in the out-going legs by conjugation (or equivalently what is called operator pushing, for example, in \cite{Pastawski:2015qua}).
An immediate consequence is that as we push an operator starting from a Cauchy slice across $T$'s, we sweep out a light-cone, either forward or backward in time, a feature already observed, for example, in the context of Multi-scale entanglement renormalization ansatz (MERA) tensor networks \cite{beny}. This ensures that information cannot flow faster than the speed of light, which is captured by the Einstein causality condition below.

\subsubsection{Operator algebra assigned to a connected subgraph $O$}

Now consider $O$ to be a connected subgraph of  $F$. Then we can pick out a set of edges all mutually space-like separated. The operator algebra $\mathcal{U}(O)$ associated to this region $O$ can be defined as the operator algebra on the maximal set $S$ of mutually space-like separated edges. Operators acting on any other legs can be pulled back to operators acting on $S$ via local unitaries. Or in other words, we will include in this operator algebra only those that can be pulled back to $S$. This defines $\mathcal{U}(O)$ that is generated by the operator algebra on $S$. To make subsequent discussion simple, when we discuss these subregions $O$, they should carry the structure of a causal diamond -- the boundary of $O$ should intersect $\partial S$. The boundary is thus separated into two pieces by $\partial S$. Each piece is related to $S$ by a unitary map.
\begin{figure}[!h]
		\centering
		\includegraphics[width=8cm]{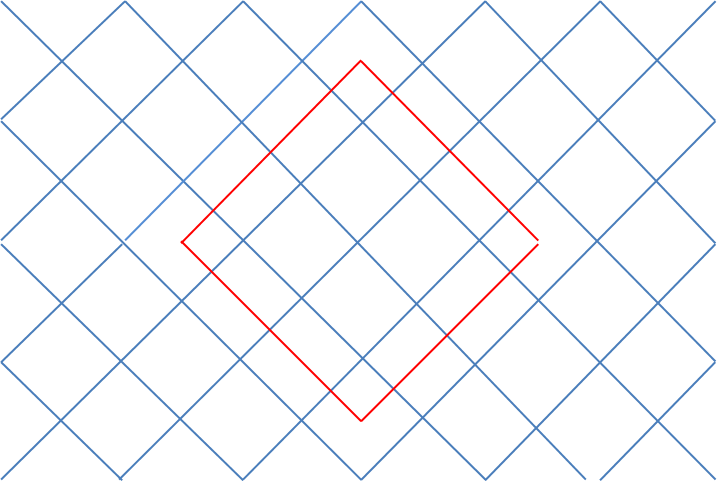}
		\caption{Causal Diamond.}
		\label{Cau}
\end{figure}\\

{\bf Observation on the tensor network}: if $O_1 \subset O_2$, then $S_1 \subset S_2$ and thus $\mathcal{U}(O_1) \subset \mathcal{U}(O_2)$. The set of algebra $\{\mathcal{U}(O)\}$ also forms a {\it net} of (C*) algebra.

\subsubsection{Cauchy surface and the time slice axiom}

\begin{figure}[!h]
		\centering
		\includegraphics[width=8cm]{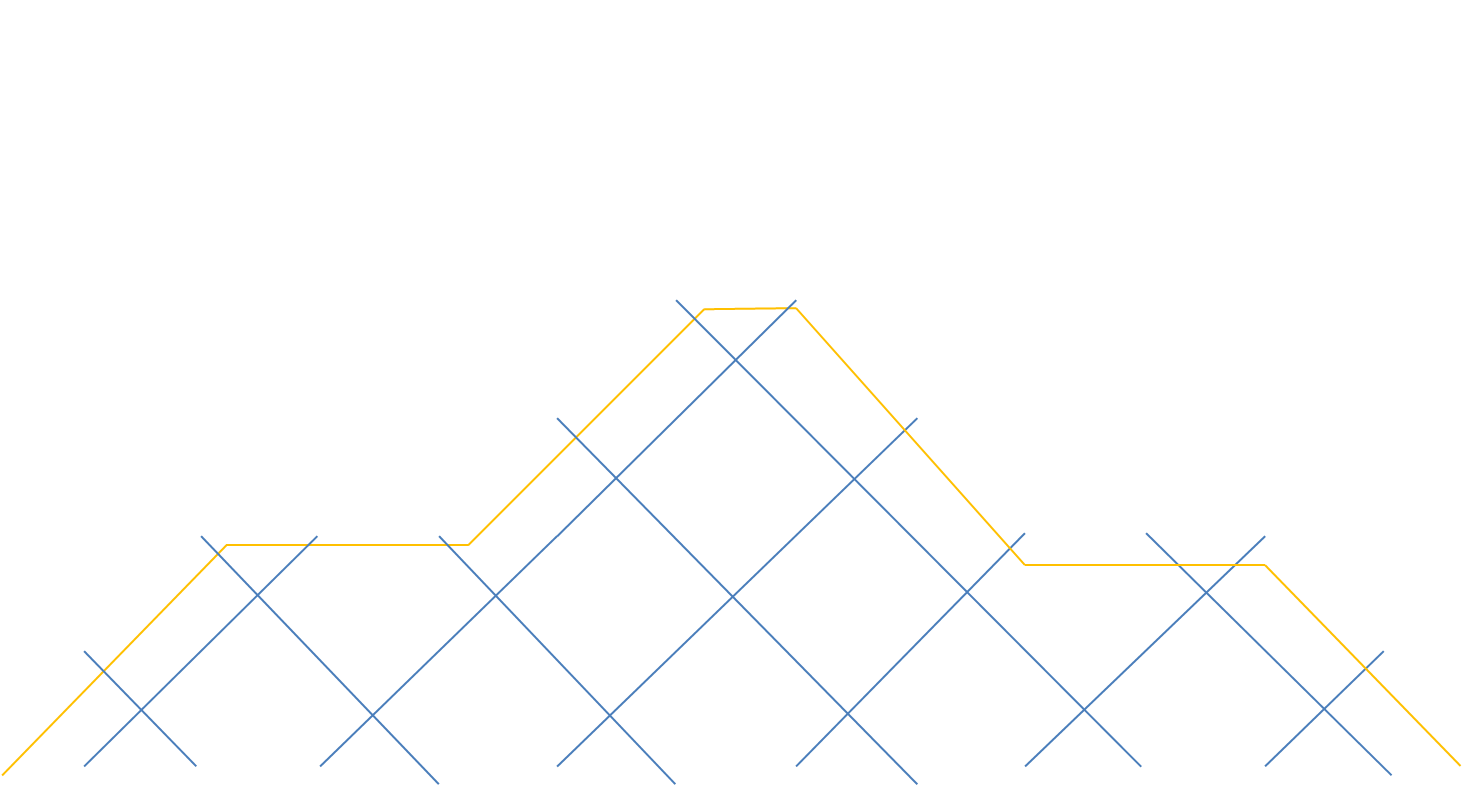}
		\caption{A typical Cauchy surface in the network.}
		\label{fig:cauchy}
\end{figure}

For an acyclic directed graph, it admits a slicing which is a set of edges, such that no any two are related causally, and that the surface does not have a boundary. Such a surface can generically be chosen as a Cauchy slice $\Sigma$. A typical Cauchy surface on the network is illustrated in figure \ref{fig:cauchy}.

A neighbourhood $N$ of a Cauchy surface $\Sigma$ can now be defined as picking up new edges connected to edges on the Cauchy surface by the tensors $T$. Since individual $T$ are all local unitaries, this defines
an algebra $\mathcal{U}(N)$ which is isomorphic to the algebra generated by the $C^*$-algebra on $\Sigma$ via the unitary maps $T$. This is also isomorphic to the algebra $\mathcal{U}(F)$, which is isomorphic to the $C^*$- algebra at any Cauchy surface $\Sigma$.

Algebras on different Cauchy surfaces are related also by isomorphism. The isomorphic map $\alpha_{1\to 2} : \mathcal{U}(\Sigma_1) \to \mathcal{U}(\Sigma_2)$ is now provided by the sets of $T$ contained between two Cauchy surfaces $\Sigma_{1,2}$ and this defines a unitary evolution.

\begin{figure}[!h]
		\centering
		\includegraphics[width=8cm]{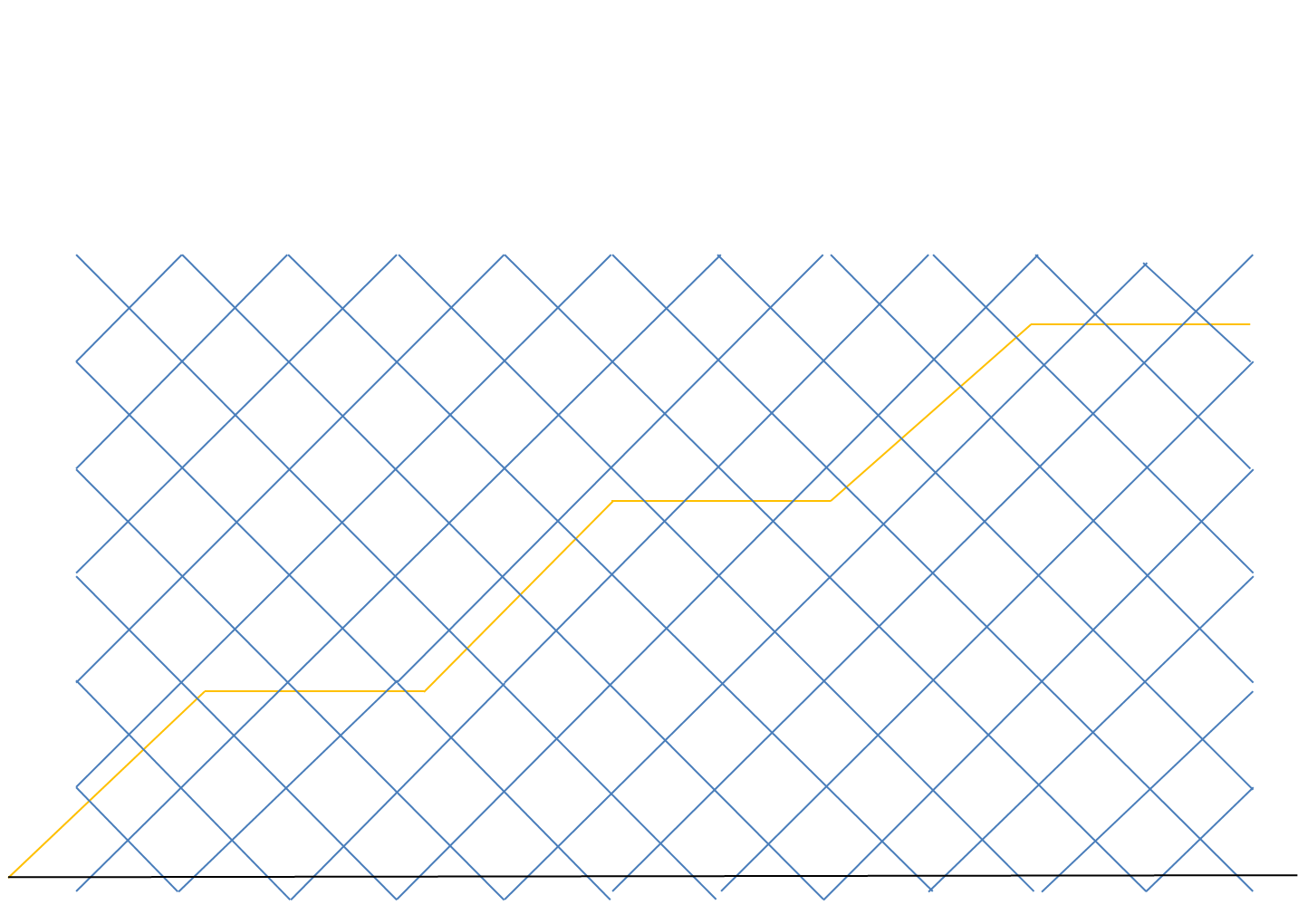}
		\caption{The set of networks between the two Cauchy surfaces : the horizontal Cauchy surface marked black, and the ladder like Cauchy surface marked yellow, is a unitary matrix, and we define that as the discrete approximation of a Lorentz transformation.}
		\label{Lor}
\end{figure}

That any two choices of Cauchy surfaces can be related by a unitary map is probably equivalent to the Stone-von Neumann theorem in finite dimensional Hilbert spaces. (See, for example, \cite{Wald:1995yp} for an explanation of the Stone- von Neumann theorem. )
Figure \ref{Lor} illustrates a unitary transformation between the horizontal surface and the slanted ladder-like surface. As we are going to see, in a homogenous network where every tensor is the same, this can be interpreted as an approximate Lorentz transformation.

A set of observers defining a ``frame'' is characterized by an ordered set of Cauchy surfaces, and these Cauchy surfaces are related by unitary transformations. If there is time translation invariance, it corresponds to the fact that the unitary matrices that map one Cauchy surface to the next remain unchanged.

\subsubsection{Einstein Locality} \label{Einstein_local}
To make actual comparison between two operator algebra $\mathcal{U}(O_{1,2})$
assigned to two different connected subgraphs $O_{1,2}$, we embed them into $\mathcal{U}(\Sigma)$ for any choice of Cauchy surface $\Sigma$, which defines an embedding map $\alpha_{1,2 \to F}:\mathcal{U}(O_{1,2}) \to \mathcal{U}(F)$ , since $\mathcal{U}(F)$ is itself isomorphic to $\mathcal{U}(\Sigma)$.

With that, we can define Einstein locality between two space-like separated regions $O_{1,2}$ using $\alpha_{1,2 \to F}$:
\be \label{einstein_causality}
[\alpha_1(\mathcal{U}(O_1)), \alpha_2(\mathcal{U}(O_2))] = 0.
\ee

As noted above, these maps $\alpha_i$ of operators are basically operators pushing across tensors $T$ along unitary directions. The above commutation relations can be phrased equivalently as follows. Consider
a Cauchy surface $\Sigma_1$ containing $S_1$. Now pull the operators in $O_2$ to $\Sigma_1$ which we denote by the map $\alpha_{2\to \Sigma_1}$. If $\alpha_{2\to\Sigma_1}(\mathcal{U}(O_2)) \subset \mathcal{U}(P_1)$, where $P_1\subset \Sigma_1$ and $S_1\cap P_1 = 0$, then (\ref{einstein_causality}) is satisfied.

Thus far, the tensor network falls short of being a "functor" mapping the category of graphs to {\bf OBS}. The reason is that it is not obvious what is the physical data that goes into defining a functor that maps different graphs to different {\bf OBS} that can be compared with a quantum field theory.  \footnote{When defining a quantum field theory, a standard procedure is to define a Lagrangian for some given set of fields. The program based on category theory defines a quantum field theory without using a Lagrangian. For a CFT, one needs a set of primaries, their conformal dimensions and their OPEs to completely specify the CFT. It is not completely clear to the authors what  the full set of data is that is needed to specify a generic QFT in this language. A tensor network probably has more data than are necessary. }

Note that to explicitly compute these commutators, we first construct Cauchy surfaces that contain each of these locations and then perform operator pushing of one operator in one of the Cauchy surfaces to the other surface.   The precise choice of the Cauchy surface is immaterial since the tensors only act locally, but existence of which is crucial. That would essentially rule out much potential confusion over whether two points are in fact connected causally.

\subsubsection{Isotony}

Isotony can be defined as follows in the tensor network. We will restrict our attention to subregions in spacetime describable by causal diamonds. The algebra attached to a causal diamond can be described as follows. We locate the space-like surface and the intersection of its causal future and causal past to define a causal diamond. The operator algebra associated to this region of space is defined as the operator algebra defined along the maximal space-like surface inside the causal diamond. Any space-like surfaces are thus related to the maximal space-like surface by isomorphisms.

Now if a causal diamond is inside another causal diamond, it is always possible to map the algebra on the space-like surface in the smaller diamond to the algebra on the maximal space-like surface in the larger diamond. The operator algebra is sure to be a sub-algebra of the algebra defined on the larger causal diamond.
Therefore, an algebra net is naturally recovered for causal diamonds.

Note that for a space-like region $A$ with a causal diamond $D(A)$, it follows that the entanglement entropy only depends on $D(A)$ but not on the specific maximal space-like surface $A$ chosen inside $D(A)$.  This is because all these maximal space-like surfaces inside $D(A)$ are all related by local unitaries which preserve entanglement entropy.

\section{Illustration based on free fermions}

In this section we would like to illustrate some of these ideas using a simple model.
We would like to construct a unitary evolution that is explicitly expressed as a tensor network of local unitaries. Then we would like to construct notions familiar in a continuous space-time in the present context, and show that they can be approximated to some extent.

\subsection{Tensor network evolution}

First of all, we have to define a model that is inspired by the free fermions but whose time evolution takes the form of a tensor network built from local unitaries. In fact, breaking up a generic unitary evolution into a product of local unitaries is frequently employed in numerical simulations or actual experiments on quantum simulations \cite{Peng:2005yg,Peng:2014kda}.
The Lieb-Robinson bound is also based on such an approximation \cite{Lieb:1972wy}. Here, however, we will take the viewpoint that the tensor network defines the model.

Our tensor network is constructed as follows.
Consider a set of fermion creation and annihilation operators $a_i, a^\dag_i$, where $i$ denotes the link where the fermion is located.
These operators satisfy the usual anti-commutator.
\be
\{a_{i},a_{j}^{\dagger }\}=\delta_{i,j}.
\ee

To construct the simplest example of a unitary evolution, we consider the quadratic Hamiltonian,
\be
H=\sum_{i=-2L+1}^{2L-1} h_{i,i+1}, \qquad  h_{i,i+1} =   \alpha\Big( a_{i}^{\dagger}a_{i+1}+ a_{i} a_{i+1}^{\dagger}\Big),
\ee
where $2L$ is the size of the spin chain. We label the vertices by $-L+\frac{1}{2}, -L+\frac{3}{2},\cdots,L-\frac{3}{2}, L-\frac{1}{2}$. Thus, the links run from $-2L+1$ to $2L-1$. We have already taken lattice spacing $a=1$ here. \par
To ensure that $H$ is Hermitian, we take $$\alpha=\frac{\tilde \alpha}{2 i}$$ for some real $\tilde \alpha$.

\begin{figure}[!h]
		\centering
		\includegraphics[width=7cm]{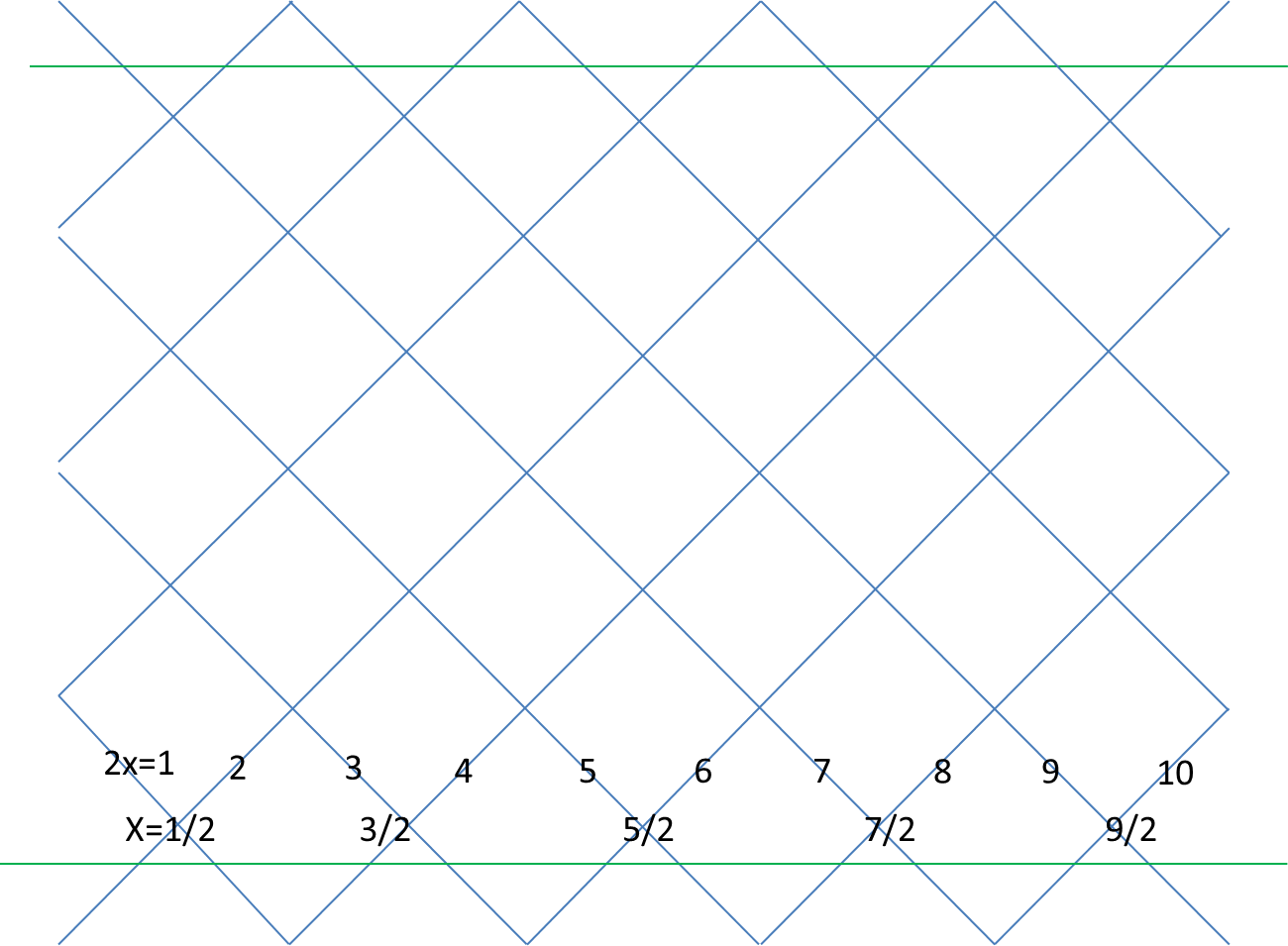}
		\caption{The collection of horizontal Cauchy surfaces, defining a particular set of ``inertial observers''. We note that we adopt a rather odd labelling for convenience with numerics, and take ``site'' numbers as half-integers $x$, while links are labelled by integers $2x, 2x+1$. }
		\label{horCau}
\end{figure}

The time evolution operator $U(\Delta t)$ over the smallest unit of time $\Delta t$ is given by
\be
U(\Delta t) = (\prod_i U_{2i, 2i+1} )\,\, (\prod_j U_{2j-1, 2j}),
\ee
where
\be
U_{i,i+1} = \exp( i \Delta t \,\,h_{i,i+1} ).
\ee
We have,
\be
U a_{2x}U^{-1}= c^2 a_{2x}+ c\,s\, a_{2x-1}-c\,s\,a_{2x+1}+ s^2a_{2x+2},
\ee
and
\be
U a_{2x+1}U^{-1}=c^2 a_{2x+1}-c\, s\,a_{2x+2}+c\,s\,a_{2x}+s^2 a_{2x-1},
\ee
where $$c= \cos(i \alpha\Delta t), \qquad s=\sin (i \alpha \Delta t),$$
and $x$ is the label of the ``site'' on a given Cauchy surface -- the vertex to which the links are connected.
For the sake of convenience in later (numerical) computations of correlation functions, {\bf  \it we have labeled these vertices by half-integers $x$, and the links $i$ by integers.
We therefore have $i=2x+1$ labelling ``even'' links, and $i=2x$ labelling ``odd'' links.} The labelling is illustrated also in figure \ref{horCau}.
Time evolution is given by repeated application of $U$. There is thus a time translation invariance over $t \to t+ 2\Delta t$.
The simplest set of Cauchy surfaces are those that are ``horizontal''.  We will take these collection of ``Cauchy slices'' to define an inertial frame.
The tensor network and the labelling are illustrated in figure \ref{horCau}.
\\

The network is invariant under translation by two links $i\to i+2$.
Therefore based on translation invariance, we expect that the eigen-modes in this inertial frame should be given by plane waves.
We will solve for the spectrum in this frame in the following.

\subsubsection{Spectra of the model}

From the translation symmetry of the network, we expect the eigen-operators to be expressible as
\be
a_{p}=\sum_{x=-L+\frac{1}{2}}^{L-\frac{1}{2}}\Big( f_{2x} a_{2x}+g_{2x+1}a_{2x+1}\Big),
\ee
for some coefficients $f_{2x}(p)$ and $g_{2x+1}(p)$ given by
\be
f_{2x}=q_{1} e^{i\,p\,x}, \qquad g_{2x+1}=q_2 e^{i\,p\,x},
\ee
where $q_{1,2}$ are independent of the site number $x$. They are to be determined by the following eigenvalue equations.
The momentum is given by
$$p=n\frac{2\pi}{2L},$$ where $n$ are integers satisfying $-L\leq n \leq L-1.$
In the limit $ L \to +\infty$, we can take $-\pi < p < \pi$.
The eigenvalue equation is obtained by demanding
\be
U\,a_{p}U^{-1}=\lambda\, a_{p}.
\ee
This implies
\begin{align}
\begin{split}
\label{ferm_evolve}
&f_{2x}\,c^2+g_{2x+1}\,c\,s+f_{2x-2}s^2-g_{2x-1}\,c\,s=\lambda\, f_{2x},\\&
g_{2x+1} c^2+g_{2x+3}s^2+f_{2x+2}\,s\,c-f_{2x}\,s\,c=\lambda\, g_{2x+1}.
\end{split}
\end{align}
For each given $p$, it gives two eigenmodes with eigenvalues
\be \label{dispersion}
\lambda_{\pm}=(c^2+s^2 \cos(p))\pm i \sqrt{(1-(c^2+s^2 \cos(p))^2)}
\ee
We note that the eigenvalues are even in the momentum $p$. In the limit $c\to 0$ the eigenvalue approaches
\be
\lim_{c\to 0} \lambda_{\pm}(p) = \exp( \pm i  |p|),
\ee
which appears as if its energy is linearly dependent on $p$. This recovers the dispersion of a relativistic free massless fermion without the doubling problem!
In the other limit where $c\to 1$, we have $\lambda(p)$ approaching a constant. The model becomes non-dispersive.
Therefore while the graph on which the model is defined remains unchanged, the effective causal structure of the system changes dramatically as the unit of time $\Delta t$ in the model is varied.

For $\lambda_{-}$, the corresponding eigen-modes are given by,
\be \label{modeminus}
\frac{q_2^-}{q_1^-}\equiv \frac{\alpha_2}{\alpha_1} =\frac{e^{\frac{i p}{2}} \left(s \sin (p)-\sqrt{(1-\cos (p)) \left(2 c^2+s^2 \cos (p)+s^2\right)}\right)}{2 c \sin \left(\frac{p}{2}\right)}.
\ee
Then for $\lambda_{+}$ the eigen-modes are,
\be \label{modeplus}
\frac{q_2^+}{q_1^+}\equiv\frac{\beta_2}{\beta_1}=\frac{e^{\frac{i p}{2}} \left(s \sin (p)+\sqrt{(1-\cos (p)) \left(2 c^2+s^2 \cos (p)+s^2\right)}\right)}{2 c \sin \left(\frac{p}{2}\right)}
\ee
In the above expressions, when we take the limit $c\rightarrow 0$, the case with positive $p$ and negative $p$ should be treated separately. We will take this limit in the subsequent analysis.

We can then define the operators corresponding to creation and annihilation of these eigen-modes.
\begin{align}
\begin{split}
&a_{p}=\mathcal{N}\sum_{x=-L+\frac{1}{2}}^{L-\frac{1}{2}} e^{i\,p\,x}\Big(\alpha_1 a_{2x}+\alpha_2 a_{2x+1}\Big)\,,\\&
b^{\dag}_{p}=\mathcal{N}\sum_{x=-L+\frac{1}{2}}^{L-\frac{1}{2}} e^{i\,p\,x}\Big(\beta_1 a_{2x}+\beta_2 a_{2x+1}\Big).
\end{split}
\end{align}
$$\mathcal{N}=\frac{1}{\sqrt{2L}}.$$
We check that,
\be
\{a_{p}, a^{\dagger}_{p}\}=1,
\ee
which implies,
\be
\alpha_{1}\alpha_{1}^{*}+\alpha_{2}\alpha_{2}^{*}=1.
\ee
Similarly for $b_{p}.$  This gives,
\be
\beta_{1}\beta_{1}^{*}+\beta_{2}\beta_{2}^{*}=1.
\ee
On the other hand, the equation (\ref{ferm_evolve}) implies
\be
\alpha_{1}\beta_{1}^{*}+\alpha_{2}\beta_2^{*}=0.
\ee
which means that
\be
\{a_{p},b^{\dagger}_{p}\}=0,
\ee
so $b^\dagger_{p}$ and $a_{p}$ behave as creation and annihilation operators respectively.  We can now define the fermion field as
\be
a_{2x}(x,t)=\mathcal{N}\sum_{n=-L}^{L-1} e^{-i\,p\, x}\Big(\frac{a_{p}e^{-i |\chi|t}}{\alpha_{1}\Big(1-\frac{\alpha_2 \beta_1}{\alpha_1\beta_2}\Big)}+\frac{b^\dagger_p e^{i |\chi|t}}{\beta_1\Big(1-\frac{\beta_2\alpha_1}{\beta_1\alpha_2}\Big)}\Big),
\ee
where
\be
\cos(\chi)=c^2+s^2 \cos(p).
\ee
Similarly,
\be \label{inversetrans}
a_{2x+1}(x,t)=\mathcal{N}\sum_{n=-L}^{L-1} e^{-i\,p\, x}\Big(\frac{a_{p}e^{-i |\chi|t}}{\alpha_{2}\Big(1-\frac{\alpha_1 \beta_2}{\alpha_2\beta_1}\Big)}+\frac{b^\dagger_p e^{i |\chi|t}}{\beta_2\Big(1-\frac{\beta_1\alpha_2}{\beta_2\alpha_1}\Big)}\Big).
\ee
We would like to define the vacuum state as,
\be
a_{p}|0\rangle=0,\qquad b_{p}|0\rangle=0.
\ee

In a discrete spacetime the notion of ``ground state'' is clearly ill-defined. If the smallest unit of time is $\Delta t$, then the energy is identified with a period i.e. $E \sim E+ 2\pi/\Delta t$.
However, mimicking the continuous situation we can look for a special state that corresponds to a ``separating vector'' -- a reference to distinguishing the ``positive/negative'' energy modes. Requiring that in the $p\to 0$ limit the ``low energy modes'' so defined by this state should recover our usual notion of low energy states (i.e. energy should decrease with decreasing momentum),  this special state would be a natural ``vacuum'' state. This is essentially the procedure that we have followed above.
We note that while a discrete evolution in Lorentzian signature does not identify a ground state without ambiguity, a Euclidean partition function would continue to project to a unique (unless otherwise degenerate) ground state.  One wonders how such an analytic continuation can be defined if time were discrete. The folklore that a quantum model in $d$ dimensions necessarily originate from a classical model in $d+1$ dimensions is by no means obvious when time becomes discrete. As we have seen above, variation of $\Delta t$ while keeping the Hamiltonian unchanged could lead to utterly different dispersion relations and subsequently causal structures.
It appears that at least in the case of integrable models however, an analytic continuation between Lorentzian and Euclidean signatures remains well defined.

\subsection{Lorentz transformation}

\begin{figure}[htbp]
\begin{center}
\begin{tabular}{cc}
\includegraphics[width=7cm]{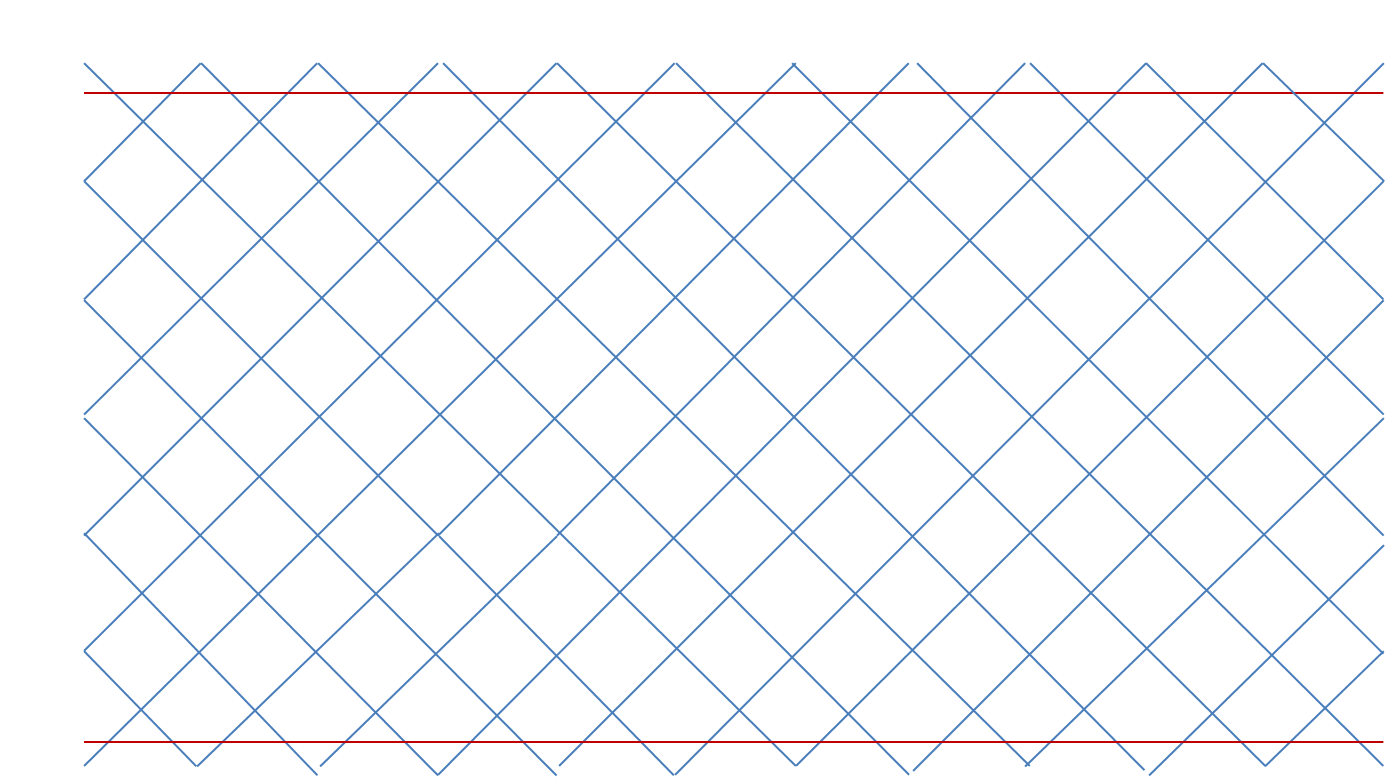}& \includegraphics[width=7cm]{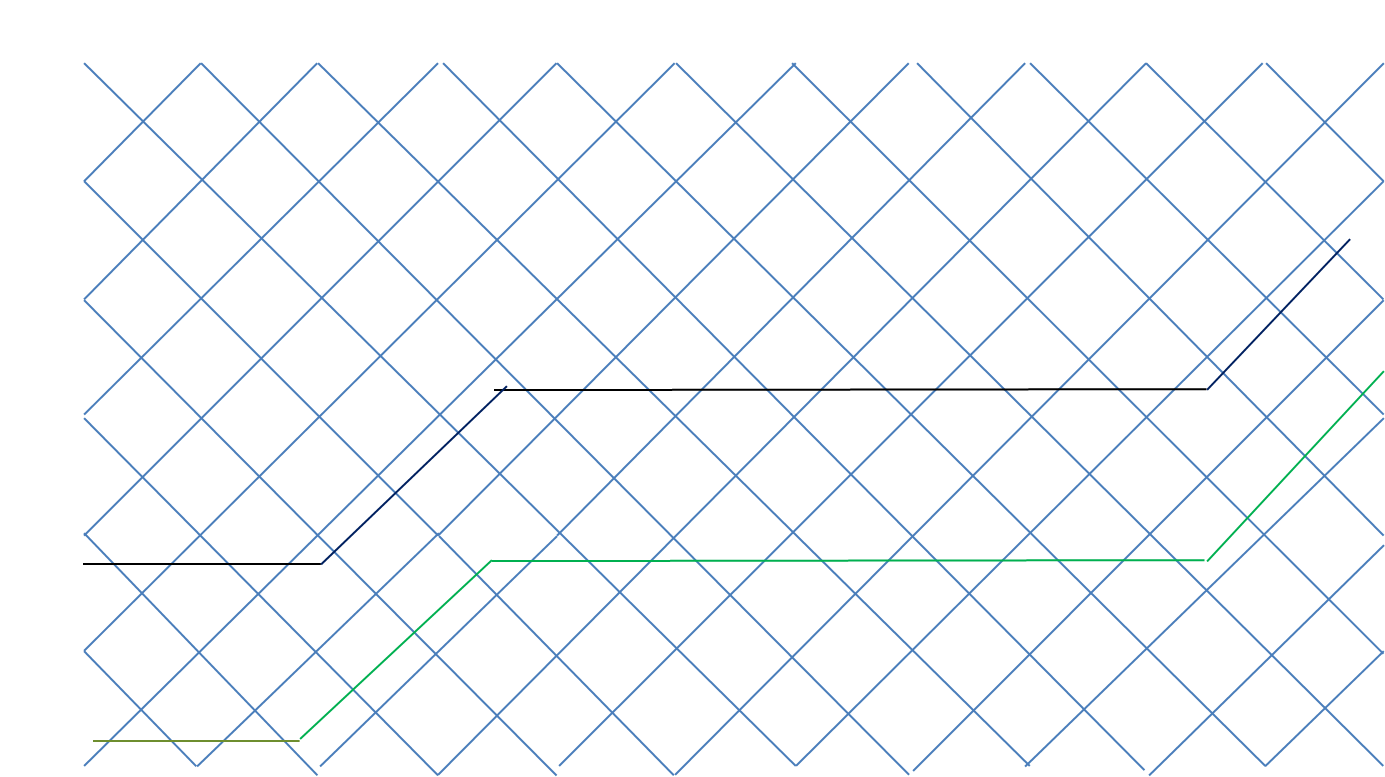} \\
(a) & (b)
\end{tabular}
\end{center}
    \caption{This diagram illustrates two independent sets of ``inertial observers'', marked by sets of Cauchy surfaces with different colours. The tensors sandwiched between each pair of Cauchy surfaces of the same colour correspond to the corresponding ``inertial time'' evolution between the pair.}
		\label{fig:inertialobservers}
		
\end{figure}

Since the system has a well defined causal structure, it is very tempting to define the notion of boost --i.e. the set of (unitary) transformations between different observers. As already explained in the previous section, an observer is defined as a collection of Cauchy surfaces. The unitary transformation connecting observers are generically given by the set of tensors sandwiched between the respective Cauchy surfaces. In general, such a transformation does not preserve the ``ground'' state. For a general set of observers, there is not even any notion of time translation invariance, such that energy is completely ill-defined. For our model, there are different observers that enjoy some degree of time translation invariance.
This is illustrated in figure \ref{fig:inertialobservers}

 If in the special case, that these different observers agree (perhaps only approximately) upon the notion of ``positive energy" modes, it means that the vacuum state remains invariant under the corresponding transformation between these observers. These observers that (approximately) share the same ground state would be the closest analogue we have for ``inertial observers'' in a continuous Lorentz invariant spacetime.

In the current tensor network we are working with, there are natural families of Cauchy surfaces corresponding to inertial observers, inspired by results in continuous space-time. Each family of Cauchy surfaces are slanted surfaces with some given slope. We note however that in a discrete space-time, these slanted surfaces with given slopes are more accurately speaking ladders. For a given slope, the corresponding Cauchy ``ladder'' is not unique, and we understand them as slightly different approximations of these inertial observers which only become indistinguishable in the long wavelength limit.

Boost transformation that relates these different observers are constructed below.

\subsubsection{Approximating the boost operator}

Consider the simplest scenario. There is a natural set of inertial observers with flat Cauchy surfaces. We consider another set of inertial observers over ladder surfaces.  We would like to construct the unitary boost operator that transforms the ladder into the flat surface. The most natural guess would correspond to the set of local unitaries sandwiched between the two Cauchy surfaces. However, after connecting the Cauchy surfaces, in general one can do further transformations that preserve the target Cauchy surface, such as translations on the surface. We therefore allow the boost operator to take the general form, corresponding to one that first translates along the time direction according to the sandwiched unitaries, which could be, depending on the situation at hand,  followed by a set of translations along the target Cauchy surface. This is the same strategy taken in \cite{Vidaltalk} in the Euclidean version when constructing transformations mapping between different (Cauchy) surfaces.

We can compute the effect these transformations have on our modes. To make the discussions most transparent and its similarity with the continuous case obvious, let us first consider the limit $c\to 0$, in which case, we recall that the dispersion relations (\ref{dispersion}) reduce to a linear one. That is directly analogous to the discussion of the free fermions in the $c\to 0$ limit in the main text.


We note that in the limit $c\to 0$, the fermion evolution (\ref{ferm_evolve}) behaves like a (fermionic) SWAP, in which even links are translated to the left by two steps  $2x +1 \to 2x-1$, and odd links are being translated to the right also by two steps $2x\to 2x+2$.  This suggests that the fermionic modes turn into a pair of chiral fermions propagating in opposite directions.
 In fact, one can directly check the expressions (\ref{modeminus}-\ref{modeplus}) to confirm that modes with positive momenta has positive (negative) energies for the odd (even) links.
 This is recapped below.
We list the eigen-operators explicitly as $c \to 0$. We note that the expression for the operator approaches a rather singular limit as $c\to 0$ where special care has to be taken when taking limits of (\ref{modeminus}-\ref{modeplus}), the correlation functions are relatively straight forward to treat. Nonetheless, it is possible to show that the modes reduce to
\begin{align}
\begin{split}
&a_{p<0} = -\mathcal{N} \sum_{x} e^{ i p x}e^{i\frac{p}{2}}a_{2x+1}, \qquad  a^\dag_{p<0} = -\mathcal{N} \sum_x e^{ -i p x}e^{-i\frac{p}{2}} a^\dag_{2x+1} \\
& b_{p\geq0} = \mathcal{N} \sum_x e^{-i p x}e^{-i\frac{p}{2}} a^\dag_{2x+1} , \qquad b^\dag_{p\geq0} = \mathcal{N} \sum_x e^{ i p x}e^{i\frac{p}{2}} a_{2x+1},
\end{split}
\end{align}
and then similarly
\begin{align}
\begin{split}
&a_{p\geq0} = \mathcal{N} \sum_{x} e^{ i p x}a_{2x}, \qquad  a^\dag_{p\geq0} = \mathcal{N} \sum_x e^{-i p x}a^\dag_{2x}, \\
& b_{p<0} = -\mathcal{N} \sum_x e^{-i p x} a^\dag_{2x}, \qquad b^\dag_{p<0} = -\mathcal{N} \sum_x e^{i p x} a_{2x},
\end{split}
\end{align}
The vacuum continues to be the one that is annihilated by $a_p$ and $b_p$.

The inverse transform would give
\be
a_{2x} = \mathcal{N}\sum_{p} e^{ -i p x} (a_{p\geq0}-b^\dag_{p<0}), \qquad  a_{2x+1} =  \mathcal{N}\sum_{p} e^{-i p x}e^{-i\frac{p}{2}} (-a_{p<0}+b^\dag_{p\geq0}).
\ee
In this case, we can inspect the effect the boost operator has on the operators $a_p$ and $b_p$.

\begin{figure}[!h]
		\centering
		\includegraphics[width=8cm]{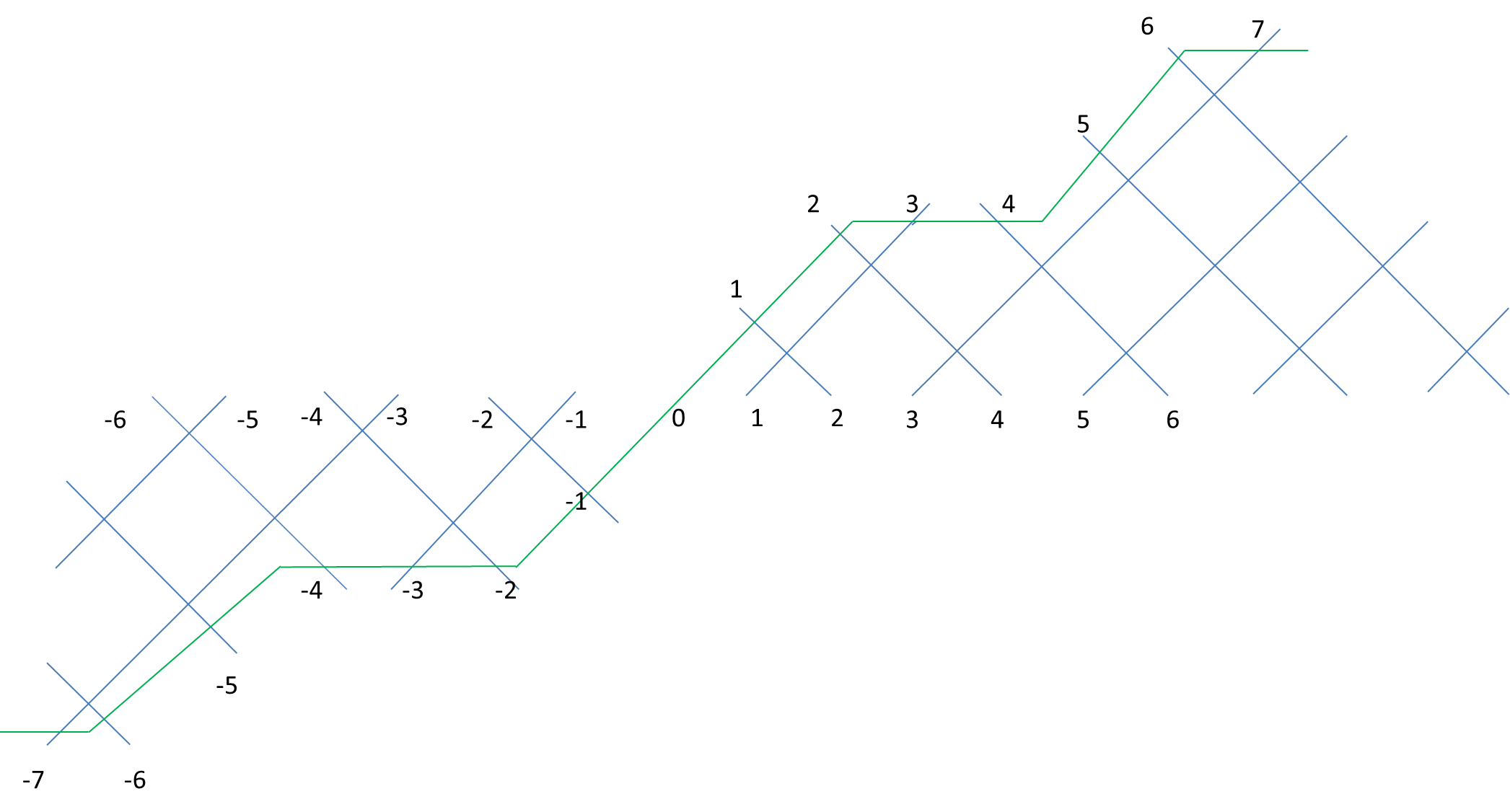}
		\caption{The explicit form of a particular boost operator}
		\label{fig:boost}
\end{figure}

The boost operator that we will illustrate in detail is shown in figure \ref{fig:boost}.
 \\	


The effect of the ladder operator $\hat B$ has on $a_{i}$ is as follows.
One can see that in figure \ref{fig:boost} for $i>0$, i.e.,$x=\frac{1}{2},\frac{3}{2},\cdots$,
\begin{align}
\begin{split} \label{boost}
&i= 2x, \qquad \Delta i = w(x+\frac{1}{2}),  \,\,\, \textrm{where}\,\, w=2 \\
&i= 2x+1,   \qquad   \Delta x =-\frac{w}{3}x-f(x),
\end{split}
\end{align}
where
\be
f(x) = \bigg\{\begin{array}{ccc}
\frac{2}{3},\quad & x-\frac{1}{2}=0 \ (\textrm{mod}\;3)\\
1,\quad &  x-\frac{1}{2}=1 \ (\textrm{mod}\;3)\\
\frac{1}{3},\quad &  x-\frac{1}{2}=2 \ (\textrm{mod}\;3)
\end{array}
\ee
For $i=0$, the operator $a_0$ is invariant under the transformation of the ladder operator $\hat{B}$. For $i<0$,
\begin{align}
\begin{split} \label{boostnegative}
&i= 2x,(x=-\frac{1}{2},-\frac{3}{2},\cdots) \qquad \Delta i = w(x-\frac{1}{2}),  \,\,\, \textrm{where}\,\, w=2 \\
&i= 2x+1,(x=-\frac{3}{2},-\frac{5}{2},\cdots)   \qquad   \Delta x =-\frac{w}{3}x-g(x),
\end{split}
\end{align}
where
\be
g(x) = \bigg\{\begin{array}{ccc}
-\frac{1}{3},\quad & x-\frac{1}{2}=0 \ (\textrm{mod}\;3)\\
0,\quad &  x-\frac{1}{2}=1 \ (\textrm{mod}\;3)\\
\frac{1}{3},\quad &  x-\frac{1}{2}=2 \ (\textrm{mod}\;3)
\end{array}
\ee

In general, we could consider also the effect of a link-dependent translation that takes
\be
i \to M i,
\ee
for some (odd) integer $M$ which would preserve the current Cauchy surface.\footnote{$M$ being odd ensures that even links and odd links remain decoupled for simple solutions. Also we
are taking the infinite size limit of the lattice for the transformation to take the same form everywhere. }  But in the current illustration, let us do without further deformation of the Cauchy surface, which does not appear to alter the physics of the final result.

Given the above expressions, one readily obtains the effect the boost operator has on the modes.
For positive links, using (\ref{boost}), we obtain
\begin{align}
\begin{split}
&\hat B a_{2x} \hat B^{-1}  = a_{2((1+ w/2)x+w/4)} =  \mathcal{N} \int_{-\pi}^\pi \frac{dp}{2\pi} \,\,e^{-i p((1+ w/2)x+w/4)} (a_{p\geq0}-b^\dag_{p<0}), \\
&\hat B a_{2x+1} \hat B^{-1} = a_{(2-\frac{w}{3})x+1-f(x)} .
\end{split}
\end{align}
Let us emphasize here that the apparent shift by $w/4$ above is an unfortunate result of our notation which takes $x$ to be half-integers.
The reader should be able to see from the figure \ref{fig:boost} that the odd links particularly, are simply scaled. i.e. $ 1\to 3, \,\, 3\to 7$ etc., with the steps scaling linearly with the link number. The even links behave similarly with a slight complication of some internal structure in the rescaling which is explained in equation (\ref{boost}) and easily recovered also from the same figure.
Since our transformation only makes sense in the thermodynamic limit where the momentum also approaches a continuous limit,  we replace the summation over $p$ by an integral over the first Brillouin zone. Note that the lattice spacing is just $a=1$ here.
For negative links, using (\ref{boostnegative}), we obtain
\begin{align}
\begin{split}
&\hat B a_{2x} \hat B^{-1}  = a_{2((1+ w/2)x-w/4)} ,\\
&\hat B a_{2x+1} \hat B^{-1} = a_{(2-\frac{w}{3})x+1-g(x)} .
\end{split}
\end{align}
To obtain the effect it has on individual $a_{p\geq0}$, we take another inverse Fourier transform to get
\begin{align}
\begin{split} \label{effect_of_boost}
\hat B a_{q\geq0} \hat B^{-1} =&    \mathcal{N}\sum_{x>0}e^{i q x}a_{2((1+ w/2)x+w/4)}+\mathcal{N}\sum_{x<0}e^{i q x}a_{2((1+ w/2)x-w/4)}\\
=&\mathcal{N}  \sum_{x>0} \, \int_{-\pi}^\pi \frac{dp}{2\pi} \,\, e^{i(q x-p((1+ w/2)x+w/4))} (a_{p\geq0}-b^\dag_{p<0})\\
&+\mathcal{N}  \sum_{x<0} \,\int_{-\pi}^\pi  \frac{dp}{2\pi} \,\, e^{i(q x-p((1+ w/2)x-w/4))} (a_{p\geq0}-b^\dag_{p<0}).
\end{split}
\end{align}
The above modes live only on the ``odd'' links.
Similar expressions can be obtained for $b_{p<0}$ and $a_{p<0}$ which depend on the ``even'' links.
Although one can expect from (\ref{boost}) that the result would be less clean.

The interesting observation is that in the thermodynamic limit where the total number of sites approach infinity, the sum over sites gives us a delta function $\delta(q- p(1+ \omega/2))$. For small $p>0$ and that $(1+\omega/2) p \le \pi$ ,
\be \label{boostap}
\hat B(\Lambda) a_p \hat B^{-1}(\Lambda) = a_{\Lambda^{-1} p}, \qquad \Lambda = e^{\eta},
\ee
where we have denoted $\Lambda=(1+ w/2) $.  We note that in the current example at hand $\Lambda=2$. We thus recover the expected transformation of the modes under a Lorentz boost. Importantly, the boost operator does not exchange annihilation operator for creation operation in this long wavelength limit.
For $\Lambda p > \pi$, one would have mapped a positive energy mode to a negative energy one.
In other words,  under the operation of our guess of a ``Boost'', it (almost, up to some lattice effect which disappears as $L\to \infty$ and $|p|$ sufficiently small i.e. continuous limit) preserves the ``positive'' and ``negative'' energy modes, which is a necessary ingredient of a Lorentz transformation in a Lorentz invariant theory that preserves the ground state.

Due to our choice of the form of the ladders, it does not treat the $p>0$ and $p<0$ modes in a completely symmetric manner, because the ladder breaks the symmetry between even and odd links. However, we associate such deviations with lattice effects.


\subsubsection{Spectra of the boost operator}

In the previous subsection, we made a crude guess of the form of the boost operator. To proceed with understanding the Unruh effect, we would like to solve for the spectrum of the boost operator, and obtain eigen-modes in the $x<0$ and $x>0$ halves of the lattice. These modes should behave like eigenmodes observed by a Rindler (accelerated) observer.

Since the boost operator does not act on the link at $i=0$, it essentially breaks up the space into two halves, the eigen-modes on the two halves of the lattice decouple.  Therefore, the form of the eigen-modes can be considered separately on the right and left half of the lattice.

We will continue to work with $c=0$ to illustrate the point, where even and odd links remain decoupled.

We begin with solving for eigen-operators on the right side of the lattice.
We first write down an ansatz for the eigenmodes. Since
\be
A^R_{\kappa} = \sum_{x>0} \psi_\kappa (x) a_{2x}.
\ee
Then we require that
\be
\hat B A^R_\kappa \hat B^{-1} = \eta_R (\kappa) A^R_\kappa.
\ee

This gives
\be
\sum_{x>0} \psi_\kappa(x) a_{2((1+ w/2)x+w/4)} = \eta_R (\kappa) \sum_{x>0} \psi_\kappa(x) a_{2 x}.
\ee

In momentum space, we have
\be
\sum_{x>0} \int_{-\pi}^{\pi} \frac{dp}{2\pi} (\psi_\kappa(x) e^{-i p((1+ w/2)x+w/4)}  - \eta^R(\kappa) \psi_\kappa(x) e^{-i p x}) \sigma_p =0 ,
\ee
where we are denoting
$$ \sigma_{p\geq0} = a_p, \qquad \sigma_{p<0} = -b_{p}^\dag$$
to avoid clutter.

Therefore the eigen wavefunction must take the form
\be
\sum_{x>0}  \int_{-\pi}^\pi \frac{dp}{2\pi} (\psi_\kappa(x)  e^{-i p((1+ w/2)x+w/4)}  - \eta^R(\kappa) \psi_\kappa(x) e^{-i p x}) = 0.
\ee

Strictly speaking, the sum over $x$ leads to a delta function in the momenta which is only defined up to $2\pi$. As we already noted, the boost operator only preserves the modes with long wavelength compared to the lattice spacing $a$. In the limit $a\to 0$, the limits of the $p$ integral would be replaced by $\pm \pi/a \to \pm \infty$.
Solutions to the equation are then given by the following:
\be \label{Rind_eigenval}
\tilde \psi^{\pm}_\kappa(p) = \Theta( \pm p ) p^{\kappa}e^{i\frac{p}{2}}, \qquad   \tilde \psi^\pm_\kappa(p)  \equiv \sum_x  \psi^\pm_\kappa(x) e^{- i p x}.
\ee
Both leads to the same eigenvalue
\be
\eta^R(\kappa) =  (1+ w/2)^\kappa.
\ee
Note that the $e^{i p/2}$ is from our fractional labels of sites $x=\frac{1}{2},\frac{3}{2},\cdots$. (In the $a\to 0$ limit this phase is really $e^{i p a/2 } \to 1$. We can neglect this factor when we consider the long wavelength mode.)
This is a solution in the rhs of the lattice where $x>0$. We note that $p$ can be both positive and negative here. Therefore, to recover the wavefunction in configuration space, we can in fact take an inverse transform, which gives
\be
\psi^\pm_\kappa(x>0)  =  \int_{-\infty}^\infty \frac{dq}{2\pi} e^{i q x} \tilde\psi^\pm_\kappa(q).
\ee
Now, one can check that
\be
\psi^+_\kappa(x>0) =  \int_0^\infty   \frac{dp}{2\pi} \,\, p^\kappa e^{i p x} =  (-i x)^{-\kappa-1} \frac{\Gamma(1 + \kappa)}{2\pi}.
\ee
Here, it is necessary that  $-1<\textrm{Re} \,\kappa<0$ while assuming $x$ is real. This computation is not regulated based on giving a small imaginary part to $x$ which would not have been a good regulator if $p<0$.
We would also like to check the result for $\psi^-_\kappa(x>0)$. It gives
\be
\psi^-_\kappa(x>0) =  \int_{-\infty}^0   \frac{dp}{2\pi}  \,\,p^\kappa e^{i p x} = (-1)^\kappa(i x)^{-\kappa-1} \frac{\Gamma(1 + \kappa)}{2\pi}.
\ee
This means that $\psi^-_\kappa (x>0) $ is proportional to $\psi^+_\kappa(x>0)$ and do not lead to a new eigenfunction.

We can work out the eigen-modes in $x<0$ in a similar manner.
As we mentioned above, the $x<0$ sector is the mirror image of the $x>0$ sector. (See figure \ref{fig:boost}.) This immediately suggests that we have a set of eigen-modes given by
\be
A^L_\xi = \sum_{x<0} \chi_\xi (x) a_{2x},
\ee
where $\chi_\xi(x)$ satisfies an entirely analogous set of eigenvalue equations (\ref{Rind_eigenval}).

We would then end up with the solutions
\be
\tilde\chi_\xi (p) \equiv \sum_{x<0} e^{-i p x} \chi_\xi (x),
\ee
where
\be
\tilde\chi^\pm_\xi(p) = p^\xi e^{-i\frac{p}{2}} \Theta(\pm p), \qquad \eta^L(\xi) = (1+w/2)^\xi.
\ee

Now this recovers
\be
\chi_\xi(x<0) =   \int_{-\infty}^\infty   \frac{dq}{2\pi}  \,\, e^{i q x} \tilde\chi_\xi (q).
\ee

An important question then arises. What values should $\kappa$ and $\xi$ takes?
Recall that we are taking these infinite ladders as a unitary evolution.
Therefore, $\kappa$ and $\xi$ should be so chosen such that we have a pure phase.

We therefore would like to have
\be\label{Renergy}
\eta^{R}(\kappa)= (1+w/2)^\kappa = e^{\eta\kappa} = e^{ - i \eta \epsilon}.
\ee
In other words, \be \kappa = - i \epsilon, \ee for positive $\epsilon$ would define positive energy modes.

Here for the $\chi_\xi(x)$ modes, by comparison with the discussion in continuous field theory, would require that we associate
\be \label{reverseRenergy}
\xi =   i \epsilon,
\ee
for positive $\epsilon$ to correspond to positive energy modes. We interpret here that the unit time evolution is evolved backward, and so $\Delta t$ has to take a negative value, thus inverting the definition of positive energies.
Here, we labor further upon some ambiguity that creeps in. There is not an exact translation symmetry with respect to the evolution between ladders of different slopes -- the analogue of evolution of Rindler observers here. As a result, there is not a precise concept of conserved energy (within the ``energy Brillouin zone'') here. But nonetheless, there is an approximate symmetry, so that we can still define $0\le \epsilon\ll1$ to be approximately conserved positive energy modes in this analogue ``Rindler'' frame. Correspondingly $-1\ll \epsilon\le 0 $ defines negative energy modes. That these ``Rindler Hamiltonian'' actually does approximate the entanglement Hamiltonian and thus bear actual resemblance to the continuous scenario with exact killing vectors would be discussed in numerical computations.

Summarizing, we have four sets of wavefunctions, two independent ones for the $x>0$ region and another two for the $x<0$ regions.
\be
\{ \psi^+_{\kappa = - i \epsilon}(x>0), \psi^+_{\kappa = i \epsilon}(x>0) \} \qquad  \{\chi^+_{\xi =  i \epsilon}(x<0), \chi^+_{\xi = -i \epsilon}(x<0) \},
\ee
where they correspond to pairs of positive and negative wavefunctions in each region for each $\epsilon$. We also note that $(\psi^+_{\kappa = i \epsilon})^*(x>0)$ is proportional to $\psi^+_{\kappa = - i \epsilon}(x>0)$ and that $(\chi^+_{\xi = -i \epsilon})^*(x<0)$ is proportional to $\chi^+_{\xi = i \epsilon}(x<0)$. Because in the same region, the wavefunction of negative energy mode should be the complex conjugate of the wavefunction of positive energy mode, we use $\psi^+_{\kappa = - i \epsilon}(x>0)$, $\chi^+_{\xi =  i \epsilon}(x<0)$ and their complex conjugate to define four sets of eigen-operators
\begin{align}
\begin{split}
A^R(\epsilon) &=\sum_{x>0} \psi^+_{-i\epsilon}(x) a_{2x}, \qquad {B^R}^\dag(\epsilon) = \sum_{x>0} (\psi^+_{-i\epsilon})^*(x) a_{2x}, \\
A^L(\epsilon) & = \sum_{x<0} \chi^+_{+ i \epsilon}(x) a_{2x}, \qquad {B^L}^\dag(\epsilon) = \sum_{x<0} (\chi^+_{+i\epsilon})^*(x) a_{2x},
\end{split}
\end{align}
where we have assumed that $\epsilon$ is positive. The complex conjugates of each operator can be obtained from the above.
Entirely analogous expressions, albeit looking less clean and simple, can be obtained for modes on the even links, which we will not dwell on in detail here.

Before we move on to a detailed discussion of the Unruh effect in the current context, we note that the boost operator constructed from an ever rising sequence of stairs are crucial towards the decoupling between left and right moving modes which can be thought of as arising from a ``Rindler horizon'' near the origin. Initially we attempted to approximate this effect with a stair with only 1 step. This is equivalent to solving a semi-infinite system with a fixed boundary condition at one end.  One can show that there is always reflection at $x=0$, leading to very different  physics.

\subsubsection{Approximating the Unruh effect}

Having worked out the eigenfunctions, we can follow a very similar path as the continuous case and look for the Bogoliubov transformation that connects eigen-modes of the ``inertial observers'' and the ``Rindler observers''.

Equation (\ref{Rind_eigenval}) is actually supplying the Bogoliubov transformations between modes in ``inertial'' frames and ``Rindler'' frames.

We would like to express $a_p$ in terms of $A^{R}(\epsilon)$ and ${B^L}^\dag(\epsilon)$. We note that
\be
\chi^{+\,*}_{i\epsilon} (x<0) = \int_{-\infty}^\infty \frac{dp}{2\pi} \,\, p^{-i \epsilon} e^{- i p x} \Theta(p) = (-1)^{1 - i\epsilon} \psi^+_{-i\epsilon} (x<0).
\ee
i.e. In the last equality, the mode correspond to $\psi^+_{i\epsilon}$, but analytically extended in the region $x<0$.
This gives
\be
A^R({\epsilon}) + (-1)^{i\epsilon-1} (A^L(\epsilon) )^\dag =  \int_{-\infty}^\infty \frac{dp}{2\pi} \sum_{x} \Theta(p) p^{-i\epsilon} e^{i p x } a_{2x} =   \int_{0}^\infty \frac{dp}{2\pi} p^{-i\epsilon} a_p.
\ee
Here, one has to make a choice of the branch-cut. If we make a choice that is
\be (-1)^{i\epsilon} =  \exp(i\pi * (i) \epsilon) =  \exp(-\pi\epsilon).
\ee
one would recover the standard result that the temperature for the half-space observer is given by \footnote{We note that the temperature is measured against our units of time which has been set to 1. }
\be
\hat T = \frac{1}{2\pi}.
\ee

This can be compared with the standard result for relativistic field theories (in 1+1 dimensions) at vanishing mass. A detailed computation can be found for example in \cite{socolovsky}
where we have,
\be
\alpha^{R}_{l>0} = \int_0^\infty \frac{dk}{2\pi} \frac{l}{k} (a_k f_l(k) + a_k^\dag f_l(-k)), \qquad f_l(k) =  k^{i l }  a^{- i  l -1} e^{\pi l /2}  \Gamma[- i  l],
\ee
where $\alpha^R_l$ are (positive energy bosonic) Rindler modes, and $a_k$ are inertial modes. The speed of light and acceleration has been set to 1.
The solution for $f_l(k)$ is, up to normalization, precisely what we have found in (\ref{Rind_eigenval}). The fact that the modes are bosonic only changes the analysis concerning normalizations. Otherwise the relationships between the modes are a direct result of Fourier transforms, and the plane wave solutions are shared between the 1+1 dimensional fermions and bosons.

What makes our analysis interesting is that the same physics is recovered purely based on the discrete recursion relations following from the discrete boost operator. We have made no reference to killing vectors or solving the Dirac equation in Rindler coordinates.

The even links can be treated in a similar manner, although the result would not have looked as clean due to our breaking of symmetries between even and odd links.

\subsection{Comparison with half-space entanglement}

What we have demonstrated in the previous subsection is that we can construct an analogue of a boost operator. Its effect on the modes in the inertial frame is very much analogous to  the expected form in the continuous case -- $a_p \to a_{\Lambda^{-1} p}$, and this action (approximately) preserves the notion of ``positive'' and ``negative modes'' that we have defined in the discrete space-time.

Then we showed that eigen-modes of this boost operator have very similar forms to the Rindler modes in continuous field theories. We demonstrated this explicitly at special values of the parameters, and showed that they closely approximate modes solved in Rindler space, and subsequently define an analogous Bogoliubov transformation between ``inertial'' and ``Rindler'' modes.

What we have not demonstrated however, is that the boost operator that we have defined is indeed the same as the entanglement Hamiltonian controlling half-space entanglement. In the case of continuous field theory with Lorentz invariance, this is guaranteed by the Bisognano-Wichmann theorem \cite{Bisognano:1976za} ( see a review for example in \cite{Haag:1992hx}). The construction relies heavily on the analytic continuation of the boost operator that rotates $t\to -t$ and $x\to -x$. As we are going to discuss in later sections inspecting more generic integrable models, such an analytic continued operator indeed exists and can be readily constructed ? it is the Euclidean version of the ?corner-transfer-matrix?.

In this section however, we will compute explicitly the half space entanglement Hamiltonian in the fermionic model for different values of $c$, and demonstrate that the modular Hamiltonian does resemble the guess of a boost operator we made in the previous subsection, therefore adding extra evidence that the boost operator that we have constructed does indeed control the half-space entanglement. In that case, the evolution generated by our boost does correspond to evolution of observers that approximately have no access to half of the space-time, closely resembling Rindler observers.

\subsubsection{The computation of the entanglement Hamiltonian}
The entanglement Hamiltonian, which is also called modular Hamiltonian, is defined by taking the log of the reduced density matrix. In the case of free fermions, given that the vacuum state we have chosen has zero fermion number, the reduced density matrix takes the following form:
\be
\rho_V=K e^{-\mathcal{H}}=K e^{-\sum_{V}\mathcal{H}_{ij}a_i^\dag a_j},
\ee
where $\rho_V$ is the reduced density matrix of region $V$, $\mathcal{H}$ is the entanglement Hamiltonian of region $V$. The normalization constant is specified as $K=\det(1+e^{-\mathcal{H}})^{-1}$.
In \cite{Casini:2009sr}, the entanglement Hamiltonian is calculated by correlation functions of local creation and annihilation operators $a_i^\dag$, $a_j$, which satisfy $\{a_i,a_j^\dag\}=\delta_{ij}$. The correlation functions are defined as
\be
\langle0|a_i a^{\dagger}_j|0\rangle=C_{ij}, \qquad \langle0|a^{\dagger}_j a_i|0\rangle=\delta_{ij}-C_{ij},
\ee
\be
\langle0|a_i a_j|0\rangle=\langle0|a^{\dagger}_i a^{\dagger}_j|0\rangle=0.
\ee
The entanglement Hamiltonian in matrix form is given as \cite{Casini:2009sr}
\be
\mathcal{H}=-\log (C^{-1}-1),
\ee
where $C$ is the matrix of correlation functions $C_{ij}$ in region $V$.\\

\begin{figure}
\begin{center}
\begin{tabular}{cc}
\includegraphics[width=7cm]{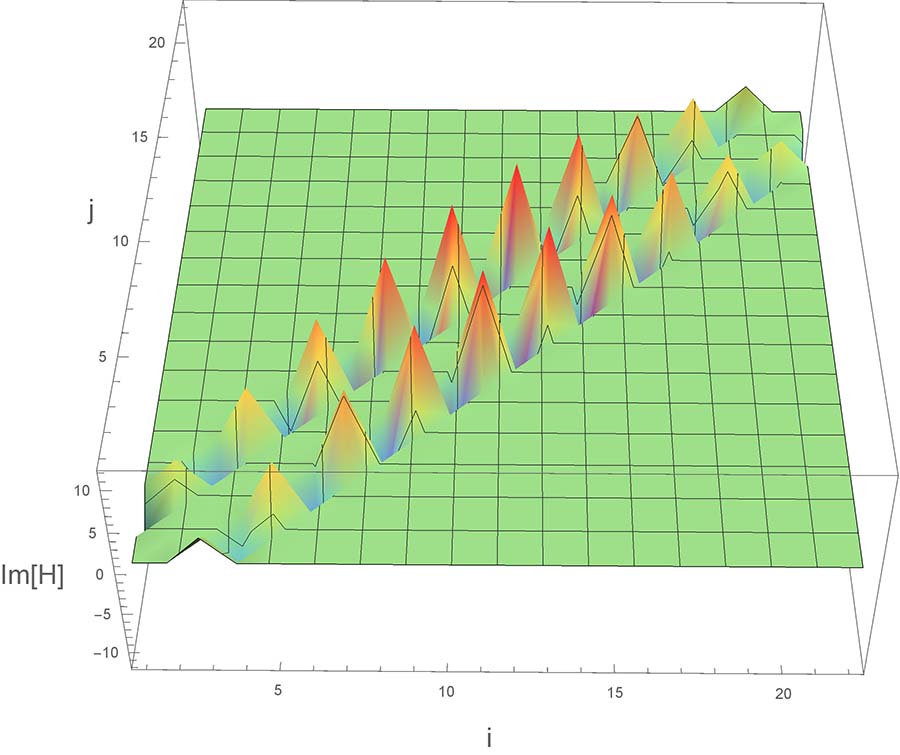}& \includegraphics[width=7cm]{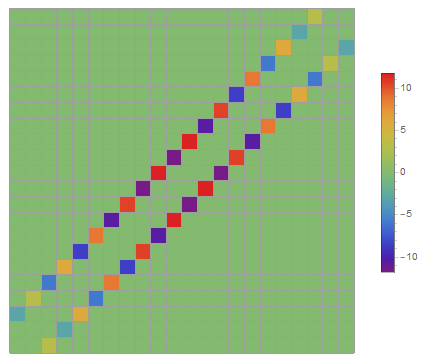} \\
(a) & (b)
\end{tabular}
\end{center}
\caption{Plots of the real part of the coefficients $(a^\dag_{i} a_{j})/i$ terms in the modular Hamiltonian with $c=0.000001$: The dominant contributions correspond to pairs of creation and annihilation operators located at positions $|i-j|=2$, where $i,j$ are the labels of links.  Figure (b) is the vertical view of figure (a) and it describes the values by colour. The parameter $c$ is chosen to be $0.000001$.
We note that neighbouring peaks appear to take opposite signs. That is because even and odd sites decouple, and their respective peaks are out of phase. }
\label{imaginary-part-modular-H-L-11-c-equals-0-000001-chop}
\end{figure}

We demonstrate a plot of the modular Hamiltonian in figure \ref{imaginary-part-modular-H-L-11-c-equals-0-000001-chop} for $c\to 0$ which is the case considered in detail in the previous section.\footnote{In our analysis, we take region $V$ as the links connected with sites labeled from $1/2$ to $L-1/2$, i.e. links from $1$ to $2L$.}

These plots suggests that the entanglement Hamiltonian takes the form
\be \label{Hc0}
\mathcal{H} = \frac{\mathcal{N}}{i} \sum_x \,\,  (\frac{ x (L - x)}{2L} + \nu_0 )(a_{2x}^\dag a_{2x+2} - a_{2x+2}^\dag a_{2x}) + (\textrm{even sublinks}) \cdots, \, \nu_0 \sim \frac{1}{10}
\ee
for $c\to 0$.

This operator in (\ref{Hc0}) have been discussed before in the literature for entanglement Hamiltonian of ground state lattice fermions with continuous time. In the computation of entanglement entropy of lattice fermionic Gaussian states, it is observed that the entanglement Hamiltonian always commute with an operator of the above form \cite{2004JSMTE..06..004P, 2009JPhA...42X4003P}.  Here, we are observing that it is itself a fair approximation of the entanglement Hamiltonian, at least in the $c\to 0$ or $c\to 1$ limit. The intermediate values of $c$ has extra complications as the translation symmetry of the ``ground state'' changes.

\begin{figure}[!h]
		\centering
		\includegraphics[width=10cm]{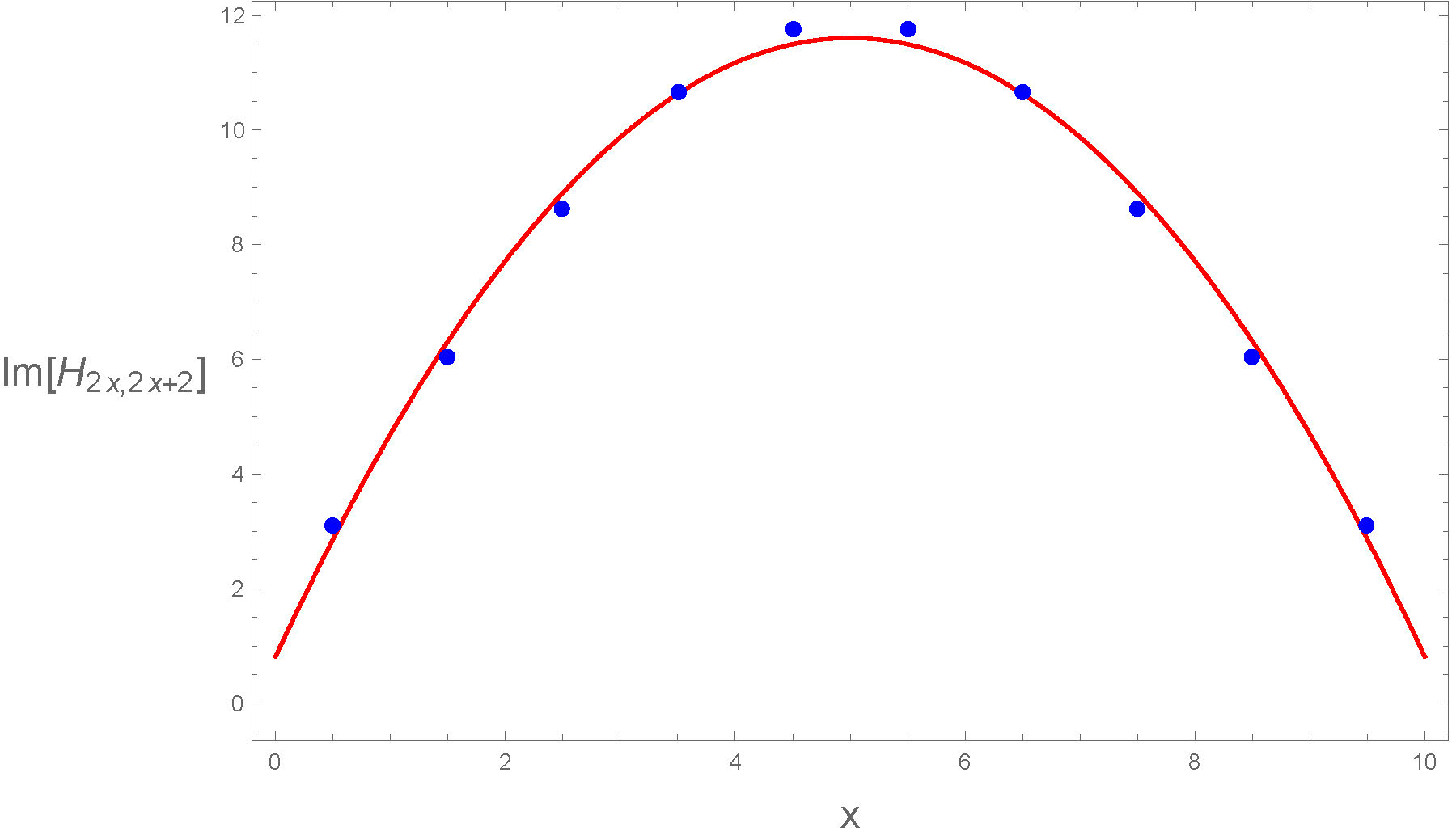}
		\caption{Fitting of peaks of imaginary part of modular hamiltonian with $c=0.000001$ and $L=11$: In this figure, we fit the peaks of imaginary part of the modular hamiltonian with a quadratic polynomial $a x^2+b x+f$, where $2x$ is the label of a link. The fitting coefficients are $a=-0.431484, b=4.31484, f=0.817762$. The parameter $c$ is chosen to be $0.000001$ and the lattice size parameter $L$ is chosen to be $11$.}
		\label{modular-fit-L-11-c-0-000001}
\end{figure}

We show the fitted curve on top of the plot of the modular Hamiltonian in figure \ref{modular-fit-L-11-c-0-000001}, showing that (\ref{Hc0}) is a close approximation of the modular Hamiltonian.

When we are sufficiently close to the entanglement boundary, one could see that it is approximately taking the form
\be \label{HEferm}
\mathcal{H} \approx  \frac{\sigma}{2i} \sum_{n} n\, (a_n^\dag a_{n+1} - a_{n+1}^\dag a_n)
\ee
for some constant $\sigma$ and
for $n$ sufficiently small and close to the boundary of the entanglement cut. We reckon this closely resembles the construction of the finite boost operator that we constructed, where the amount of time evolution grows linearly with the distance from the entanglement cut located at $x=0$. In a continuous Lorentz invariant theory, the vacuum entanglement Hamiltonian of half-space is famously known to be given by the boost operator which is given by $K= \int dx\, x T_{00}$ at the $t=0$ slice.  Therefore, the tensor network is naturally recovering this result albeit only approximately. Our naive guess of the boost appears to match the actual entanglement Hamiltonian reasonably well, for different values of $c$, not restricted only to $c\to 0$, even though we solved the spectra of the boost operator only in that limit.

In the discussion below, we will inspect the commutation relations between the entanglement Hamiltonian and the creation/annihilation operators $a_p$, and show that the algebra closely resembles our naive guess.


\subsubsection{Computing the algebra of the entanglement Hamiltonian}
The half-space entanglement Hamiltonian is only defined within the positive half line. We can complete the operator into one that acts on all of space by considering $\mathcal{H}-\mathcal{\bar{H}}$ where $\mathcal{\bar{H}}$ is the entanglement Hamiltonian of the complement of the region corresponding to $\mathcal{H}$, in this case therefore, the ``negative half region''.  We consider the commutator $[\mathcal{H}-\mathcal{\bar{H}},a_p]$, i.e. links connected on sites $x=-L+\frac{1}{2},\cdots,-\frac{1}{2}$.

To make comparison with the boost operator that we constructed explicitly, we will particularly try to obtain the algebra of the entanglement Hamiltonian obtained above for $c\to 0$. Also,  in the limit that $L$ is very large, we assume that we can ignore the corrections to the entanglement Hamiltonian that arises from the other entangling boundary in a finite system.
\footnote{There is subtlety with boundary conditions when dealing with the infinite system limit. In obtaining the reduced density matrix of any segment in a finite system, no matter how large the subsystem is taken to be within an even larger lattice, the pattern of the entanglement Hamiltonian is always sensitive to both boundaries. Our discussion here therefore has to be taken as a sanity check that the emerging pattern does have a set of recognizable physics.}

In this limit, the entanglement Hamiltonian therefore takes the form
\be \label{completeH}
\Gamma \equiv (\mathcal{H} -  \mathcal{\bar H})_{c\to 0, L \to \infty} =  \frac{\mathcal{N}}{2i} \sum_{x=-\infty}^\infty (x - \nu) (a_{2x}^\dag a_{2(x+1) }- a_{2(x+1)}^\dag a_{2x}),
\ee
for some appropriate normalization $\mathcal{N}$, and $\nu$ is some constant.

We thus have
\be
[\Gamma, a_p] = \mathcal{N} \sum_{x=-\infty}^\infty \,\, a_{2x} e^{i x p}( x\sin p  + \frac{1}{2i}  ((1-\nu) e^{-ip} + \nu e^{ip} ) ).
\ee

The $x \sin p$ term suggests that  $\Gamma$ acts as $ p \partial_p a_p$ in the leading long wave-length limit, which is, up to a constant shift $\nu$, the infinitesimal version of the boost transformation that we introduced in (\ref{boostap}).

%


\subsection{(Anti-)Commutation relations and Correlation functions -- visualizing the light cone}


As a further check to the emergence of Lorentz invariance particularly in the $c\to 0$ limit in which the Unruh effect can be solved analytically, we would like to inspect both the (anti-)commutation relations and two point correlation functions evaluated wrt the ``ground state'' that we have defined here.
We would like to inspect the causal structures intrinsic to these tensor network and their manifestation in the (anti-)commutation relations and correlation function.
It is also instructive to see how the boost symmetry manifests themselves in the correlation functions, at least in some limits such as $c\to 0$ and $c\to 1$.

The anti-commutation relations and correlation functions can be readily obtained. We relegate some of the details to the appendix, while keeping only the main results.
The complete expressions are presented in the appendix.

\begin{figure}
\begin{center}
\begin{tabular}{cc}
\includegraphics[width=6.5cm]{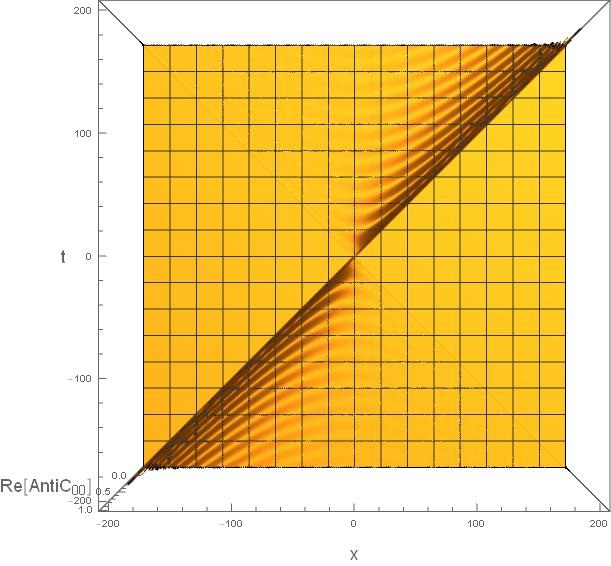}& \includegraphics[width=6.53cm]{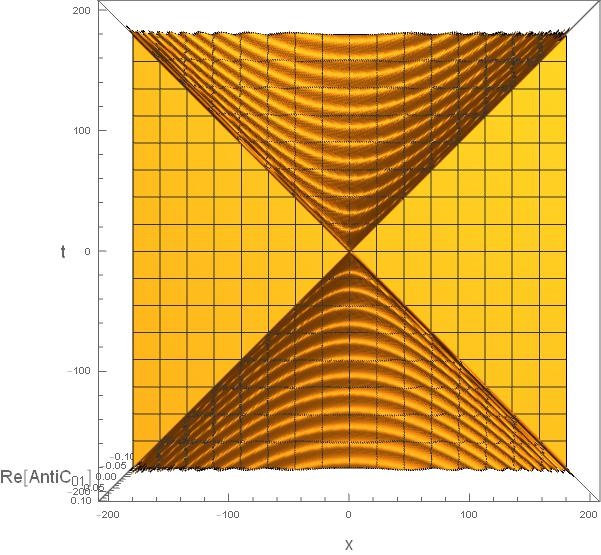} \\
(a) & (b)\\
\includegraphics[width=6.5cm]{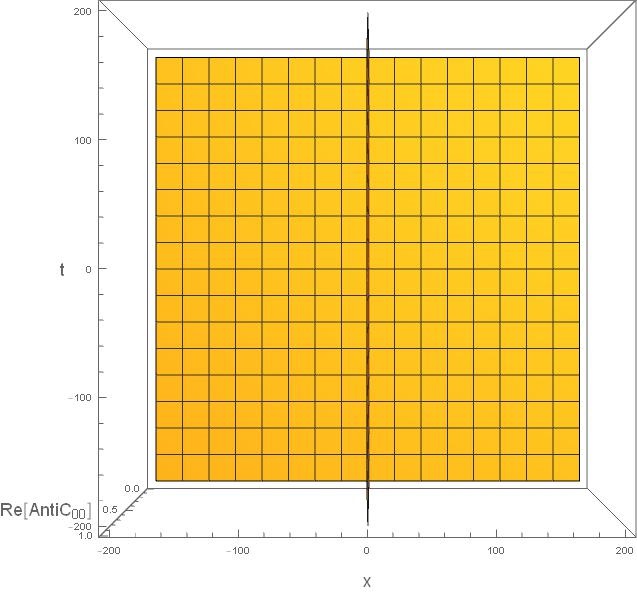}& \includegraphics[width=6.54cm]{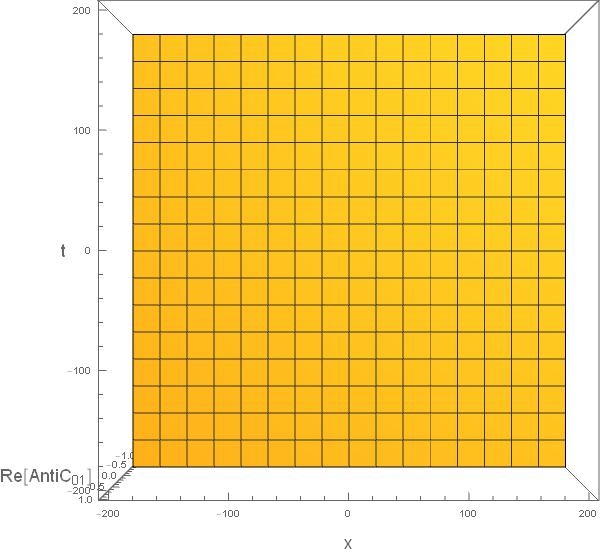}\\
(c) & (d)\\
\end{tabular}
\end{center}
\caption{These are anti-commutators of fermions. (a) Anti-commutators $\{a_{2x}(x,0),a_{2y}^\dagger(0,t)\}$ at $c=0.1$. (b) Anti-commutators $\{a_{2x}(x,0),a_{2y+1}^\dagger(0,t)\}$ at $c=0.1$. (c) Anti-commutators $\{a_{2x}(x,0),a_{2y}^\dagger(0,t)\}$ at $c=1$. (d) Anti-commutators $\{a_{2x}(x,0),a_{2y+1}^\dagger(0,t)\}$ at $c=1$, which is 0 in the whole spacetime. }
\label{anticommutator}
\end{figure}

(Anti-)Commutation relations are the most effective measure of causal structure, which is part of the Einstein locality axiom as discussed in section \ref{Einstein_local}.
In figure \ref{anticommutator}, we can see that the anti-commutators are exactly vanishing outside of the (effective) light cone for two choices of parameters corresponding to $c\to 0$ which displays emergent Lorentz invariance, and $c\to 1$ where the model becomes non-dispersive. At $c=0$, since the dispersion relation is linear, the invariance under boost translates into an invariance of the anti-commutators under a simple transformation in $(x,t)$ coordinates: namely the Lorentz transformation. The hyperbolas marking ``equipotential'' lines are clearly visible in figure \ref{anticommutator}. There can be more symmetries in the dynamics than the graph would have suggested for special choices of these evolution tensors.
As $c\to 1$ the evolution is non-dispersive with the light cone closing up. While the topology of the graph determines the maximal size of the light-cone, their effective size depends on the actual evolution dictated by the choice of tensors populating the tensor network graph.

We have also looked into correlation functions which are contained in the appendix.
There are various special limits in which the results are particularly illuminating. Again we focus on the $c\to0$ limit, where each tensor behaves like a swap between the neighbouring fermionic sites.
We reproduce here the limit where $c\to0$.
\begin{align}
\begin{split}\label{c0lim}
\langle0|a_{2x}(x,0)a^{\dagger}_{2y}(y,t)|0\rangle &=\mathcal{N}^2\sum_{p\geq0}e^{-i\,p\,(x-y)}e^{i\,|p| t}, \\
\langle0|a_{2x+1}(x,0)a^{\dagger}_{2y+1}(y,t)|0\rangle&=\mathcal{N}^2\sum_{p<0}e^{-i\,p\,(x-y)}e^{i\,|p| t}.
\end{split}
\end{align}

It is noteworthy that in (\ref{c0lim}), only half of the modes contribute. It is a direct consequence that in this limit, half of the links are decoupled from each other. Half of the links contribute to modes moving right i.e. have ``positive energy'' for $p>0$;  the other half contribute to left moving modes, with $p<0$ having positive energies. At the other end, where $c\to 1$, the time dependence completely drops out. The ground state however, is chosen to have non-trivial entanglement, since it is still annihilated by half of the momentum eigenmodes taken as annihilation operators, as their eigen-energies tend to $0^-$ in the non-dispersive limit.

We note also that when we further take the  limit $L\to \infty$ in the $c\to 0$ regime, the correlation functions reduce to the result of a free massless fermion in $1+1$ d.

\begin{figure}
\begin{center}
\begin{tabular}{cc}
\includegraphics[width=6.5cm]{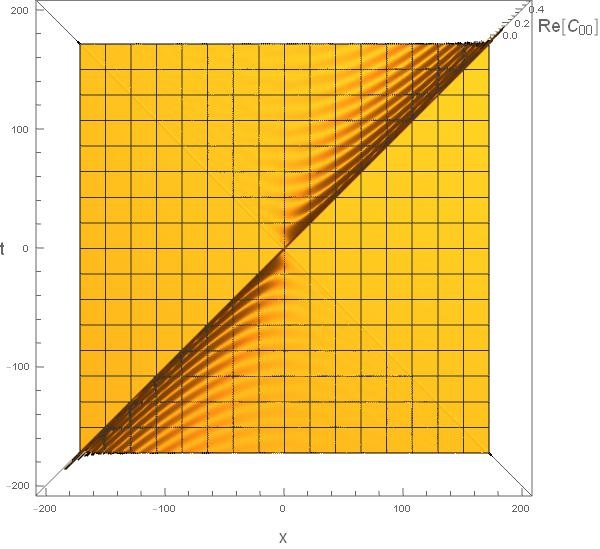}& \includegraphics[width=6.5cm]{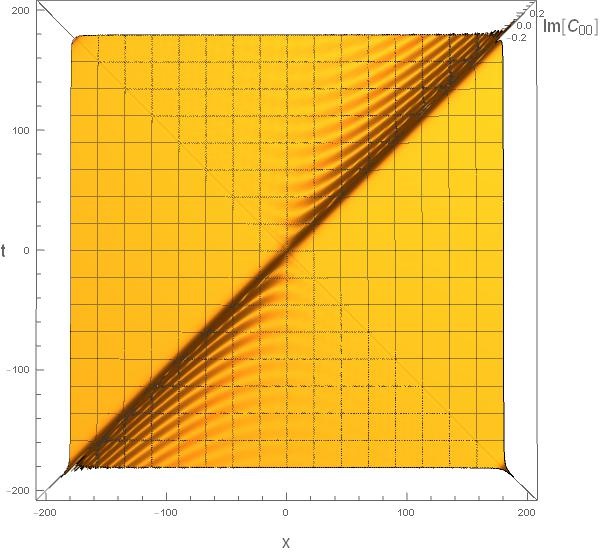} \\
(a) & (b)\\
\includegraphics[width=6.5cm]{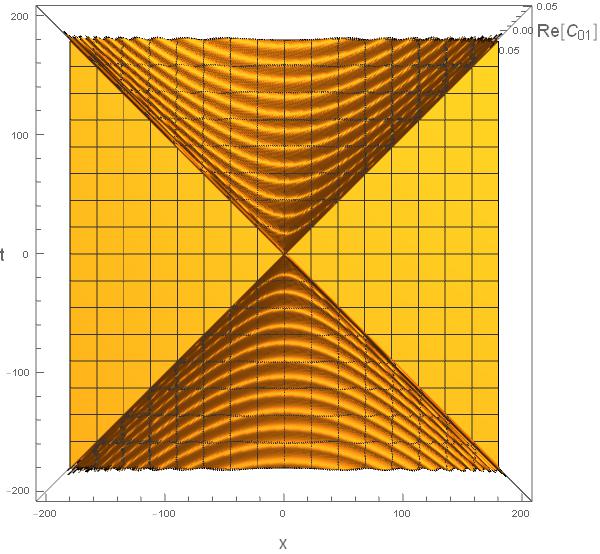}& \includegraphics[width=6.5cm]{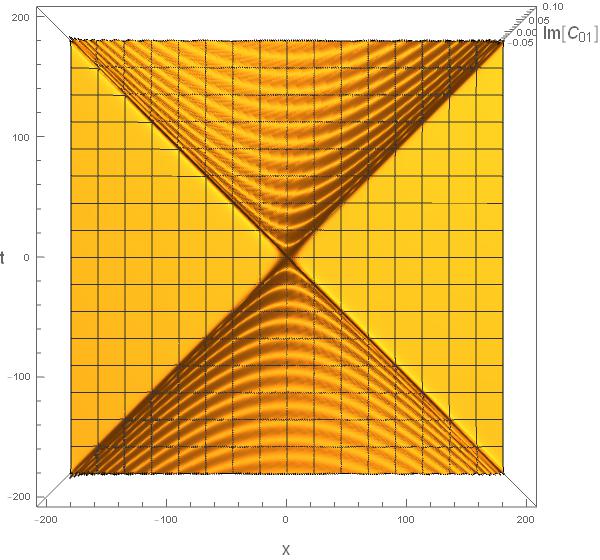}\\
		(c) & (d)\\
\end{tabular}
\end{center}
\caption{These are correlation functions evaluated at $c=0.1$. (a) Real part of correlation functions $\langle0|a_{2x}(x,0)a_{2y}^\dagger(0,t)|0\rangle$ . (b)Imaginary part of correlation functions $\langle0|a_{2x}(x,0)a_{2y}^\dagger(0,t)|0\rangle$. (c)Real part of correlation functions $\langle0|a_{2x}(x,0)a_{2y+1}^\dagger(0,t)|0\rangle$. (d)Imaginary part of correlation functions $\langle0|a_{2x}(x,0)a_{2y+1}^\dagger(0,t)|0\rangle$.}
\label{figc=0}
\end{figure}

\begin{figure}[htbp]
\begin{center}
\begin{tabular}{cc}
\includegraphics[width=6.5cm]{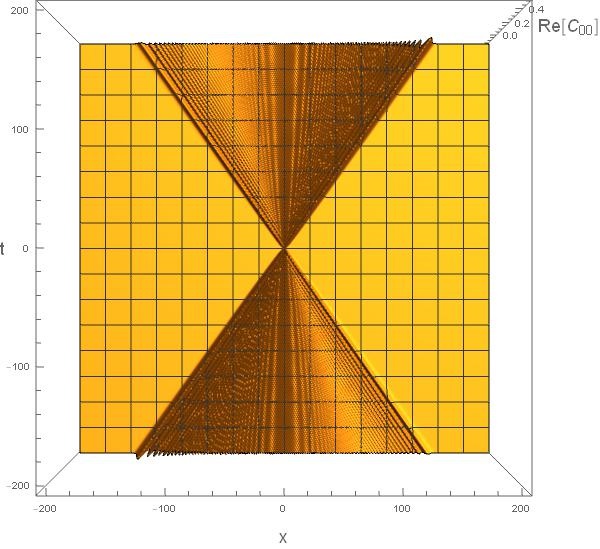}& \includegraphics[width=6.5cm]{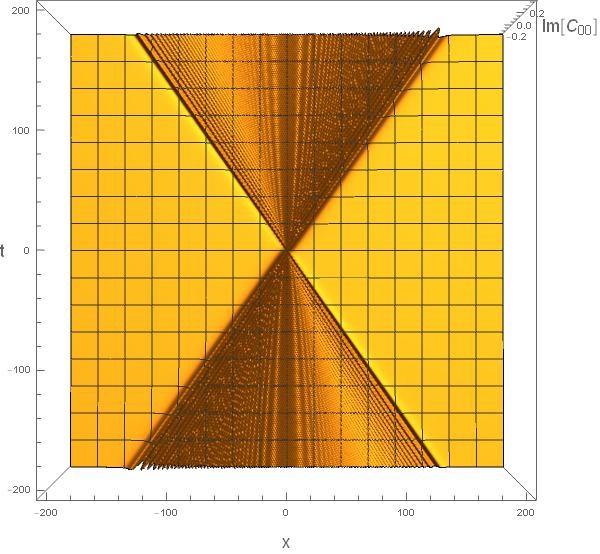} \\
(a) & (b)\\
\includegraphics[width=6.5cm]{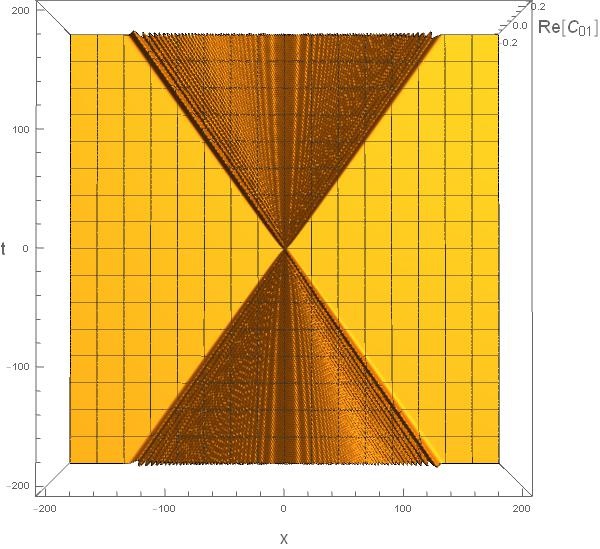}& \includegraphics[width=6.5cm]{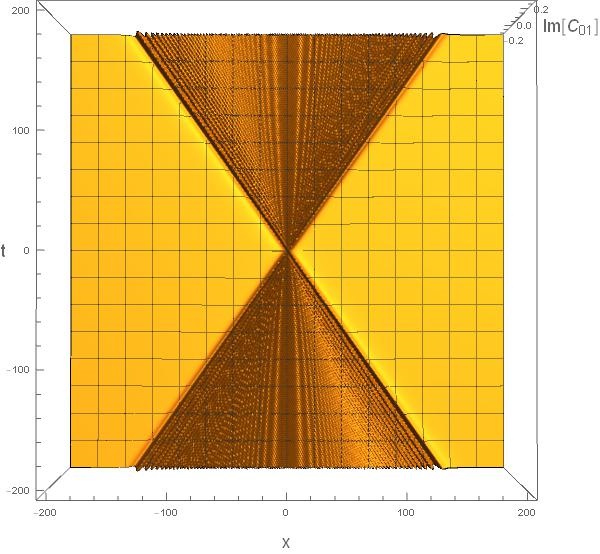}\\
		(c) & (d)\\
\end{tabular}
\end{center}
\caption{These are correlation functions evaluated at $c=s= 1/\sqrt{2}$. (a) Real part of correlation functions $\langle0|a_{2x}(x,0)a_{2y}^\dagger(0,t)|0\rangle$ . (b)Imaginary part of correlation functions $\langle0|a_{2x}(x,0)a_{2y}^\dagger(0,t)|0\rangle$. (c)Real part of correlation functions $\langle0|a_{2x}(x,0)a_{2y+1}^\dagger(0,t)|0\rangle$. (d)Imaginary part of correlation functions $\langle0|a_{2x}(x,0)a_{2y+1}^\dagger(0,t)|0\rangle$.}
\label{figc=s}
\end{figure}

\begin{figure}[htbp]
\begin{center}
\begin{tabular}{cc}
\includegraphics[width=6.5cm]{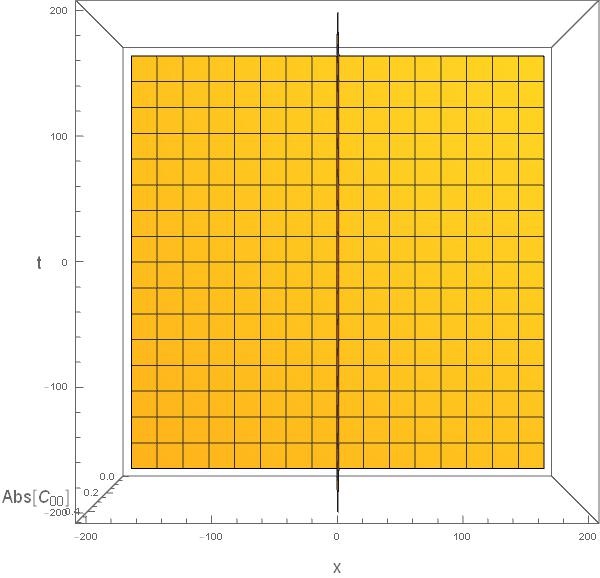}& \includegraphics[width=6.5cm]{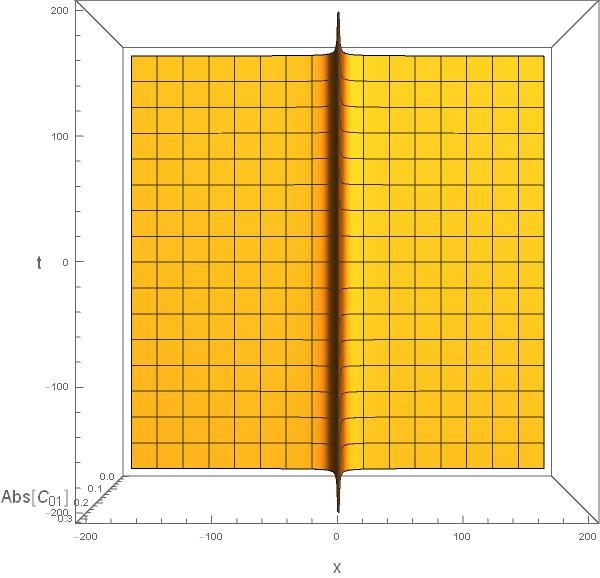} \\
(a) & (b)\\
\end{tabular}
\end{center}
\caption{These are correlation functions evaluated at $c= 1$. (a) Absolute value of correlation functions $\langle0|a_{2x}(x,0)a_{2y}^\dagger(0,t)|0\rangle$. (b)Absolute value of correlation functions $\langle0|a_{2x}(x,0)a_{2y+1}^\dagger(0,t)|0\rangle$.}
\label{figc=1}
\end{figure}

In figure \ref{figc=0}, \ref{figc=s} and \ref{figc=1} we present the correlation functions corresponding to $c=0.1$, $c=s=1/\sqrt{2}$ and $c=1$ respectively, for comparison.
We use the notations $C_{00}\equiv\langle0|a_{2x}(x,0)a_{2y}^\dagger(0,t)|0\rangle$ and $C_{01}\equiv\langle0|a_{2x}(x,0)a_{2y+1}^\dagger(0,t)|0\rangle$.
The lattice is finite, where we take $L=200$.
We can again see the light cone clearly in the figures, demonstrating the causal structure that follows immediately from the construction of the tensor network by local unitary. Moreover, as already mentioned above, a given topology of the tensor network graph gives only the upper bound of the size of the light cone. The precise values that the tensor takes control the actual size of the light cone. At $c=1$ the light cone has shrunk to vanishing size as expected of a non-dispersive theory.
The regions outside of the light-cones are generically not vanishing. They are decaying with a power law (see for example (\ref{continuousc01}, \ref{continuousc02}, \ref{continuousc03}, \ref{continuousc11}, \ref{continuousc12}) and their structures are not so much visible when plotted next to the interior of the light-cone where they take much greater values.

\subsection{Comments on Bosons}

This section might have conveyed the message that the network was built in from the beginning a discretization of the continuous path-integral and it is always possible, at least with some notion of a long-wavelength limit, to recover the physics of the continuous field theory.  There is however a reason why the paper focused on a discussion of Gaussian models of fermions rather than bosons. The authors have begun the journey with the hope of obtaining a tensor network of free bosons. Following almost exactly the same strategy as we have described for the fermions in this section, it is found that we invariably end up with a pair of eigen-modes, one satisfying the usual commutator $[a_p, a^\dag_q] = \delta_{p,q}$, while the other pair satisfying the commutator with the wrong sign.  This is somehow reminiscent of the issue encountered when quantizing bosons behind the black hole horizon. (See for example \cite{Papadodimas:2012aq}.) We have not understood the physical reasons for such a behaviour other than the technical reasons why it ended up that way. It may be suggesting that there is potential obstruction to supporting particular kinds of operator algebra in a given network. This is not unheard of, as in the case of chiral fermions which are known to be impossible to be simulated on a discrete (spatial) lattice. Discretising both space and time in the case of a tensor network might suffer additional obstruction which is an interesting subject in its own right. We relegate the details of our (failed) attempt to model free bosons in a tensor network in the appendix. Rather than lamenting an obstruction, the example emphasizes that the discretized model is intrinsically different from the continuous ones.

\section{Generalization to Integrable models }
In the previous section, we have worked with a tensor network that corresponds to a system of effectively free fermions i.e. the time evolution is based on a quadratic Hamiltonian.
In this section, we would like to generalize our consideration beyond completely free theories. To retain some degree of analytic control, we will focus on a family of integrable models in 1+ 1 dimensions, namely the XYZ model.  Much of our discussion however continues to hold for more general (integrable) models.
Integrable model is a vast subject. It is impossible to give a complete account of this subject. Our perspective is one based heavily on the set of classical statistical models which will be taken as Euclidean continuation of our quantum tensor network models. One classic reference on these classical lattice model is \cite{Baxter:1982zz}.
Our notation is mainly inherited from the review of \cite{Faddeev:1996iy}.  It takes a somewhat more modern perspective compared to \cite{Baxter:1982zz} and sets up the model directly using Lax operators $L_{n,f}(v)$ as building blocks. The Lax operators satisfy
\be \label{YB}
R_{f_1,f_2} (a-b) L_{n,f_1}(a) L_{n,f_2}(b) = L_{n,f_2}(b) L_{n,f_1}(a) R_{f_1,f_2}(a-b),
\ee
\begin{figure}[!h]
		\centering
		\includegraphics[width=8cm]{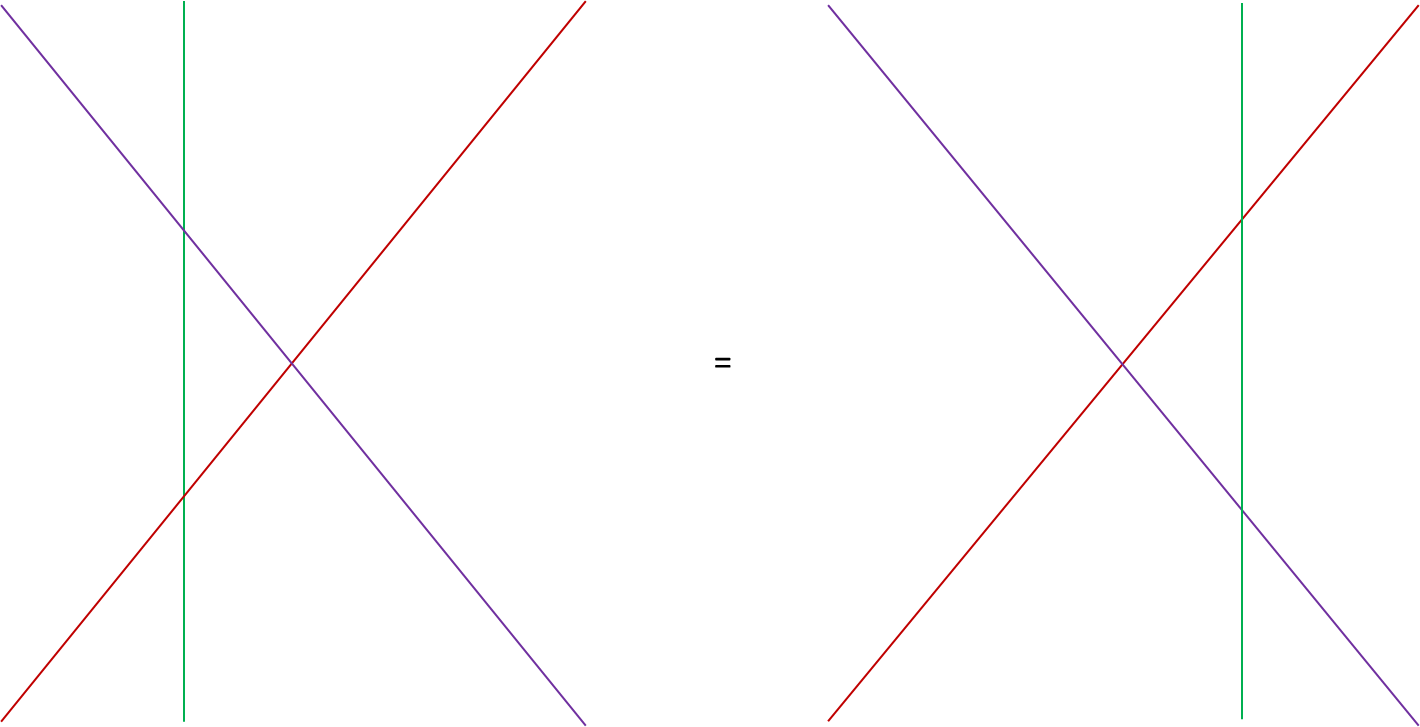}
		\caption{A diagrammatic illustration of the Yang-Baxter equation (\ref{YB}). }
		\label{Yang-Baxter equation}
\end{figure}
where subscripts $f_i$ denote auxiliary spaces on which the respective matrices act, and $n$ denotes the physical Hilbert space at site $n$. The parameters $a,b$ are the spectral parameters. $R$ denotes the ``R matrix'' satisfying the Yang-Baxter equation. Generically, one can take  $R = L$ as a specific solution.
The solutions are obtained through the (homogenous) matrices $T_{N,f}(v)$, which is defined as
\be \label{homotransfer}
T_{f}(v) = (L_{N,f}(u)L_{N-1,f}(u) \cdots L_{1,f}(u)), \,\,\,T(v) = \textrm{tr}_f (L_{N,f}(u)L_{N-1,f}(u) \cdots L_{1,f}(u))
\ee
where the lattice has $N$ physical sites. Taking the trace wrt the auxiliary space $f$ defines the transfer matrix $T(v)$ which is equivalent to the ``row-to-row'' transfer matrices in  the classical models (in periodic boundary condition) reviewed for example in \cite{Baxter:1982zz}.
Since $T(v)$ evaluated at different $v$ commutes, they define a set of commuting operators, which underlines integrability. The quantum Hamiltonian $H$ is often taken as
\be
\ln\{T(u) T^{-1}(0)\} \sim 1 - i u H + \mathcal{O}(u^2).
\ee
The energy eigenstates are constructed using $T_{N,f}(u)$.
The above relations are generic. In the particular families of XYZ models, each physical site accommodates a spin 1/2.
The auxilliary space $f$ can be chosen to be two dimensional, matching the dimensionality of the physical Hilbert space at each site.  In this case it was shown that
\be \label{transfer2}
T_{f}(u) = \left(\begin{tabular}{cc} $A(u)$ & $B(u)$ \\
$C(u)$ & $D(u)$\end{tabular}\right),
\ee
and $B(v)$ can be used to constructing eigenstates. The generic XYZ algebraic Bethe ansatz looks somewhat cumbersome. These can be found for example in \cite{Baxter:1982zz} .
The equations simplify significantly if we focus on the XXZ models. To get a feel of the schematics, we have then
\be \label{ABA}
|\{u_i\} \rangle = \prod_i B(u_i) |0\rangle,
\ee
where $|0\rangle$ is the reference state corresponding to all spin up in the physical sites \cite{Baxter:1982zz,Davies:1989zz,Faddeev:1996iy}.
The important point is that the parameter $v_i$ here controls the momenta and energy of the eigenstates.
In fact
\be \label{momentum}
\exp(i p(u)) =   \frac{\sinh(\lambda-u)}{\sinh(u)},
\ee
where $p(u)$ is the momentum carried by the ``excitation'' $B(u_i)$. The parameter $\lambda$ is a constant depending on the model itself.
The meaning of these parameters defining the XYZ families of integrable models are explained in the appendix  \ref{integrable_mod}.
\subsubsection{Analytic continuation of the spectral parameter} \label{continue}
Caution has to be taken here about the spectral parameter. Here, we adopt the choice of parametrization in Faddeev, so that $T(v)$ defines a unitary evolution. The parametrization in \cite{Baxter:1982zz, Davies:1989zz,Thacker:1985gz} is based on the statistical model however.  One such example is the 6-vertex model, which is related to the quantum XXZ model is reviewed in the appendix. Generically for square lattices one can define the so called row-to-row transfer matrix, which is denoted as $T^E(v)$. It is related to the quantum transfer matrix defined above by
\be \label{analytic_continue}
T(u) = T^E(iu).
\ee
When we work with such ``Euclidean'' versions of any operators we will include the superscript explicitly.
This connection between the Lorentzian and Euclidean signatures in a discretized world is noteworthy.
The usual procedure asserts that a quantum Lorentzian theory in continuous space-time can be connected to a classical theory in one higher dimension by analytic continuation --
the Lorentzian time $t$ is continued to a Euclidean time $\tau_E$ via
\be
\tau_E = i t.
\ee
However, in a discrete space-time, how such a continuation should be defined to connect the Lorentzian quantum model with the classical model becomes less clear. At least in the context of integrable models, the spectral parameter takes up the role of continuous time, allowing one to continue between signatures. It is worth understanding in greater depth whether a continuation can be defined more generically.
\subsection{Inhomogenous Algebraic Bethe ansatz vs the tensor network}
To make connections with tensor networks, we need to construct a tensor network with a well defined causal structure. The row-to-row transfer matrix makes the connection with local unitaries somewhat obscure. We would like to construct an integrable tensor network made up of local unitaries that takes a similar form as the fermions considered in the previous section.
It turns out that this problem has in fact been considered in the literature \cite{Faddeev:1992xa,Faddeev:1996iy}, although that was well before the dawn of the notion of tensor networks. Operators known as the inhomogeneous transfer matrix were considered.  It takes the following form \cite{Faddeev:1996iy}:
\be \label{inhomotransfer}
T_f(u,v) = L_{2N,f}(u + v) L_{2N-1,f} (u - v) \cdots L_{2,f}(u + v) L_{1,f} (u-v),
\ee
where there are $2N$ sites here, and $f$ denotes an auxiliary space. $L_{i,f}(\lambda)$ are the Lax operators that satisfy the relation
\be
R_{a_1,a_2}(w- u)L_{n,a_1}(w) L_{n,a_2} (u) = L_{n,a_2} (u) L_{n,a_1}(w)  R_{a_1, a_2} (w-u),
\ee
where $R_{a_1 a_2}$ is a set of $R$ matrix that acts in the auxiliary spaces $a_1$ and $a_2$. This is the star-triangle relation that underlies integrability.
Now define
\be
U_+ = \textrm{tr}_f T_f(w, w), \qquad U_- = \textrm{tr}_f T_f(-w, w).
\ee
Suppose one identifies
\be
U_+ = \exp(- i (H-P)/2), \qquad U_- = \exp(i(H+P)/2),
\ee
then
\begin{align}\begin{split} \label{integrable_evolution}
&\exp(-i H) = U_+ U_-^{-1} = V\prod l_{2n, 2n-1}(2w) V^{-1} \prod l_{2n,2n-1}(2w) \\
&= \prod l_{2n+1,2n}(2w) \prod l_{2n,2n-1} (2w),
\end{split}
\end{align}
where
\be \label{Ltol}
L_{n,f}(u) = P_{n,f} l_{n,f}(u),
\ee
for $P_{n,f}$ corresponding to a SWAP operator between the spaces $n$ and $f$,
and $V$ is a shift operator that takes $n\to n+1$.
One can see that (\ref{integrable_evolution}) indeed recovers a network that is exactly analogous to the free fermion model that we considered in the previous section. Compare for example with the evolution of ``horizontal'' Cauchy surfaces illustrated in figure \ref{horCau}.
The eigen-modes of this evolution can be solved again by the Algebraic Bethe Ansatz explained in detail in \cite{Faddeev:1996iy}, similar to the homogenous case .
In the following however, this is not the path that we are going to follow. The reason being that as already demonstrated in the previous section, to discuss Unruh effect we find it useful to obtain the form of the eigen-wavefunctions as well, and the Algebraic Bethe Ansatz is not particularly  convenient for the purpose.
We would instead like to work with coordinate Bethe Ansatz. These models that we work with, particularly the XYZ family of models, are well known classical lattice statistical models whose partition function has been studied in depth in the literature. A comprehensive review can be found in \cite{Baxter:1982zz}. The XYZ model can be recovered from the classical 8-vertex lattice model. The definition of the 8 vertex model is reviewed in the appendix. In particular, the definition of the model is summarized in figure \ref{fig:8vertex}.
We also note that the inhomogenous transfer matrix defined above is essentially generating the so called ``diagonal-to-diagonal'' transfer matrix, as opposed to the usual ``row-to-row'' transfer matrix. This corresponds to rotating the classical integrable lattice model by 45 degrees (which can be readily illustrated in the simplest limit of the XXX model).
We will explore the change of basis between the row-to-row and diagonal-to-diagonal picture in section \ref{basischange}.

We solved the eigen-wavefunctions in a way directly analogous to the case of the fermions near the ferromagnetic ground state $|0\rangle$. Since the actual computation is very similar to the previous section, we have relegated the details to appendix \ref{integrable_mod}, where eigen-functions and the form of correlation functions for the XXZ model are explicitly computed. It is noteworthy that the excitations above the reference state $|0\rangle$ cannot be made ``positive definite'' in the way of the fermions discussed in the previous section. It necessrily contains both ``positive'' and ``negative'' excitations. (See the discussion near (\ref{Lambda}). )
On the other hand, boost operators can be constructed in an exactly analogous manner as in the case of fermions. All the computations at least for single spinon states have a direct analogue in the integrable model. We note that there are limits of the parameters where the dispersion relations simplify, and the correlation functions again resemble the case in the previous sections recovering the usual invariance along hyperbolas. (See figures \ref{abs-part-c00-integrable-L200-b-equals-0-99} and \ref{abs-part-c01-integrable-L200-b-equals-0-99} and the discussion nearby.)

As we are going to discuss in the next section, the way we have constructed the boost operator is a (Lorentzian) approximation to the boost operator that can be constructed in an integrable model based on the corner-transfer-matrix. The ferromagnetic ground state is indeed invariant under such a boost. That it is a direct product state however, can be traced to the observation we made above -- that excitations around the state cannot be made positive definite (in the ``first Brillouin zone'' for energies). We will discuss the relation of this fact with the Reeh-Schlieder theorem in section \ref{ReehSchlieder}.

\subsection{Corner transfer matrix } \label{sec:CTM}
\begin{figure}[!h]
		\centering
		\includegraphics[width=8cm]{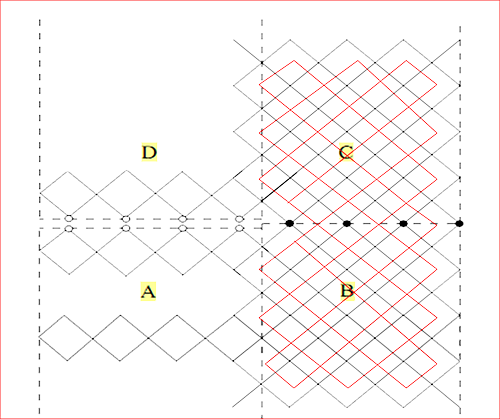}
		\caption{This figure illustrates the meaning of the corner transfer matrix. Consider for example the matrix $A.B$. The incoming incoming indices correspond to degrees of freedom sitting on the white dots adjacent to region $A$, and the outgoing indices correspond to the degrees of freedom sitting on the black dots. The matrix elements of a given matrix is defined by fixing the degrees of freedom at the boundary of the matrix while summing over all intermediate dofs weighted by the statistical weight defining the model. Matrix multiplication correspond to the weighted sum over all the degrees of freedom sitting at the boundary separating the two region. In this case, the boundary between $A$ and $B$ is marked by one of  the dashed line.  The tensor network is the dual graph marked in red. Picture courtesy  cond-mat/9810174}
		\label{fig:CTM}
\end{figure}

The corner transfer matrix (CTM) was introduced by Baxter \cite{Baxter:1982zz}. The corner transfer matrix is illustrated in the picture figure \ref{fig:CTM}
It was observed that the corner transfer matrix has a discrete spectrum even in the thermodynamic limit, and that in that limit its spectrum can be exactly solved in various models, such as the XYZ model, making it a very powerful tool.
It has been noted that the CTM can be used to compute entanglement entropies of integrable model.  The partition function $Z$ of the statistical model can be expressed in terms of the CTM's. It is shown that the reduced density matrix of half-space can be expressed as \cite{Baxter:1982zz}
\be \label{halfspaceCTM}
\rho_{\textrm{half}} = \mathcal{N} (A .B.C.D)_{\{\sigma\}, \{\sigma'\}},
\ee
up to some normalization $\mathcal{N}$. Each of these matrices  are also marked in figure \ref{fig:CTM}.
 \\	
The partition function $Z$ is given by $\textrm{tr}(\rho_{\textrm{half}}) = \textrm{tr}(A.B.C.D)$.
Spectra of the reduced density matrix and their corresponding entanglement entropy, of the XYZ model for example, have been discussed in detail \cite{Baxter:1982zz}.
So now we have two loose ends to be tied together at least in the context of integrable model in which exact solutions are more readily available..
On the one hand, we have half-space reduced density matrix $\rho_{\textrm{half}}$ which can be solved exactly. On the other hand, we have constructed naive versions of boost operators in the previous section, which appears, at least in some analytically controllable limits, to illustrate the physics of the Unruh effect. This in turn should be a manifestation of half-space entanglement. Here, we would like to discuss the connections between the boost operator and half-space entanglement in the broader context of integrable models, and demonstrate that such a connection is generic. The boost operator can indeed be defined, and whose algebraic structure has been explored. In fact, our naive construction is a generalization of the boost operator that has been constructed in the past that preserves lattice symmetries, allowing for some sensible approximations when the boost is not an exact symmetry of the space-time lattice.  The CTM defines rotation in Euclidean signature by 90 degrees in the row to row basis. Up to a basis change, it is indeed the same constructions of the boost operator as we have, apart from the fact that we allow for more general angles of rotation/boost by choosing the step size of ladders. Our construction does not correspond to exact symmetries of the space-time lattice, but only approximate ones. Moreover, our construction is based on Lorentzian signature in the ``diagonal-to-diagonal'' basis, which also differs from the usual ``row-to-row'' basis. We will discuss how the analytic continuation to a Lorentzian signature is performed, and how the basis transformation is performed, at least perturbatively in the spectral parameter $\lambda$.
\subsubsection{CTM and Lorentz transformation }
\begin{figure}[!h]
		\centering
		\includegraphics[width=8cm]{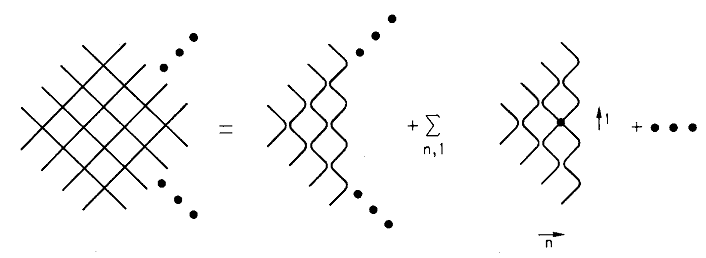}
		\caption{Expansion of the corner transfer matrix around u=0. Picture courtesy \cite{Thacker:1985gz}.}
		\label{fig:expand_CTM}
\end{figure}

It was observed in the early days that the CTM is related to the notion of a boost operator \cite{Thacker:1985gz,Tetelman}.  They have constructed the generator of a boost operator explicitly and obtained the commutation relations that they satisfy.
For concreteness, we will explore the families of 8-vertex models, which is related to the XYZ models, and reduces to the XXZ and XXX models in some limits explained in the appendix near equation \ref{XXZpar}.
By setting the boundary conditions of the Lax operator $L^E_{n,f}$ to satisfy the boundary conditions
\be
L^E_{n,f}(u=0) = P_n,
\ee
it is readily shown graphically  (see figure \ref{fig:expand_CTM}), that
\be \label{smalluCTM}
A^E(u) = 1-u K + \mathcal{O}(u^2), \qquad K = \sum_{n=0}^{\infty} n H_{XYZ}(n,n+1),
\ee
where
\be
H_{XYZ} (n,n+1) = - \frac{1}{2} (J_x \sigma^x_n \sigma^x_{n+1} + J_y \sigma^y_n \sigma^y_{n+1} + J_z \sigma^z_n \sigma^z_{n+1}).
\ee
Using properties of the CTM, where
\be
A^E(v) B^E(u) = X^E(v-u),
\ee
and that
\be
B^E(u) = A^E(\lambda - u),
\ee
where $\lambda$ is a parameter defining the 8-vertex model  (see figure \ref{hamiltonian})
and ogether with the small spectral parameter expansion (\ref{smalluCTM}), it implies that \cite{Thacker:1985gz,Tetelman}
\be
{\hat A}^E (u) = \exp( - u K).
\ee
Here ${\hat A}^E$ is customarily defined as $A^E(u)/a_0$, where $a_0$ is the largest eigenvalue of $A^E$.
The matrix is known to generate a 90 degree rotation in the Euclidean plane. This tells us that $A^E$ are precisely the Euclidean version of the boost operator.
To obtain inversion of $(x,t) \to (-x,-t)$ which is a 180 degree rotation in the Euclidean plane and which is the ingredient necessary in the Bisognano-Wightman theorem, it is given by $(A^E). (B^E)^t $, using figure \ref{fig:CTM}.
It is shown that the Lax operator and $H_{XYZ}(n,n+1)$ satisfy the following commutation relations
\be \label{eq:commutator}
[H_{XYZ}(n,n+1), L^E_{n,f}(v) L^E_{n+1,f}(v)] = {L^E}'_{n,f}(v) L^E_{n+1,f} (v) - L^E_{n,f}(v) {L^E}'_{n+1,f}(v),
\ee
where  ${L^E}'_{n,f}(v)$ denotes derivative of the (Euclidean) Lax operator wrt $v$.
This relation follows from the Yang-Baxter equation, and can thus be generalized to generic integrable model, by identifying
the Hamiltonian with
\be
H(n,n+1) = L'_{n,n+1}(0) P_{n,n+1},
\ee
where the derivative is taken wrt to the spectral parameter $v$.

Now similar to the case of the free fermions, one can define a ``completed" modular Hamiltonian (\ref{completeH}), and get
\be
K_c \equiv K - \bar K = \sum_{n=-\infty}^{\infty} n H_{XYZ}(n,n+1).
\ee
i.e. We simply complete the boost operator by extending the sum from $n=-\infty \to +\infty$.
Using (\ref{eq:commutator}) gives
\be \label{shiftv}
[K_c, T^E_{N,f}(v)] = \partial_v T^E_{N,f}(v),
\ee
where $T_{N,f}(v)$ are the transfer matrices defined in (\ref{homotransfer}).
Therefore, the operator $K_c$ shifts the spectral parameter in $B(u)$ defined in (\ref{transfer2}).  Using (\ref{momentum}), one sees that the shift of the spectral parameter $v$ under the effect of the generator $K_c$ is to vary $p$.

There is a generalization of the Lorentz algebra \cite{Thacker:1985gz}. By defining
\be
\ln T^E(u) = \sum_{n=0}^\infty \frac{u^n}{n!} C_n,
\ee
we have
\be
[K_c, C_n] = i C_{n+1}, \qquad [C_n, C_m] = 0,
\ee
and that $C_0 = P$ i.e. the translation operator shifting globally by 1 lattice site,  and $C_1= H$.
It would reduce to the usual Lorentz algebra when $C_{2n} = P$ and $C_{2n+1} = H$.  This is argued to occur in appropriate continuum limit \cite{Thacker:1985gz}.
We also note that, as already alluded to in section \ref{continue}, the rotation operator $A^E$ that generates rotation in the Euclidean plane can be continued to a Lorentzian boost $\hat{A}$ by taking $u\to i u$.
There is an interesting lesson here. $\hat A(u)$ as a tensor network has been fixed, as shown in figures \ref{fig:CTM} and \ref{fig:expand_CTM} . Therefore, for a partition function defined at some fixed value of $u$, $A^E(u)$ generates a rotation on the network by 90 degrees. On the other hand, the value of $u$ also plays the role of the angle that is actually rotated. If we consider ${ A}^E(v)$ at some other value of $v$, such as $v= u/2$, then ${{A}}^E(u/2)$ is approximating the operator that generates a rotation by 45 degrees on the network defined at $u$.
This can be readily checked in the small $u$ limit discussed in section \ref{basischange}.
 In other words, any given effective angle that is actually rotated can either be approximated by the ladder structure that we have introduced, or varying the variable $u$, allowing it to deviate from the value that defines the tensor network.

\subsubsection{A comment on the commutator $[K_c,T_{N,f}(u)]$} \label{sec:comcommute}

In the previous section, we reviewed the construction of the boost generator $K_c$, and that the Yang-Baxter equation (\ref{YB}) directly provides a way to construct a boost generator that shifts the spectral parameter appearing in the transfer matrix $T^E$.

This can be compared with the effect of our boost operator in equation (\ref{boostap}), which looks very similar.
There, it appears that while the commutator is producing the desired term $\partial_p a_p$, it contains an extra term. We would like to inspect the mechanism of equation (\ref{shiftv}) that generates a simple shift in the spectral parameter. As we are going to see -- this is the subtlety that comes with an infinite space. It involves and infinite $v$ dependent normalization, and also a push of the discrepancy towards infinity. We do not know as yet whether this makes physical sense, but we present the detailed mechanism.

In order to see this, we will again restrict our attention to the XXX model where the expressions simplify.

In the XXX limit which is reviewed also in the appendix \ref{integrable_mod}, we take $\lambda = \pi + \epsilon$, keeping $\epsilon/u$ fixed as $u,\epsilon \to0$.
We define the new spectral parameter to be $\tilde u = u/\epsilon$.
\be
K_{c, \,\, XXX} = \sum_{n= -N/2}^{N/2} n P_{n,n+1}.
\ee
where $N$ is the number of sites and one has to take $N\to \infty$ to recover (\ref{shiftv}).

The transfer matrix $T_{N,f} = \prod_{i=-N/2}^{N/2} {L_{i f}}$, and we would like to obtain the single spinon state by
\be
|v\rangle = (T_{N,f})_{12} |0\rangle,
\ee
where the subscript $12$ denotes the 12 component in the auxiliary space of $T_{N,f}$, corresponding to operator $B(v)$ in (\ref{transfer2}).
The reference state is the all spin up state described in (\ref{ABA}).

Now one can expand the form of $|\tilde u\rangle$. To that end, let us rewrite the (Lorentzian form of the) Lax operator as
\be \label{laxXXX}
L_{nf}(\tilde u) \equiv L^E_{nf}(i \tilde u) = \left(\begin{tabular}{cc}
$\tilde u + \frac{i}{2} \sigma_z(n) $& $\frac{i}{2} \sigma_- (n)$ \\
$\frac{i}{2}\sigma_- (n) $& $\tilde u- \frac{i}{2}\sigma_z(n) $
\end{tabular}\right), \qquad \sigma_\pm = \sigma_x \pm i \sigma_y.
\ee
We note that $\sigma_+$ annihilates the reference state $|0\rangle$. Also $\sigma_z(n)|0\rangle = |0\rangle$.

Therefore for any given lattice of $N$ sites, the single spinon state is given by
\be
|\tilde u\rangle = \sum_n (\tilde u- i/2)^{n-1} \sigma_-(n) (\tilde u+ i/2)^{N-n-1} |0\rangle,
\ee
where $\sigma_-$ appears at exactly one site $n$, and $n$ is summed over all sites. In this limit we have
\be
\exp(i p) = \frac{\tilde u-i/2}{\tilde u+i/2},
\ee
and that $(\tilde u+i/2)^N/(\tilde u-i/2)$ behaves like normalization for any given $N$.

Now naively we find
\begin{align}
\begin{split}
&[K_{c,\,\,XXX}, B(v)]|0\rangle \\
&= \mathcal{N}(p) \sum_{n,m} n [P_{n,n+1} , \sigma_-(m)] \exp(i p m) |0\rangle \\
&= \mathcal{N}(p)\sum_n  \exp(i p n) \sigma_-(n) (2n (\cos p - 1) + (1- \exp(- i p)) |0\rangle \\
& = \mathcal{N}(p)  \sum_n \exp( i p n) \sigma_-(n) (- 4n\sin^2(\frac{p}{2}) + 2 i \exp(- i p/2) \sin(\frac{p}{2}) ) |0\rangle,
\end{split}
\end{align}
where
\be \label{eq:normal}
\mathcal{N}(p) = (\tilde u(p)- i/2)^N (\tilde u(p)+i/2)^{-1}.
\ee
Formally the corner transfer matrix is defined for $N\to\infty$, where $N$ is the number of sites.
We keep $N$ explicit for now.
One can see that very similar to the case of free fermions considered in the previous section, a naive computation of the commutation relation does not appear to be consistent with (\ref{shiftv}) ---  the term that is not linearly dependent on $n$ does not have the right form so that together this is an over derivative in $p$. On the other hand, the result (\ref{eq:commutator}) should guarantee that this is true.

To reconcile the two, we note the following. The way (\ref{eq:commutator}) leads to (\ref{shiftv}) is based on cancellation between
\begin{displaymath}
n [H_{n,n+1}, L_{n,f} L_{n+1,f}]\end{displaymath}
and
\begin{displaymath}
(n+1) [H_{n+1,n+2}, L_{n+1,f} L_{n+2,f}].\end{displaymath}

Now consider $n [P_{n,n+1} , \sigma_-(m)] \exp(i p m) |0\rangle$ above. As we noted above, two values of $m$ contribute -- $m=n$ and $m=n+1$.
Explicitly, we have
\begin{align}
\begin{split}
&\mathcal{N}(p)\,\, n [P_{n,n+1} , \sigma_- (n) \exp(i p n) + \sigma_- \exp(i p (n+1))] |0\rangle  \\
&= \mathcal{N}(p) \,\,n (\exp (i p n) (\exp(i p )-1) \sigma_n - \exp(i p (n+1)) (1- \exp(- i p )) \sigma_{n+1})|0 \rangle.
\end{split}
\end{align}

Rather than doing a re-parametrization in the re-summation above, we consider adding and subtracting the term
\begin{displaymath}
(n+1) [\exp (i p n) (\exp(i p )-1) \sigma_n] \end{displaymath} and
\begin{displaymath}- (n-1) [ \exp(i p (n+1) (1- \exp(- i p )) \sigma_{n+1}]. \end{displaymath}
The term
\begin{displaymath} -(n+1) [\exp (i p n) (\exp(i p )-1) \sigma_n]
\end{displaymath}
would combine with
\begin{displaymath}
n (\exp (i p n) (\exp(i p )-1) \sigma_n)
\end{displaymath}
to give
\begin{displaymath}
-  (\exp (i p n) (\exp(i p )-1) \sigma_n.
\end{displaymath}
Similarly
\begin{displaymath}
- n\exp(i p (n+1)) (1- \exp(- i p )) \sigma_{n+1}
\end{displaymath} combine with
\begin{displaymath}
(n-1) \exp(i p (n+1)) (1- \exp(- i p )) \sigma_{n+1}
\end{displaymath} to give
\begin{displaymath}
- \exp(i p (n+1)) (1- \exp(- i p )) \sigma_{n+1}.
\end{displaymath}
On the other hand, the left over terms, namely
\begin{displaymath}
(n+1) [\exp (i p n) (\exp(i p )-1) \sigma_n] - (n-1) [ \exp(i p (n+1) (1- \exp(- i p )) \sigma_{n+1}]
\end{displaymath}
can be further combined with addition and subtraction of
\begin{displaymath}
-(n+2)  [\exp (i p n) (\exp(i p )-1) \sigma_n] + (n-2) [ \exp(i p (n+1) (1- \exp(- i p )) \sigma_{n+1}] . \end{displaymath}
This can be continued indefinitely. Then as we collect all the terms involving $\sigma_-(n)$, we would obtain a better and better approximation of the term
\begin{displaymath}\partial_{\tilde u} [ \mathcal{N}(p(\tilde u)) \exp ( i p(\tilde u) n) \sigma_-(n) |0\rangle]
\end{displaymath}
in the limit $N\to \infty$.
Of course here there are two subtleties. First the normalization $\mathcal{N}(p)$ defined in (\ref{eq:normal}) plays a crucial role in the algebra, and yet its value is not well-defined in the large $N$ limit. Secondly, we are adding and subtracting terms to push the discrepancy of the result from a total derivative to infinity. At present, we only present the mechanism that led to (\ref{shiftv}) at the spinon level. This mechanism should also be equally  applicable to the free fermion case. This is evidence that the our boost operator, which generates finite rescaling in $p$ is indeed consistent with the infinitesimal transformation generated by the entanglement Hamiltonian, despite appearance. Whether this is consistent with sensible and physical boundary conditions are to be investigated and clarified.

\subsubsection{CTM in the  diagonal-to-diagonal- basis} \label{basischange}

The CTM is expressed in terms of the ``row-to-row'' basis in the above.
However, our construction of the tensor network so that the time evolution is explicitly made up of local unitaries are more conveniently expressed in terms of the ``diagonal-to-diagonal'' basis, which is the same lattice rotated by 45 degrees.

\begin{figure}[!h]
		\centering
\begin{tabular}{cc}
		\includegraphics[width=8cm]{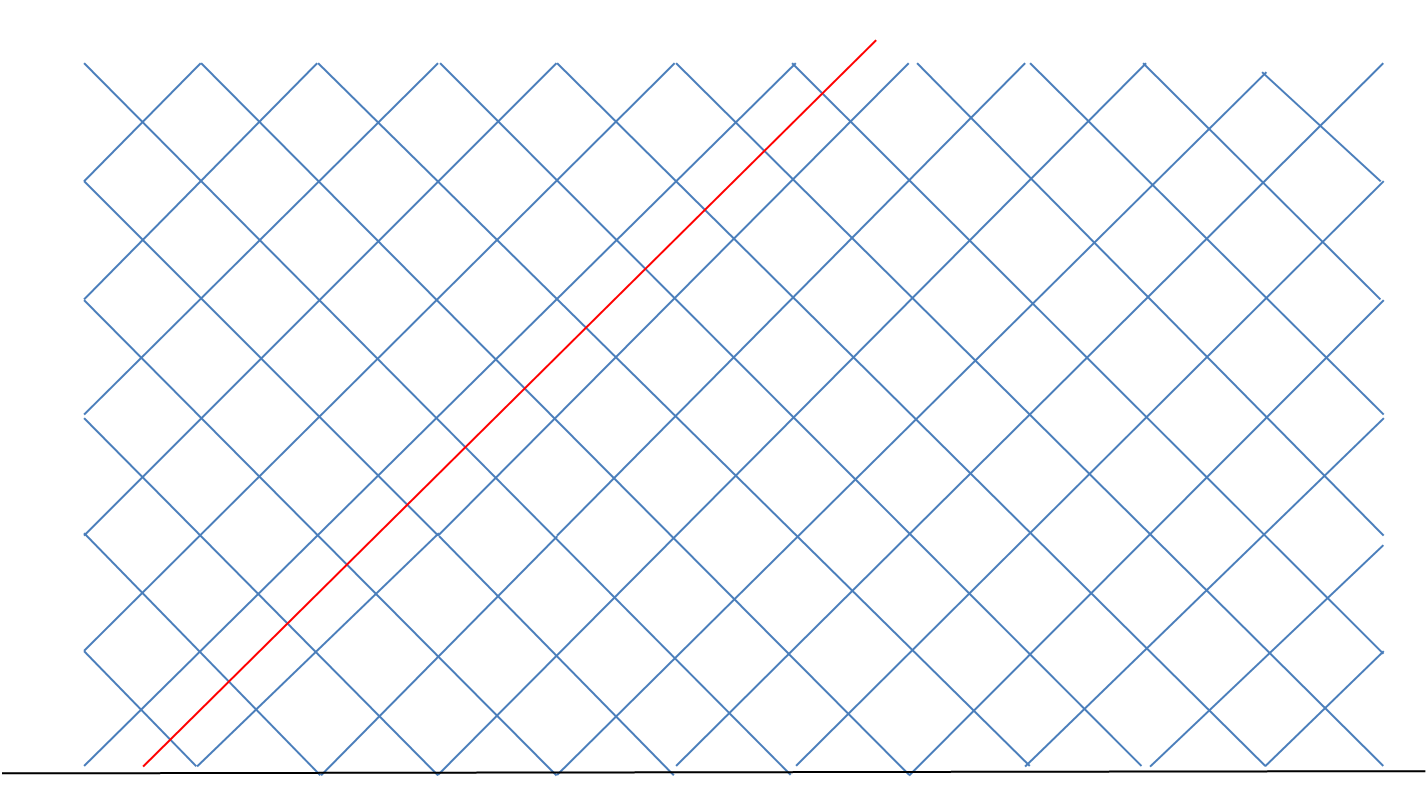} & \includegraphics[width=3cm]{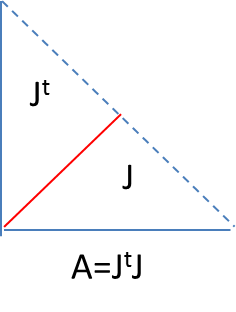} \\
		(a) & (b)
\end{tabular}
		\caption{Illustration of the set of tensors sandwiched between the Cauchy surface and the horizontal Cauchy surface gives the transformation $J^{E}(u)$ that effects a basis change between the diagonal-to-diagonal and row-to-row basis.}
		\label{basis_change}
\end{figure}

The story developed above for row-to-row transfer matrices can be translated to the ``diagonal-to-diagonal'' basis. To that end, we need to obtain a rotation matrix $J^E(u)$ that rotates by 45 degrees. This can be constructed. This is basically given by the tensor network sandwiched between the Cauchy surface at 45 degrees to the horizontal Cauchy surface in the diagonal-to-diagonal basis. This is shown in figure \ref{basis_change}.

{\bf  \underline{Small $u$ limit}}\\

In the small $u$ limit, one can show that
\be
J^E(u) = 1 - u \tilde{K} + \mathcal{O}(u^2) , \qquad \tilde{K}  = \sum_{n=1}^{\infty} n (H_{2n-1,2n} + H_{2n, 2n+1}) ,
\ee
Indeed $\tilde{K}$ is almost equal to $K/2$, corroborating that variations in the the form of the network and variations of the value of $u$ are (almost) interchangeable.
One should also note that
\be
A^{E}(u) = {(J^E(u))}^t J^E(u).
\ee
$J^E$ would be responsible for transformation between the ``row-to-row`` basis and the ``diagonal-to-diagonal" basis.

Given any operator $\mathcal{O}^{rr}$ in the ``row-to-row" basis we have
\be
\mathcal{O}^{dd} = J^{E\,\,-1}(u) \mathcal{O}^{rr} J^E(u).
\label{row-diag}
\ee

The differential equation (\ref{shiftv}) would be modified under this basis transformation, since $J(v)$ contains $v$ dependence.
Replacing $T^{rr}(u) = J^E(u) \mathcal{O}^{dd} J^{E\,\,-1}(u) $ in the equation, it reduces to
\be
\partial_v \mathcal{O}^{dd}(v) = [J^{E\,\,-1}(u) K J^E(u), \mathcal{O}^{dd}(v)].
\ee
The effective boost operator in the diagonal-to-diagonal basis is thus given by $J^{E\,\,-1}(u) K J(u)$. In the limit that $u$ is small,
one can readily show that $J$ is given by the identity matrix to leading order. Therefore,
\be
J^{E\,\,-1}(u) K J^E(u)  \approx K
\ee

{\bf  \underline{Large $u$ limit}} \\

We can also inspect the large $u$ limit. This can be computed very readily in the case of the XXX model.
Things simplify tremendously because the Lax operator already discussed in (\ref{laxXXX}) and also the $l_{n,f}$ operators defined in (\ref{Ltol}) simplifies to
\be
L_{n,f}(\tilde u) = \tilde u \mathcal{I}_{n,f} + i P_{n,f},  \qquad l_{n,f} = i \mathcal{I}_{n,f} + u P_{n,f}.
\ee
As shown in (\ref{integrable_evolution}), each cross in the tensor network is simply given by $l_{n,f}$ up to an overall normalization. The boost operator $K$ in (\ref{smalluCTM}) reduces to
\be
K_{XXX} = - \mu \sum_n \,\, n (P_{n,n+1} ) , \qquad \mu \equiv J_x = J_y = J_z.
\ee
Note that we have dropped a constant term since that only changes the overall normalization of the CTM and do not contribute to any of the commutation relations.

The boost operator in the diagonal-to-diagonal basis can be computed by conjugation by $J$ defined in (\ref{basis_change}) as in (\ref{row-diag}).
In the large $\tilde u$ limit, this can be computed readily.
The leading term in the large $\tilde u$ expansion is contributed by the replacement $l_{n,f} \to P_{n,f}$.

This immediately gives
\be
J^{E \,\,-1}K_{XXX} J^E =  - \mu \sum_n \,\, n (P_{2n, 2(n+1)} )  + \mathcal{O}(\tilde u^{-1}).
\ee

This appears like a singular transformation. Despite building the tensor network by local unitaries, it appears that half of  the degrees of freedom (the odd sites) are missing.
This is because odd sites are thrown towards infinity $n\to \infty$. In fact $\tilde u \to\infty$ limit is the infinite boost limit which appears to decouple the left and right moving modes.

One could readily obtain a few sub-leading terms, expanding in $1/\tilde u$ and find that they involve growingly non-local terms for higher orders of $1/\tilde u$.
In both limits however, the boost generator takes on a simple local form in the diagonal-to-diagonal basis.


\subsection{The eigen-modes of the corner transfer matrix}

In the discussion of free fermions, we present the eigen-wavefunctions of the candidate boost operator that we proposed. Here, we would like to discuss the eigen-modes of the boost operators and compare them with the results following from our guesses. There has been considerable amount of work exploring the entanglement entropy following from the reduced density matrices constructed from the CTM as in (\ref{halfspaceCTM}). \cite{1999AnP...511..153P, 1997LNP...478..167N, Ercolessi:2009kc, Calabrese:2010rx}.   It is found for example, that near the critical points, the resultant entanglement entropies recovers a log divergence and whose coefficient matches with the expected central charge of the corresponding CFT \cite{Ercolessi:2009kc}.

Here, we will particularly focus on the XXZ model, in which the eigen-wavefunctions have been solved explicitly \cite{Davies:1989zz, Frahm:1991hr}, in addition to the eigenvalues.

The set of solutions take a very simple form by noting the following. Eigenstates are given by
\be
|\{u_i\}\rangle = \prod_i B(u_i) |0\rangle,
\ee
where $B(u)$ is the operator defined in (\ref{transfer2}).

Here, the rapidity parameter $u$ is related to the actual lattice momentum by
\be \label{rapidity_defn}
e^{ip} = \frac{e^{\lambda} z - 1}{e^\lambda - z}, \qquad \frac{1}{2}\ln z = u- \lambda/2,
\ee
where $\lambda, u$ are those that appeared in equation (\ref{momentum}).
It is observed in \cite{Thacker:1985gz, Tetelman}, and recollected in (\ref{shiftv}) that the boost operator $K$ generates a shift in $u$. Therefore the eigenstates transform as
\be
K |\{z_i\}\rangle = \sum_{j}\partial_{z_j} | \{z_i\} \rangle.
\ee
This suggests that the correct eigenstates of $K$ is given by Fourier transforming in the rapidity $z$. For the ``single spinon state'' for example,
\be \label{K_eigenstate}
|l\rangle = \int dz \,\, z^l \, B(z) |0\rangle, \qquad K |l \rangle = l | l \rangle.
\ee
We would like to comment on the relation of this solution with the fermionic solution.

Now, consider the critical limit which is explained in the appendix B, in which $|\Delta| \to 1$. In that case, $\lambda \to 0$, and we can expand  (\ref{rapidity_defn}). If we in addition also assume that $z $ is small, we get
\be
e^{ip} = \frac{(1+ x) z - 1}{1+x  - z} \approx -1 + \lambda + 2 \lambda z + \mathcal{O}(\lambda^2, z^2).
\ee
This means that $p = 1/i  \ln (-1 (1- \lambda - 2\lambda z))$.
This says that $p$ has a background value determined by the coupling of the model $\lambda$,
\be
p_b = \pi +i \lambda + \mathcal{O}(\lambda^2)
\ee
and a dependence on the rapidity which is a variable that changes the momentum
\be
\delta p (z) = 2 i \lambda z.
\ee

One notes  that $\delta p$ depends linearly on $z$.
Therefore the eigenstates (\ref{K_eigenstate}) would take the form
\be
 | l  \rangle = \int \frac{dz}{z}  z^l  B(z)  |0\rangle.
\ee
which is a power in $\delta p(z) $. This can be compared with (\ref{Rind_eigenval}), where we find $l \leftrightarrow \kappa$.\footnote{We note that the background value of the momentum has to do with the basis change. }

Here we note the following. The reference state of the ferromagnetic phase $|0\rangle$ as introduced in (\ref{ABA}) is a direct product state in configuration space. However, it is an exact eigenstate of the boost operator $K$. Since $K$ can be defined up to some constants, it means that this state can be taken as invariant under boost, {\it even though it contains no entanglement}.

\subsection{A comment on the Reeh- Schlieder theorem} \label{ReehSchlieder}

This can be contrasted with the case of the fermions, where we demonstrated that the ground state is highly entangled, and at the same time (approximately) invariant since the boost operator (approximately) preserves the notion of positive/negative energy.

In AQFT, there is an important theorem, namely the Reeh-Schlieder theorem, that guarantees that in a translation invariant theory with a positive definite energy operator, the ground state is cyclic and separating with respect to any algebra $\mathcal{A}(O)$ associated to any subregion $O$. In particular, that means
 that no local operator can annihilate the state \cite{Haag:1992hx}. This implies that the ground state is highly entangled, which is the crucial ingredient of the Unruh effect.

This is clearly not the case for the reference state $|0\rangle$ in the ferromagnetic phase for the 6-vertex model. i.e. This state $|0\rangle$ is not a cyclic and separating vector in the Hilbert space.  In fact, it is a direct product state with no entanglement at all. At the same time when we solve for ''excitations'' around this reference state, we find that the spectra always involves both one eigenvalue and its complex conjugate. (The details of the states can be found in the appendix.)
This means that there is no natural notion of ``positive energy'' states in this case, which can be contrasted with the fermion ground state that we constructed.

Of course, in the tensor network, the notion of ``positive energy'' is itself ambiguous. Since the time evolution is discrete in units of $\Delta t$, energy is only defined up to $2\pi/\Delta t$. Moreover, in any finite (spin) system such as the 6-vertex model considered here, the Hilbert space is finite dimensional and so energy is lower bounded by definition. In such cases, is there anything we can learn from the Reeh-Schlieder theorem as guidance to the entanglement structure of the ground state?

Here, we note that the proof of the Reeh-Schlieder theorem made use of the following. The assumption of a positive energy operator allows one to decompose any (local) operators into the following \cite{Haag:1992hx}
\be \label{decompose}
Q(t,x) = Q^+ + Q^- + Q^0, \qquad  Q^-|\Omega\rangle = Q^{+\,*}|\Omega\rangle = 0,
\ee
where $Q^{\pm, 0}$ corresponds to positive, negative and zero modes under a Fourier transfrom wrt time.  i.e.

Translation invariance requires that eigen-operators can be further decomposed as sum over operators with definite momenta. Therefore
\be \label{FTQ}
Q_i = \int d^dp \, e^{- i p_0 t + i p_i x^i} \tilde{Q}_i(p_0, p_i), \qquad \tilde{Q}_+(p_0, p_i) |\Omega\rangle = \tilde{Q}^*_-(p_0, p_i) |\Omega\rangle  = 0.
\ee
These $Q_i(p)$ cannot be local in space as operators with definite momentum. Therefore, (\ref{FTQ}) implies that the ground state is annihilated by roughly half of all the operators built with definite momenta. This guarantees that the vacuum is highly entangled, and conceivably (although we haven't produced a rigorous proof, examples are easily constructed) that the reduced density matrix can be inverted.

This boundedness of the spectrum turns into properties of correlation functions, in which it is observed that correlation functions involving any local operator $Q$'s
\be
F(x_1, x_2 \cdots) = \langle \Psi  Q(x_1) Q(x_2) \cdots | \Omega\rangle,
\ee
can be extended to an analytic function over a tubular region in complex coordinates
\be
z_1 = x_1 + i \eta_1, \qquad z_2 = x_1 - x_2 - i \eta_2 ,   \,\,\cdots z_n = (x_{n-1} - x_n) - i \eta_n \,\,\cdots.
\ee
Here we denote $x_a = (t_a, x_a^i)$, and $\eta_i$ are $d$-vectors lying in the forward cone.

The computation at real $x_i$ becomes the boundary values of this analytic function. This analytic extension to an analytic function over an extended region was crucial towards showing that $|\Omega\rangle$ is cyclic and separating.
As we see above, this analyticity of correlation functions is directly related to entanglement of the ground state, although in a discrete system the procedure of such analytic extension would become obscure even though the cause of entanglement is still very much applicable.

\subsection{A comment on the anti-ferromagnetic case}

Before we end, let us comment on the antiferromagnetic case. As it is well known the antiferromagnetic case (corresponding to $\Delta <-1$) has a continuous limit near the critical point where the model admits a description as a non-linear sigma model that is Lorentz invariant \cite{Baxter:1982zz,Faddeev:1996iy}. The ground state of the anti-ferromagnetic state is built up from the $|0\rangle$ state by populating the reference state by spinons $B(z_i)$ until exactly half of the spins are flipped. Naively, such a state has no hope of being invariant under the boost operator $K$, since $K$ shifts all the rapidities $z_i$ uniformly when commuting through the $B(z_i)$. The rescue, as proposed in \cite{Frahm:1991hr}, is that such a shift is not consistent with the Bethe ansatz, assuming that we first take periodic boundary conditions before taking the thermodynamic limit. The Bethe ansatz takes the form of a non-trivial integral equation constraining the density of states which would be violated if we shift the rapidities $z_i$ uniformly, thus violating the boundary condition we started out with. It is believed that some non-trivial interplay of restoring the Bethe ansatz and the shift of $z_i$ should eventually leave the ground state invariant. This has not been shown directly, although \cite{Frahm:1991hr} took an alternative route in demonstrating that the true ground state is a $K$ eigenstate perturbatively order by order in $1/|\Delta|$. We believe this amazing interplay is closely related to restoring the cyclic and separating property of the ground state in the antiferromagnetic state. A thorough exploration however is beyond the scope of the current paper.

%
%
%

\section{Conclusions and Outlook}

In this paper, our goal is to explore the description of a Lorentzian space-time via tensor networks. This is the first step towards building a controllable tensor network description of more exotic backgrounds, such as more generic curved backgrounds and perhaps ones mimicking AdS spaces and AdS black holes, which would shed light on a covariant understanding of the error correcting code/tensor network description of the AdS/CFT correspondence which has achieved a lot of successes in static spacetimes \cite{Pastawski:2015qua,Hayden:2016cfa}.

We approached the problem by first comparing the framework of algebraic quantum field theory with the tensor network, and find that the tensor network can very naturally fit into that picture, allowing one to formulate interesting questions about the tensor networks in very much the same language as is used in the AQFT.
With inspirations from AQFT, we define notions of causality, Cauchy surfaces and different frames, and also specify the unitary transformations relating these observers.

In the second half of the paper, we consider explicit toy models based on fermions. In particular, as a first exercise, we would like to illustrate that physics of the Minkowski space can be  captured to some extent in these simple settings, allowing one to explore questions such as the Unruh effect, which is a close relative of Hawking temperature and Hawking radiation in black holes. In specific limits the dispersion relations of our models show clear signature of Lorentz invariance -- the dispersion relation becomes linear without the doubling problem.
We constructed a boost operator appropriate for the discrete spacetime and solve its spectra explicitly in the limit where dispersion approaches a linear one, and demonstrate that they look completely parallel to modes observed by Rindler observers.  We also give support to this construction of the boost by comparing it with the actual half-space entanglement Hamiltonian which in the continuous case should indeed be equal to the boost operator.

Causal structure is subsequently studied based on anti-commutation relations, invoking the Einstein locality axiom. One observes the emergence of light cone that does not necessarily coincide with the same choice of graph. Not surprisingly, the effective light cone is controlled by the evolution tensors, which supply the ``meat'' of space-time as the graph supplied a skeleton.

Finally, we generalize these constructions to  integrable models. We find that our naive guesses of the boost operators in the free fermion system is basically an approximation of an operator known to the literature as the corner transfer matrix, which is found to be a close analogue of the Lorentz boost operator in lattice models. This gives extra support to the methods pursued and we hope to generalize our constructions to curved backgrounds, and to higher dimensions in a future publication.

Recently, we also note that the further comprehensive studies in this subject has appeared since, including \cite{Vidal2} and \cite{Qi2} that supply complementary perspectives.

\section*{Acknowledgements}
We would like to thank Xiao-liang Qi for suggesting to look at free fermion correlation functions instead of bosons. We would like to thank Hank Thacker, Yong-Shi Wu and Guifre Vidal for very insightful discussions and suggestions.
We would like to thank Muxin Han, Ce Shen, Gabriel Wong and Jieqiang Wu for discussions and comments. We thank Chen-Te Ma for a critical and meticulous reading of our manuscript. We would also like to thank Jiawen Yan for processing figures.
LYH and ZY would like to thank the Tsinghua Sanya International Mathematics Forum
for hospitality during the workshop and research-in-team program ``Black holes, Quantum
Chaos, and Solvable Quantum Systems'', during which part of this work was undertaken.
LYH acknowledges the support of Fudan University and the Thousands Young Talents Program. LC acknowledges support from  China postdoctoral Science Foundation (Grant No.2016M591593).
AB like to thank Prof. Tadashi Takayanagi for useful discussions. AB is supported by JSPS  Grant-In Aid within JSPS fellowship (17F17023).

%
%
%
%
%
%

\appendix

\section{Free bosons}

In this appendix, we will construct the tensor network that consist of free bosons.

 As the case of free fermions,  we consider a set of bosonic creation and annihilation operators $a^{\dagger}_{n}$ and $a_{n}$. which satisfy the usual commutation relation
 \be
  [a_{n}, a_{n'}^{\dagger}]=\delta_{n,n'}.
\ee
 The Hamiltonian that generates the time evolution is
\be
H=\sum^{2L-1}_{n=-2L+1}h_{n-1,n} =\sum^{2L-1}_{n=-2L+1} \frac{1}{2i}\left(a^{\dagger}_{n-1} a_{n}-a_{n}^{\dagger}a_{n-1}\right).
\ee
Again, over sufficiently small unit of time $\Delta t,$ the time evolution operator $U(\Delta t)$ can be well approximated as follows,
\be
U(\Delta t) = (\prod_i U_{2i, 2i+1} )\,\, (\prod_j U_{2j-1, 2j}),
\ee
where
\be
U_{i-1,i} = \exp( i \Delta t \,\,h_{i-1,i} )
\ee

Following the same procedure as in the fermion case, the eigen-operators  are given by
\bea
a_{p}=\sum_{n=-L+\frac{1}{2}}^{L-\frac{1}{2}}\Big( \lambda^+_1 \e^{i p n} a_{2n-1}+\lambda^+_2 \e^{i p n}a_{2n}\Big), \\
b_{p}=\sum_{n=-L+\frac{1}{2}}^{L-\frac{1}{2}}\Big( \lambda^-_1 \e^{i p n} a_{2n-1}+\lambda^-_2 \e^{i p n}a_{2n}\Big),
\eea
which follow from the eigen-equations $ U^{\dagger}a_p U=E a_p $ and $ U^{\dagger}b_p U=E^* a_p $. Furthermore, the explicit form for these eigen-equations are
\bea
\frac{\lambda^+_1}{ \lambda^+_2}=\frac{\e^{-i p}cs-cs}{c^2+s^2\e^{-i p}-E}=\frac{E-\e^{i p}s^2-c^2}{cs \e^{i p}-cs},\\
\frac{\lambda^-_1}{ \lambda^-_2}=\frac{\e^{-i p}cs-cs}{c^2+s^2\e^{-i p}-E^*}=\frac{E^*-\e^{i p}s^2-c^2}{cs \e^{i p}-cs}
\eea
where $c$ and $s$ denote $\cos\Delta t/2$ and $\sin\Delta t/2$ respectively .
It is straightforward to see that
\bea
\frac{\lambda_1^+}{\lambda_2^+}\cdot\frac{\lambda_1^{-*}}{\lambda_2^{-*}}=-1.
\eea
From this one can immediately conclude that
\bea
  [a_{p}, b_{p'}]=  [a_{p}, b_{p'}^{\dagger}]=0.
\eea

Beside this the following relation has to hold,
\bea
\lambda_1^{+} \lambda_1^{+*}+\lambda_2^+ \lambda_2^{+*}=1
\eea
to get,
 \bea
  [a_{p}, a_{p'}^{\dagger}]=\delta_{p,p'}
  \eea
  But for the other pair we end up with ,
  \bea
  \lambda_1^- \lambda_1^{-*}+\lambda_2^-\lambda_2^{-*}=1
  \eea
  and
  \bea
  ~~~[b_{p}^{\dagger}, b_{p'}]=-\delta_{p,p'}.
 \eea
 We always end up with the opposite sign for this commutator, unlike for the fermionic case where both the signs are correct.

\section{(Anti-)Commutators and correlation functions of the fermionic tensor network model}
We relegate computations of the correlation function of the fermions into the appendix.

\be\label{c00f}
\begin{aligned}
\langle0|a_{2x}(x,0)a^{\dagger}_{2y}(y,t)|0\rangle=&\mathcal{N}^2\sum_{p}e^{-i\,p\,(x-y)}e^{i\,|\chi | t}\left(\frac{c^2}{2(2c^2+s^2\cos p+s^2)}\right.\\
&\left.+\frac{(s \sin p +\sqrt{(1-\cos p)(2c^2+s^2\cos p+s^2)})^2}{4(1-\cos p)(2c^2+s^2\cos p+s^2)}\right),
\end{aligned}
\ee
\be
\langle0|a_{2x}(x,0)a^{\dagger}_{2y+1}(y,t)|0\rangle=-\mathcal{N}^2\sum_{p}e^{-i\,p\,(x-y)}e^{i\,|\chi | t}\frac{e^{i\frac{p}{2}}c\sqrt{(1-\cos p)(2c^2+s^2\cos p+s^2)}}{2\sin \frac{p}{2}(2c^2+s^2\cos p+s^2)},
\ee
\be
\langle0|a_{2x+1}(x,0)a^{\dagger}_{2y}(y,t)|0\rangle=-\mathcal{N}^2\sum_{p}e^{-i\,p\,(x-y)}e^{i\,|\chi | t}\frac{e^{-i\frac{p}{2}}c\sqrt{(1-\cos p)(2c^2+s^2\cos p+s^2)}}{2\sin \frac{p}{2}(2c^2+s^2\cos p+s^2)},
\ee
\be\label{c11f}
\begin{aligned}
\langle0|a_{2x+1}(x,0)a^{\dagger}_{2y+1}(y,t)|0\rangle=&\mathcal{N}^2\sum_{p}e^{-i\,p\,(x-y)}e^{i\,|\chi | t}\left(\frac{c^2}{2(2c^2+s^2\cos p+s^2)}\right.\\
&\left.+\frac{(s \sin p -\sqrt{(1-\cos p)(2c^2+s^2\cos p+s^2)})^2}{4(1-\cos p)(2c^2+s^2\cos p+s^2)}\right),
\end{aligned}
\ee

In the limit $c\rightarrow 0, s\rightarrow 1$, we have
\be
\langle0|a_{2x}(x,0)a^{\dagger}_{2y}(y,t)|0\rangle=\mathcal{N}^2\sum_{p\geq0}e^{-i\,p\,(x-y)}e^{i\,|p| t},
\ee
\be
\langle0|a_{2x}(x,0)a^{\dagger}_{2y+1}(y,t)|0\rangle=\langle0|a_{2x+1}(x,0)a^{\dagger}_{2y}(y,t)|0\rangle=0,
\ee
\be
\langle0|a_{2x+1}(x,0)a^{\dagger}_{2y+1}(y,t)|0\rangle=\mathcal{N}^2\sum_{p<0}e^{-i\,p\,(x-y)}e^{i\,|p| t}.
\ee
When $p$ is summed, the results become
\be
\langle0|a_{2x}(x,0)a^{\dagger}_{2y}(y,t)|0\rangle=\frac{1}{2L}\frac{1-e^{-i\pi(x-y-t)}}{1-e^{-i\frac{\pi}{L}(x-y-t)}},
\ee
\be
\langle0|a_{2x}(x,0)a^{\dagger}_{2y+1}(y,t)|0\rangle=\langle0|a_{2x+1}(x,0)a^{\dagger}_{2y}(y,t)|0\rangle=0,
\ee
\be
\langle0|a_{2x+1}(x,0)a^{\dagger}_{2y+1}(y,t)|0\rangle=\frac{1}{2L}\frac{e^{i\pi(x-y+t)}-1}{1-e^{-i\frac{\pi}{L}(x-y+t)}}.
\ee
When we take the limit $L\rightarrow\infty$, the above correlation functions become
\be
\langle0|a_{2x}(x,0)a^{\dagger}_{2y}(y,t)|0\rangle=\frac{1-e^{-i\pi(x-y-t)}}{i 2\pi(x-y-t)},
\ee
\be
\langle0|a_{2x}(x,0)a^{\dagger}_{2y+1}(y,t)|0\rangle=\langle0|a_{2x+1}(x,0)a^{\dagger}_{2y}(y,t)|0\rangle=0,
\ee
\be
\langle0|a_{2x+1}(x,0)a^{\dagger}_{2y+1}(y,t)|0\rangle=\frac{e^{i\pi(x-y+t)}-1}{i 2\pi(x-y+t)}.
\ee
In the limit $c\rightarrow 1, s\rightarrow 0$, we have
\be
\langle0|a_{2x}(x,0)a^{\dagger}_{2y}(y,t)|0\rangle=\frac{1}{2}\mathcal{N}^2\sum_{p}e^{-i\,p\,(x-y)},
\ee
\be
\langle0|a_{2x}(x,0)a^{\dagger}_{2y+1}(y,t)|0\rangle=-\mathcal{N}^2\sum_{p}e^{-i\,p\,(x-y)}\frac{e^{i\frac{p}{2}}\sqrt{2(1-\cos p)}}{4\sin \frac{p}{2}},
\ee
\be
\langle0|a_{2x+1}(x,0)a^{\dagger}_{2y}(y,t)|0\rangle=-\mathcal{N}^2\sum_{p}e^{-i\,p\,(x-y)}\frac{e^{-i\frac{p}{2}}\sqrt{2(1-\cos p)}}{4\sin \frac{p}{2}},
\ee
\be
\langle0|a_{2x+1}(x,0)a^{\dagger}_{2y+1}(y,t)|0\rangle=\frac{1}{2}\mathcal{N}^2\sum_{p}e^{-i\,p\,(x-y)}.
\ee
There is no time dependence in the correlation functions in the limit $c\rightarrow 1, s\rightarrow 0$, as expected of a theory with trivial dispersion relation. \\

Summing $p$, the results become
\be
\langle0|a_{2x}(x,0)a^{\dagger}_{2y}(y,t)|0\rangle=\frac{1}{2}\delta_{xy},
\ee
\be
\langle0|a_{2x}(x,0)a^{\dagger}_{2y+1}(y,t)|0\rangle=-\frac{1}{4L}\frac{2-e^{-i\pi(x-y-\frac{1}{2})}-e^{i\pi(x-y-\frac{1}{2})}}{1-e^{-i\frac{\pi}{L}(x-y-\frac{1}{2})}}
=-\frac{1}{2L}\frac{1}{1-e^{-i\frac{\pi}{L}(x-y-\frac{1}{2})}},
\ee
\be
\langle0|a_{2x+1}(x,0)a^{\dagger}_{2y}(y,t)|0\rangle=-\frac{1}{4L}\frac{2-e^{-i\pi(x-y+\frac{1}{2})}-e^{i\pi(x-y+\frac{1}{2})}}{1-e^{-i\frac{\pi}{L}(x-y+\frac{1}{2})}}
=-\frac{1}{2L}\frac{1}{1-e^{-i\frac{\pi}{L}(x-y+\frac{1}{2})}},
\ee
\be
\langle0|a_{2x+1}(x,0)a^{\dagger}_{2y+1}(y,t)|0\rangle=\frac{1}{2}\delta_{xy}.
\ee

When we take the limit $L\rightarrow\infty$, the above correlation functions become
\be
\langle0|a_{2x}(x,0)a^{\dagger}_{2y}(y,t)|0\rangle=\frac{1}{2}\delta(x-y),
\ee
\be
\langle0|a_{2x}(x,0)a^{\dagger}_{2y+1}(y,t)|0\rangle=-\frac{2-e^{-i\pi(x-y-\frac{1}{2})}-e^{i\pi(x-y-\frac{1}{2})}}{4\pi i(x-y-\frac{1}{2})}
=-\frac{1}{2\pi i(x-y-\frac{1}{2})},
\ee
\be
\langle0|a_{2x+1}(x,0)a^{\dagger}_{2y}(y,t)|0\rangle=-\frac{2-e^{-i\pi(x-y+\frac{1}{2})}-e^{i\pi(x-y+\frac{1}{2})}}{4\pi i(x-y+\frac{1}{2})}
=-\frac{1}{2\pi i(x-y+\frac{1}{2})},
\ee
\be
\langle0|a_{2x+1}(x,0)a^{\dagger}_{2y+1}(y,t)|0\rangle=\frac{1}{2}\delta(x-y).
\ee

\section{More details on integrable models} \label{integrable_mod}
We will illustrate in detail how to obtain eigen-wavefunctions in the 6-vertex model, and compute some of the correlation functions in detail illustrating the lightcone effect.

\begin{figure}[!h]
		\centering
		\includegraphics[width=8cm]{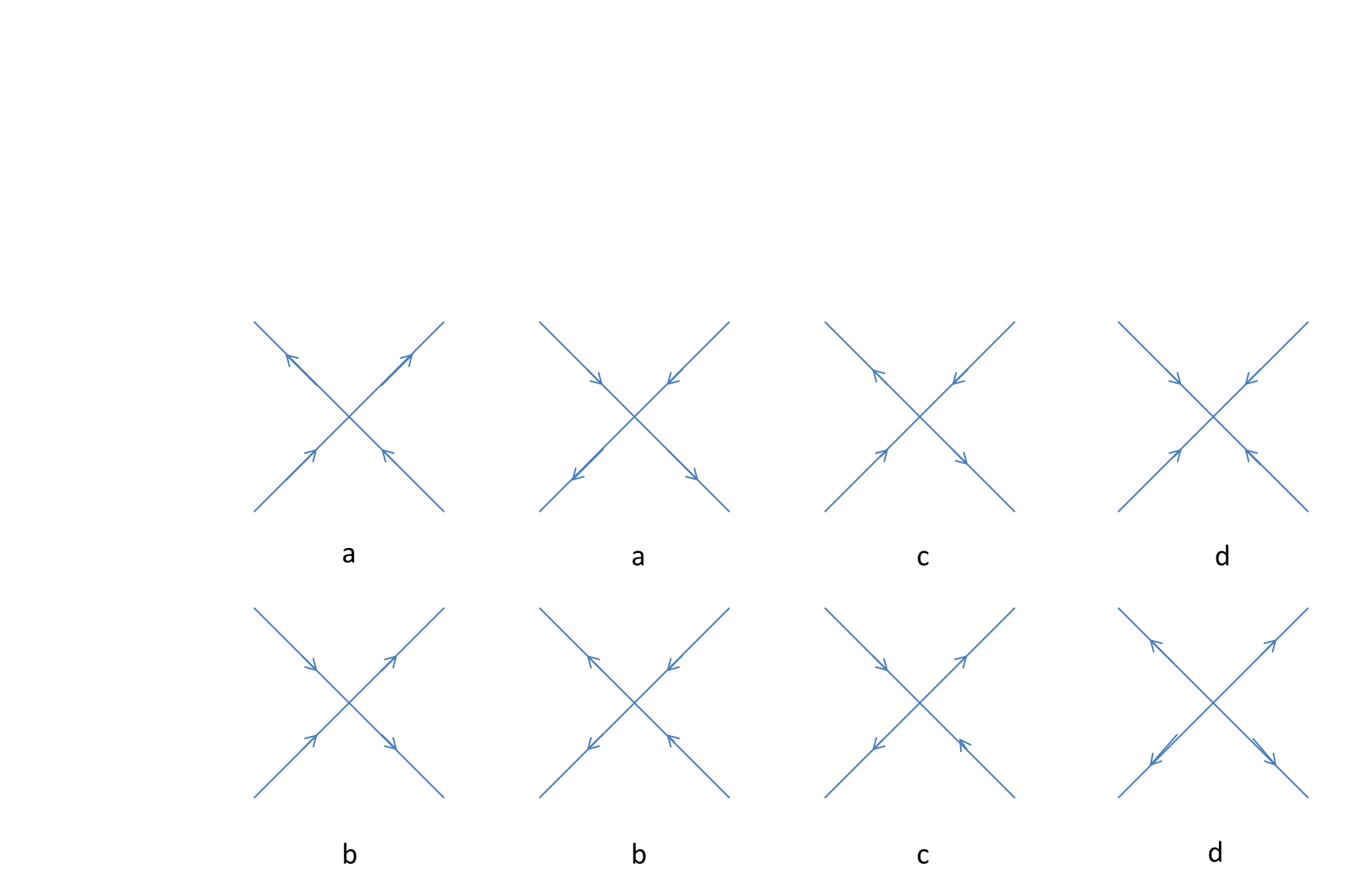}
		\caption{The 8-vertex model.}
		\label{hamiltonian}
\end{figure}

We consider an integrable model. The classical statistical model is defined as in figure \ref{hamiltonian}, which gives the assigned weights to each local configuration.
\\	

The partition sum is the weighted sum of all configurations. The transfer matrices defined for example in (\ref{transfer2}, \ref{analytic_continue}) can be read-off from the statistical model as follows. In those cases, they correspond to the ``row-to-row'' transfer matrices. It is illustrated in
\begin{figure}[!h]
		\centering
		\includegraphics[width=8cm]{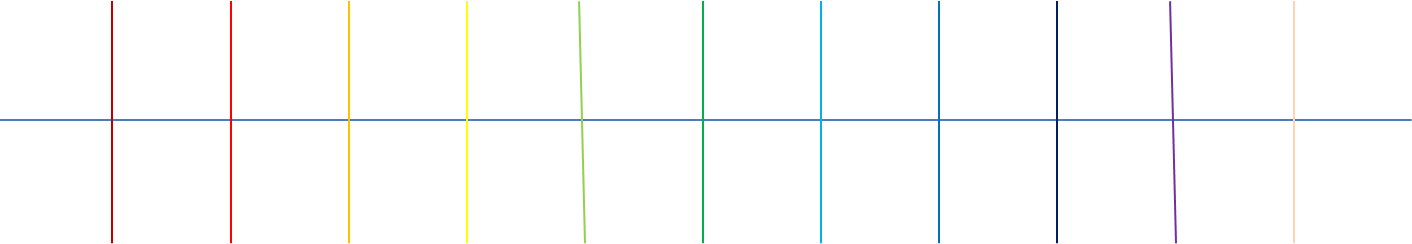}
		\caption{``row-to-row'' transfer matrices.}
		\label{transfer matrix}
\end{figure}	

The incoming index correspond to configurations of links at the bottom of the layer, and the out-going index correspond to the configurations of links at the top layer.  Each element of the matrix is obtained by doing the weighted sum over links connecting vertices in the middle for given fixed boundary conditions at the top and bottom of the layer.

The condition for integrability is that transfer matrices $T$ and $T^{'}$ commute, which leads to a parametrisation of the weights given as the following:
\be \label{BaxterparamXYZ}
a:b:c:d=snh(\lambda-u):snh(u):snh(\lambda):k\cdot snh(\lambda)snh(\lambda-u)snh(u)
\ee
k and $\lambda$ are fixed constants associated with the model and u is variable.
These are elliptic functions defined as
\be
snh(u) = -i k^{-1/2} \frac{H(iu)}{\Theta(iu)},
\ee
where $H$ and $\Theta$ are theta functions. The details of these functions can be found in \cite{Baxter:1982zz}, chapter 15.
When $k\to 0$, $snh(u) \to \sin u$.
In this limit, the 8-vertex (or XYZ) model then reduces to the 6 vertex (XXZ) model. i.e. (\ref{BaxterparamXYZ}) becomes
\be \label{XXZpar}
a = \sin(\lambda - u), \qquad b = \sin u, \qquad c= \sin \lambda, \qquad \Delta = -\cos\lambda.
\ee
If we further replace $\lambda = \pi + \epsilon $ while taking $\epsilon , u\to 0$ and $\epsilon/u$ fixed,  we recover the XXX model.

\begin{figure}[!h]
		\centering
		\includegraphics[width=8cm]{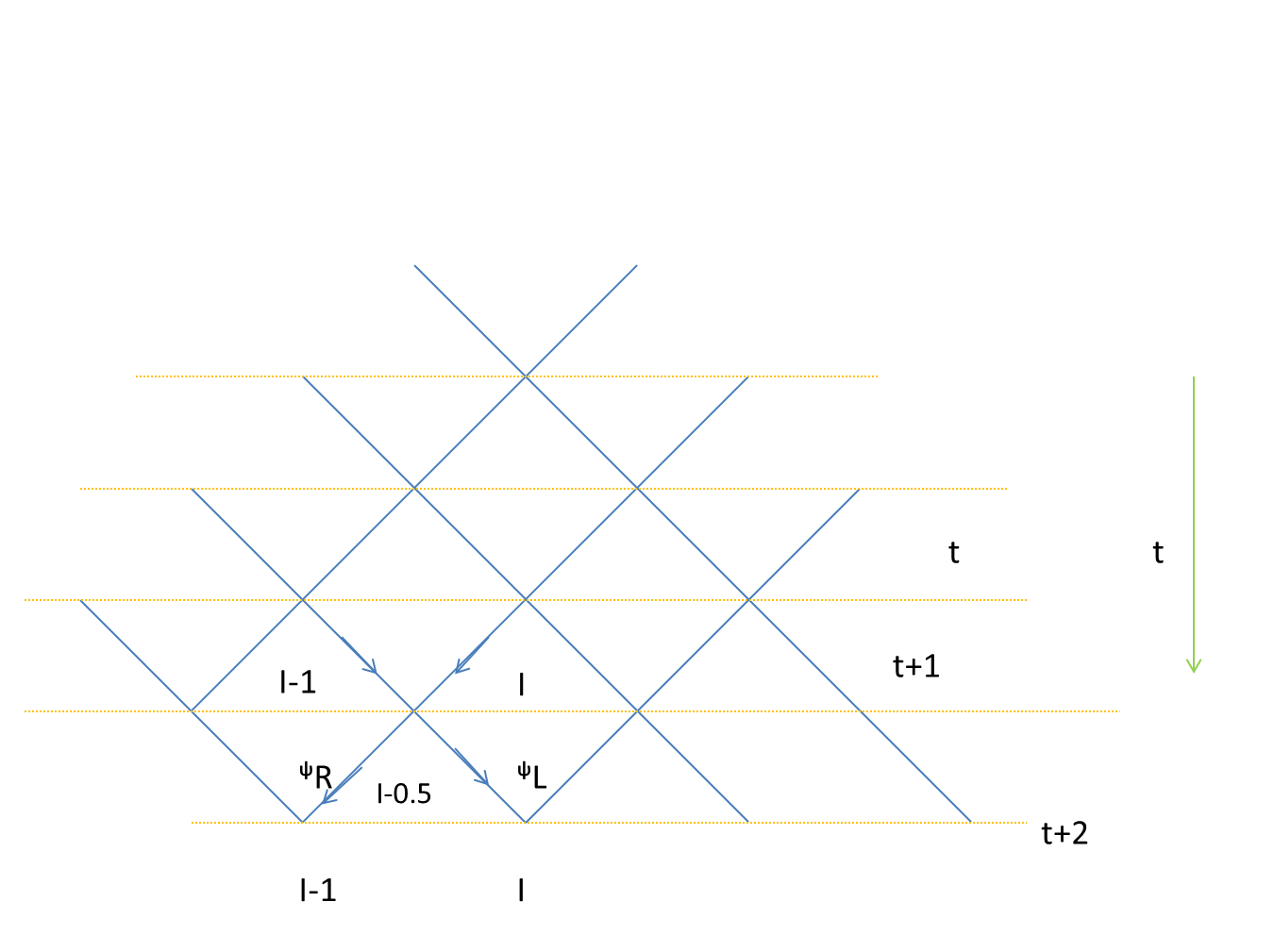}
		\caption{The lattice of the integrable model. The picture illustrates the definition of the model, where in the classical partition function sum over classical configurations of these links, each pattern of link connected to a vertex is assigned some weight. }
		\label{fig:8vertex}
\end{figure}

In our tensor network construction based on local unitaries, we are interested not in the row-to-row transfer matrix, but the ``diagonal-to-diagonal'' transfer matrix, obtained by rotating the square lattice by 45 degrees. Each individual vertex can now be viewed as a matrix with incoming indices from the bottom pair of links, and outgoing indices from the top pair of links.
Now, requiring that each such local transformation to be a unitary matrix, we have
\be
|a|^2=1,
\ee
\be
|b|^2+|c|^2=1
\ee
and
\be
bc^*+b^*c=0.
\ee
Here we put $a=1$, $b$ a real positive number and $c$ a pure imaginary number $i|c|$.\\
We consider the lattice as in figure \ref{fig:8vertex}. The lattice repeats itself every two layers of evolution.
\\
We would like to solve for eigen wavefunctions based on the coordinate Bethe Ansatz.  Eigen modes of the diagonal-to-diagonal transfer matrix obtained via the coordinate Bethe Ansatz can be found in \cite{Truong:1983kz}. Our parametrization of the lattice is somewhat different from his, and we will solve it from scratch, borrowing heavily his strategy. We will then show that indeed our solution could have been obtained from his via appropriate reparametrization.
It is well known that the 6-vertex partition function satisfies a conversation of arrows. Namely for each of the 6 vertices shown in figure \ref{hamiltonian}, it preserves the number of ``down'' arrow across the vertex. Therefore,
one can take the convention that a down arrow represents a particle, and consider ``particle excitations'' over a reference state with all arrows up.	\\
We have the eigen equations
\be
\frac{b^2}{a^2}\psi_L(I-1)+\frac{bc}{a^2}\psi_R(I-1)+\frac{c^2}{a^2}\psi_L(I)+\frac{bc}{a^2}\psi_R(I)=\Lambda \psi_L(I),
\ee
\be
\frac{bc}{a^2}\psi_L(I-1)+\frac{c^2}{a^2}\psi_R(I-1)+\frac{bc}{a^2}\psi_L(I)+\frac{b^2}{a^2}\psi_R(I)=\Lambda \psi_R(I-1).
\ee
We have the ansatz
\be
\psi_L(I)=\alpha e^{i p I},
\ee
\be
\psi_R(I)=\beta e^{i p I}.
\ee
The eigen equations become
\be\label{eigen1}
\frac{b^2}{a^2}\alpha e^{-i p}+\frac{bc}{a^2}\beta e^{-i p}+\frac{c^2}{a^2}\alpha+\frac{bc}{a^2}\beta=\Lambda \alpha,
\ee
\be
\frac{bc}{a^2}\alpha e^{-i p}+\frac{c^2}{a^2}\beta e^{-i p}+\frac{bc}{a^2}\alpha+\frac{b^2}{a^2}\beta=\Lambda \beta e^{-i p}.
\ee
Solving these two equations, we have eigenvalues
\be\label{Lambda}
\Lambda_{\pm}=b^2 \cos p+c^2\pm i \sqrt{2b^2 \cos^2\frac{p}{2}(b^2-b^2 \cos p-2c^2)},
\ee
which we note is again an even function in $p$.
For $\Lambda_+$, we get
\be
\psi_L(I)=\alpha_+ e^{i p I},
\ee
\be
\psi_R(I)=\beta_+ e^{i p I},
\ee
where
\be \label{solr1}
r_1\equiv\frac{\alpha_+}{\beta_+}=\frac{i e^{-i\frac{p}{2}}}{2c}\left(\sqrt{2}\sqrt{b^2-b^2\cos p-2c^2}-2b \sin \frac{p}{2}\right).
\ee
For $\Lambda_-$, we get
\be
\psi_L(I)=\alpha_- e^{i p I},
\ee
\be
\psi_R(I)=\beta_- e^{i p I},
\ee
where
\be \label{solr2}
r_2\equiv\frac{\alpha_-}{\beta_-}=\frac{i e^{-i\frac{p}{2}}}{2c}\left(-\sqrt{2}\sqrt{b^2-b^2\cos p-2c^2}-2b \sin \frac{p}{2}\right).
\ee
We define
\be\label{p}
|p_\pm\rangle=\mathcal{N}\sum_{x=-L+\frac{1}{2}}^{L-\frac{1}{2}}e^{ipx}(\alpha_\pm\sigma(2x)+\beta_\pm\sigma(2x+1))|0\rangle,
\ee
where
\be
\mathcal{N}=\frac{1}{\sqrt{2L}},
\ee
\be
p=n\frac{2\pi}{2L}, \quad n=-L,-L+1,\cdots, L-1.
\ee
Imposing normalization condition
\be \label{norm}
\langle p_\pm|p_\pm\rangle=1,
\ee
gives
\be
|\alpha_\pm|^2+|\beta_\pm|^2=1.
\ee
From
\be
\langle p_+|p_-\rangle=0,
\ee
we get
\be
\alpha_+^*\alpha_-+\beta_+^*\beta_-=0.
\ee
Similarly, from
\be
\langle p_-|p_+\rangle=0,
\ee
we get
\be
\alpha_-^*\alpha_++\beta_-^*\beta_+=0.
\ee

Here, we note some important difference from the free fermion case which underlies the fact that the reference state here is a direct product state with no entanglement.
One can see that in the  limit $c\to 0$, the dispersion relation (\ref{Lambda}) also approaches the linear one as in the case of the free fermion. This is not surprising.
It is well known that the critical point occurs at $|\Delta| =1$. Comparing with (\ref{XXZpar}), we find that it matches precisely with the $c\to 0$ limit where we recover linear dispersion.
Relativistc feature would naively be recovered as well, as in the case of the fermions. But this is not so. We note that the reference state is a direct product state and so it could not possibly resemble the highly entangled structure of a relativistic ground state. This is captured by the fact that the excitations around the reference ``ground state'' contain both positive and negative energy ones, unlike the case of fermions.

\subsection{Correlation functions}


We define
\be
\sigma(2x,t)|0\rangle=\mathcal{N}\sum_{p}e^{-ipx}\frac{\beta_-e^{i E_p t}|p_+\rangle-\beta_+e^{-i E_p t}|p_-\rangle}{\alpha_+\beta_--\alpha_-\beta_+}
\ee
and
\be
\sigma(2x+1,t)|0\rangle=\mathcal{N}\sum_{p}e^{-ipx}\frac{\alpha_-e^{i E_p t}|p_+\rangle-\alpha_+e^{-i E_p t}|p_-\rangle}{\alpha_-\beta_+-\alpha_+\beta_-}.
\ee
With (\ref{p}), the above two equations become
\be
\begin{aligned}
\sigma(2x,t)|0\rangle=\mathcal{N}^2\sum_{p,z}\frac{e^{ip(z-x)}}{\alpha_+\beta_--\alpha_-\beta_+}&\left[(\alpha_+\beta_-e^{iE_p t}-\alpha_-\beta_+e^{-iE_p t})\sigma(2z)|0\rangle\right.\\
&+\left.(\beta_+\beta_-e^{iE_p t}-\beta_+\beta_-e^{-iE_p t})\sigma(2z+1)|0\rangle\right]
\end{aligned}
\ee
and
\be
\begin{aligned}
\sigma(2x+1,t)|0\rangle=\mathcal{N}^2\sum_{p,z}\frac{e^{ip(z-x)}}{\alpha_-\beta_+-\alpha_+\beta_-}&\left[(\alpha_+\alpha_-e^{iE_p t}-\alpha_+\alpha_-e^{-iE_p t})\sigma(2z)|0\rangle\right.\\
&+\left.(\alpha_-\beta_+e^{iE_p t}-\alpha_+\beta_-e^{-iE_p t})\sigma(2z+1)|0\rangle\right].
\end{aligned}
\ee
We also have
\be
\langle0|\sigma(x)\sigma(y)|0\rangle=\delta_{xy}.
\ee
From the above equations, we can get the correlation functions
\be\label{c00in}
\begin{aligned}
\langle0|\sigma(2y,0)\sigma(2x,t)|0\rangle&=\mathcal{N}^2\sum_{p}e^{ip(y-x)}\frac{\alpha_+\beta_-e^{i E_p t}-\alpha_-\beta_+e^{-i E_p t}}{\alpha_+\beta_--\alpha_-\beta_+}\\
&=\mathcal{N}^2\sum_{p}e^{ip(y-x)}\frac{r_1e^{i E_p t}-r_2e^{-i E_p t}}{r_1-r_2},
\end{aligned}
\ee
\be
\begin{aligned}
\langle0|\sigma(2y+1,0)\sigma(2x,t)|0\rangle&=\mathcal{N}^2\sum_{p}e^{ip(y-x)}\frac{2 i \beta_+\beta_-\sin(E_p t)}{\alpha_+\beta_--\alpha_-\beta_+}\\
&=\mathcal{N}^2\sum_{p}e^{ip(y-x)}\frac{2 i\sin(E_p t)}{r_1-r_2},
\end{aligned}
\ee
\be
\begin{aligned}
\langle0|\sigma(2y,0)\sigma(2x+1,t)|0\rangle&=\mathcal{N}^2\sum_{p}e^{ip(y-x)}\frac{2 i \alpha_+\alpha_-\sin(E_p t)}{\alpha_-\beta_+-\alpha_+\beta_-}\\
&=\mathcal{N}^2\sum_{p}e^{ip(y-x)}\frac{2 i\sin(E_p t)}{\frac{1}{r_1}-\frac{1}{r_2}}
\end{aligned}
\ee
and
\be\label{c11in}
\begin{aligned}
\langle0|\sigma(2y+1,0)\sigma(2x+1,t)|0\rangle&=\mathcal{N}^2\sum_{p}e^{ip(y-x)}\frac{\alpha_-\beta_+e^{i E_p t}-\alpha_+\beta_-e^{-i E_p t}}{\alpha_-\beta_+-\alpha_+\beta_-}\\
&=\mathcal{N}^2\sum_{p}e^{ip(y-x)}\frac{r_2e^{i E_p t}-r_1e^{-i E_p t}}{r_2-r_1}.
\end{aligned}
\ee
From (\ref{eigen1}), we have the relations
\be
r_1\equiv\frac{\alpha_+}{\beta_+}=\frac{bc(1+\frac{1}{z})}{\Lambda_1-\frac{b^2}{z}-c^2}
\ee
and
\be
r_2\equiv\frac{\alpha_-}{\beta_-}=\frac{bc(1+\frac{1}{z})}{\Lambda_2-\frac{b^2}{z}-c^2},
\ee
where
\be
z=e^{ip},\quad \Lambda_1=e^{i E_p},\quad \Lambda_1=e^{-i E_p}.
\ee
As already noted in the previous section, in the limit $b\rightarrow 1$ and $c\rightarrow 0$, the energy approaches
\be
E_p=|p|.
\ee
The correlation functions become
\be
\langle0|\sigma(2y,0)\sigma(2x,t)|0\rangle=\frac{1}{2L}\frac{2i\sin(\pi(y-x-t))}{e^{i\frac{\pi}{L}(y-x-t)}-1},
\ee
\be
\langle0|\sigma(2y+1,0)\sigma(2x,t)|0\rangle=i\frac{|c|}{2L}\sum_p e^{ip(y-x+\frac{1}{2})}\frac{\sin(pt)}{\sin\frac{p}{2}},
\ee
\be
\langle0|\sigma(2y,0)\sigma(2x+1,t)|0\rangle=i\frac{|c|}{2L}\sum_p e^{ip(y-x-\frac{1}{2})}\frac{\sin(pt)}{\sin\frac{p}{2}},
\ee
\be
\langle0|\sigma(2y+1,0)\sigma(2x+1,t)|0\rangle=\frac{1}{2L}\frac{2i\sin(\pi(y-x-t))}{e^{i\frac{\pi}{L}(y-x-t)}-1}.
\ee
 With (\ref{c00in}) to (\ref{c11in}), we plot the correlation functions of integrable model with different parameters. We use the notations $C_{00}\equiv\langle0|\sigma(0,0)\sigma(2x,t)|0\rangle$ and $C_{01}\equiv\langle0|\sigma(0,0)\sigma(2x+1,t)|0\rangle$. In the following figures, we take $L=200$. We can see the light cone clearly in the figures.
\begin{figure}[!h]
		\centering
		\includegraphics[width=10cm]{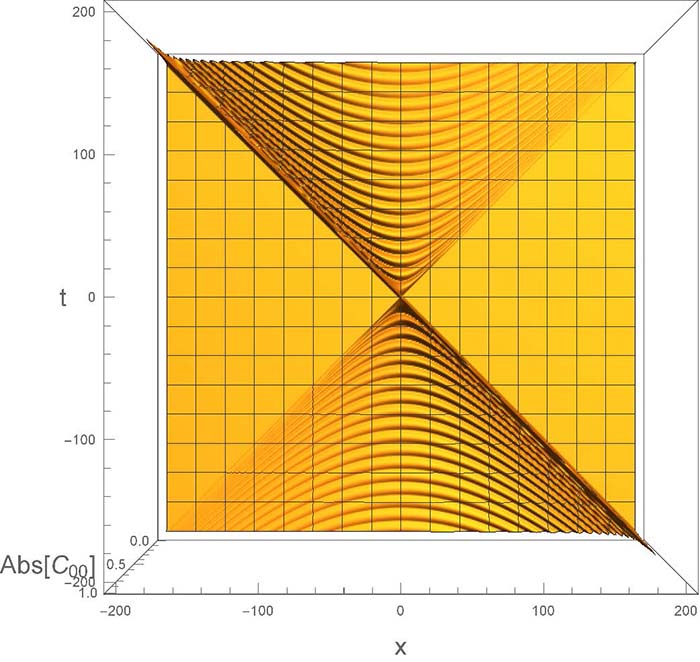}
		\caption{Absolute value of correlation functions $\langle0|\sigma(0,0)\sigma(2x,t)|0\rangle$ with $b=0.99$}
		\label{abs-part-c00-integrable-L200-b-equals-0-99}
\end{figure}\\
\begin{figure}[!h]
		\centering
		\includegraphics[width=10cm]{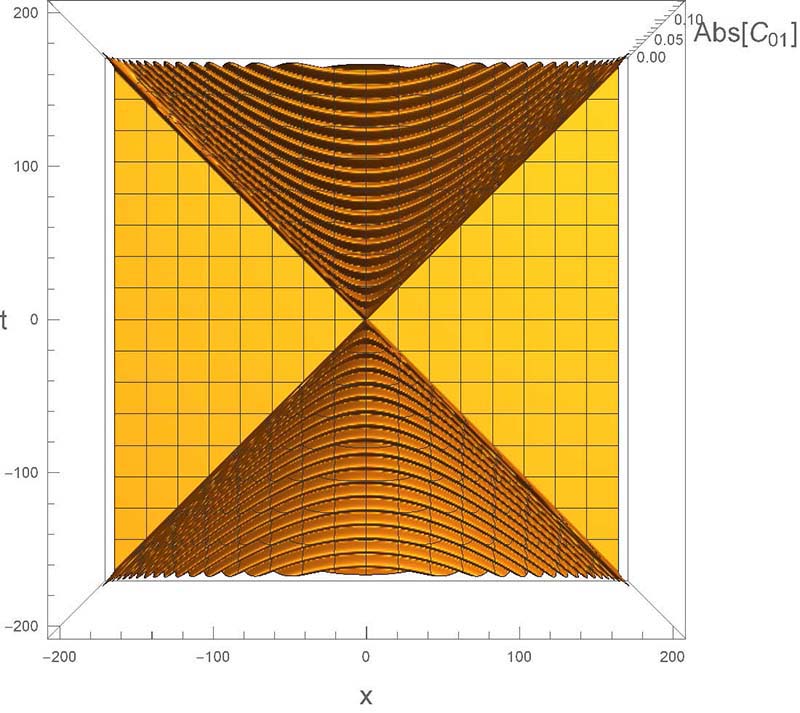}
		\caption{Absolute value of correlation functions $\langle0|\sigma(0,0)\sigma(2x+1,t)|0\rangle$ with $b=0.99$}
		\label{abs-part-c01-integrable-L200-b-equals-0-99}
\end{figure}\\
\begin{figure}[!h]
		\centering
		\includegraphics[width=10cm]{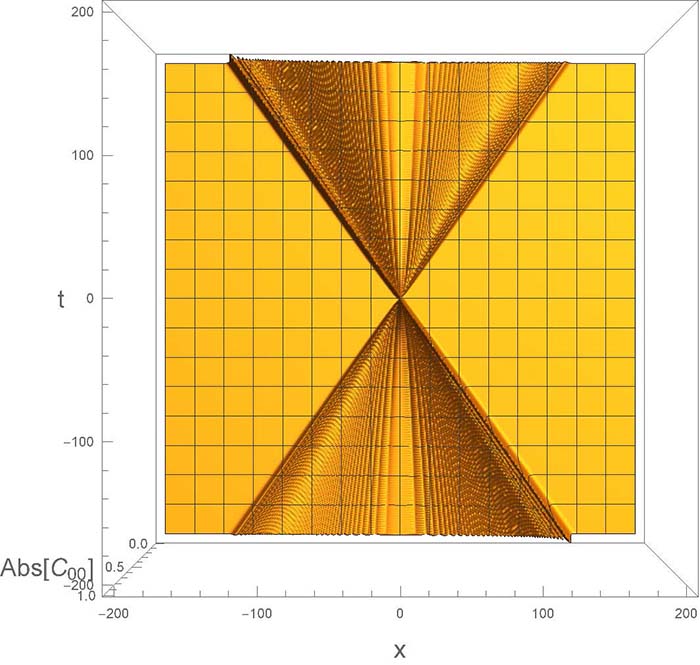}
		\caption{Absolute value of correlation functions $\langle0|\sigma(0,0)\sigma(2x,t)|0\rangle$ with $b=1/\sqrt{2}$}
		\label{abs-part-c00-integrable-L200-b-equals-c}
\end{figure}\\
\begin{figure}[!h]
		\centering
		\includegraphics[width=10cm]{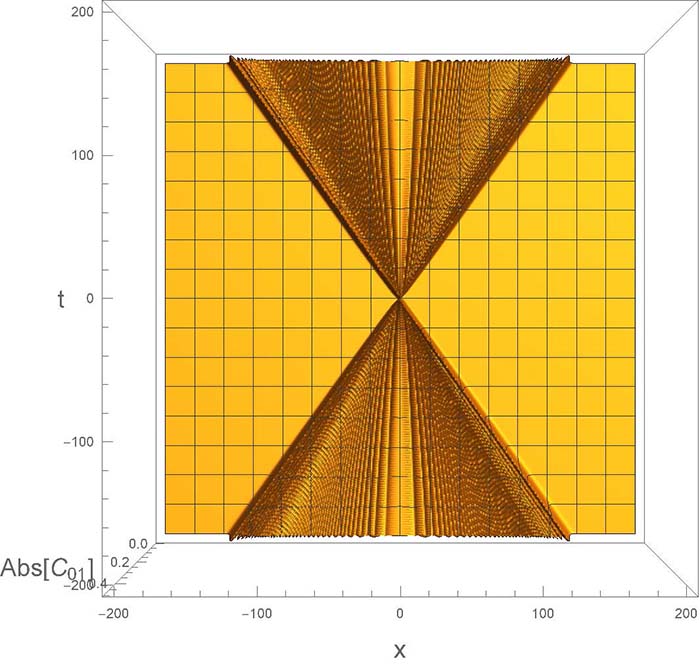}
		\caption{Absolute value of correlation functions $\langle0|\sigma(0,0)\sigma(2x+1,t)|0\rangle$ with $b=1/\sqrt{2}$}
		\label{abs-part-c01-integrable-L200-b-equals-c}
\end{figure}\\
\begin{figure}[!h]
		\centering
		\includegraphics[width=10cm]{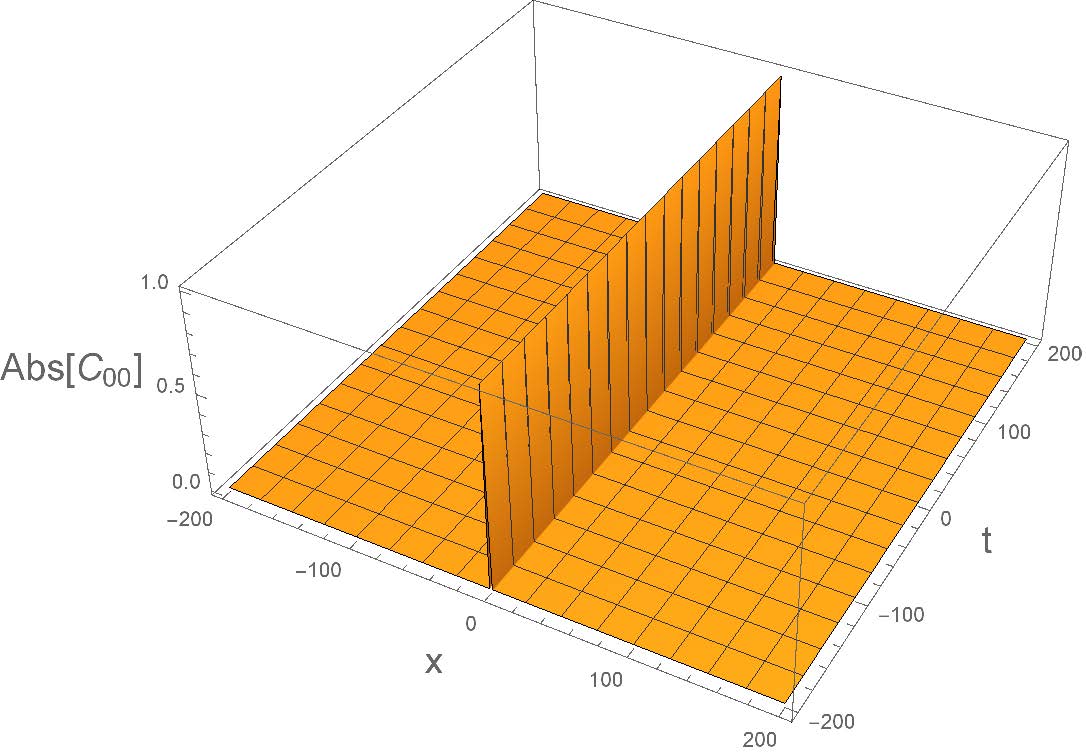}
		\caption{Absolute value of correlation functions $\langle0|\sigma(0,0)\sigma(2x,t)|0\rangle$ with $b=0$}
		\label{abs-part-c00-integrable-L200-b-equals-0}
\end{figure}\\

\begin{figure}[!h]
		\centering
		\includegraphics[width=10cm]{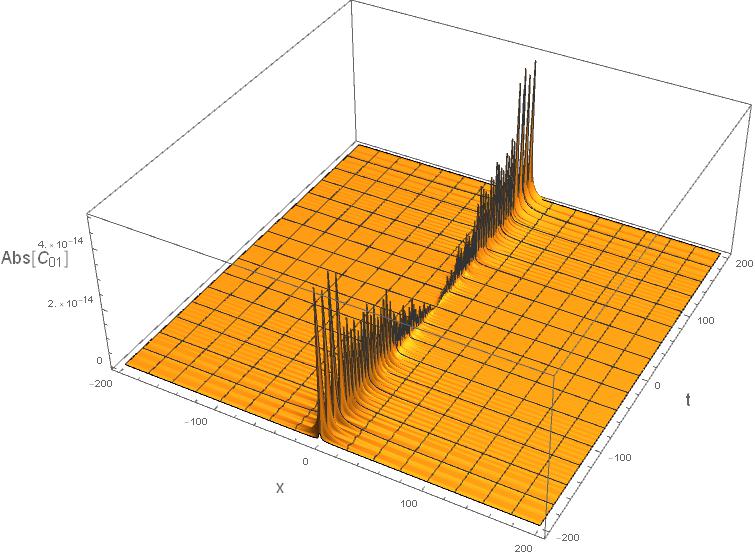}
		\caption{Absolute value of correlation functions $\langle0|\sigma(0,0)\sigma(2x+1,t)|0\rangle$ with $b=0$}
		\label{abs-part-c01-integrable-L200-b-equals-0}
\end{figure}
The light cone is very clear in the figures. It becomes small as the parameter $b$ becomes small.
One major difference that led to a final answer that does not preserve Lorentz invariance arises from the sum over all $p$, whereas half of the $p$ modes
were canceled out in the free fermion theory.

\subsection{Half-space Hamiltonians and their solutions} \label{app:halfH}

Here we would like to discuss also solutions of integrable models with boundaries. In the main text, we have taken the ``boost'' operator as a Hamiltonian, and discussed the corresponding eigenstates. There is something quite interesting, reflecting the fact that there is some sort of a horizon at the boundary. We noted, particularly in the critical limit, that the eigen-wavefunction (\ref{Rind_eigenval}) contains only right moving modes -- since the sign of the momentum is locked with that of the energy. This is the same observation as in the case of continuous (massless free) field theories where holomorphic (or right moving) positive energy modes defined on the complete real line are decomposed as holomorphic modes on half spaces, and similarly for the anti-holomorphic modes. There is no mixing between the left and right moving when we decompose the full space modes into half-space ones.

This can be contrasted with actually solving a generic Hamiltonian defined only on the half plane.  Such scenarios have been considered in the literature before (see for example \cite{Sklyanin:1988yz}).  For a generic integrable model characterized by a solution to the Yang-Baxter equation, a Hamiltonian that has a boundary has to be treated with extra care: extra conditions have to be imposed, in order that the model remains integrable. The model with boundary is characterized by an extra matrix, often called $K_{\pm}$, in addition to the Lax operators $L_{n,f}(v)$. These matrices satisfy extra algebraic equations, which is the boundary analogue of the YB equation.
\begin{figure}[!h]
		\centering
		\includegraphics[width=8cm]{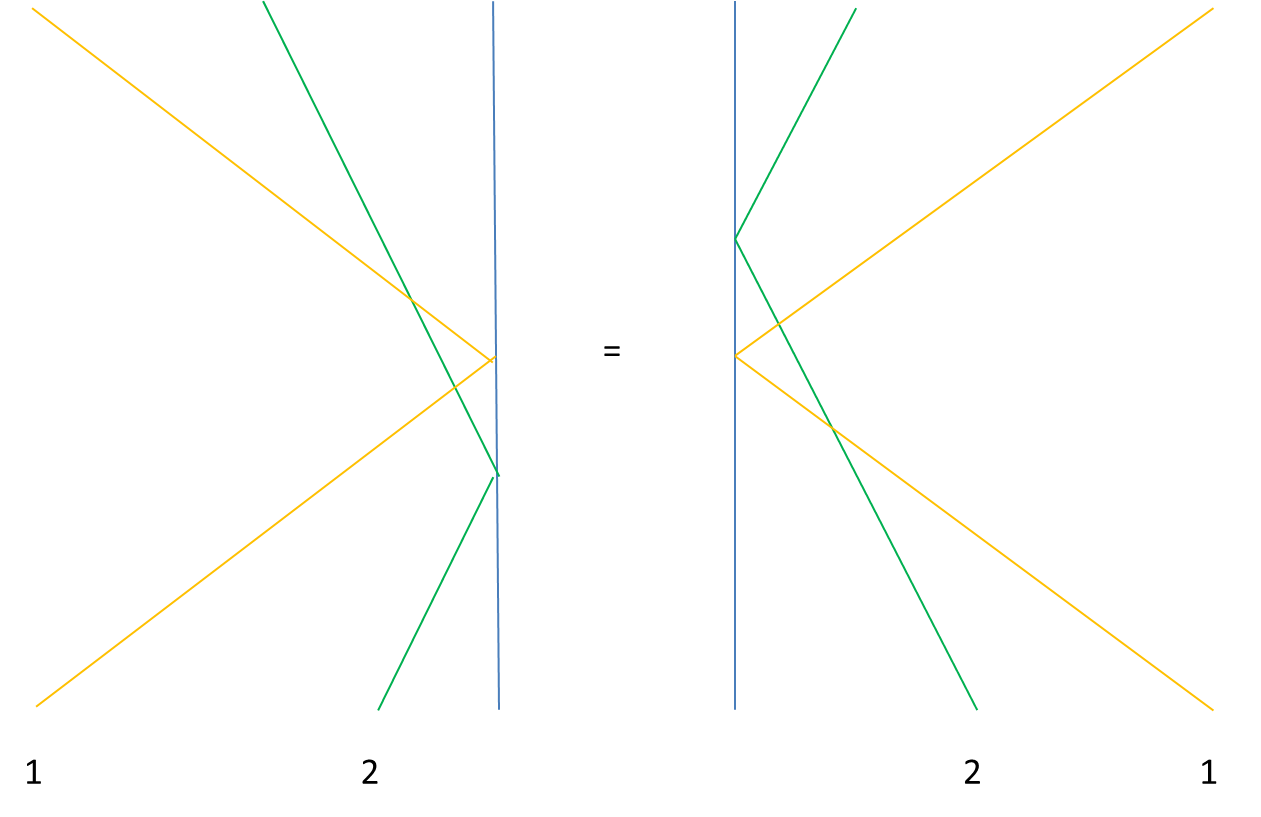}
		\label{reflection equation}
\end{figure}

i.e.
\begin{align}
\begin{split}
&R_{12}(\pm v_1\mp v_2) K^1_{\mp}(v_1) R_{12}(\pm v_1 \pm v_2 - 2 \Theta(\mp) \eta) K^2_{\mp}(v_2) = \\
& K^2_{\mp}(v_2) R_{12}(\pm v_1 \pm v_2 - 2\eta \Theta(\mp)) K^1_{\mp}(v_1) R_{12}(\pm v_1 \mp v_2),
\end{split}
\end{align}
where $\Theta(x)$ is the Heaveside -Theta function that vanishes for $x<0$, and $\eta$ is a model dependent parameter, charecterized by the relation of the $R$ matrix
\be
R^T_{12}(v) R^T_{12}(-v-2\eta) = \tilde \rho(v),
\ee
for some scalar function $\tilde \rho$. The superscripts/subscripts denote the tensor space the operators act on.

We can again take the XXZ model as an example. To construct a tensor network here, we can adopt exactly the same strategy as described in (\ref{inhomotransfer}) where we introduce an inhomogeneous transfer matrix. In the presence of a boundary, it would then take the form
\be
t(v,w) = \textrm{tr}_f(K_+(v) T(v,w) K_-(v) \hat{T}(v,w),
\ee
where
\begin{align}
\begin{split}
T(v,w) &= L_{2N}(v+w) L_{2N-1}(v-w) \cdots L_2(v+w) L_1(v-w), \\
 \hat T(v,w) &= L_{1}(v-w) L_{2}(v+w) \cdots L_{2N-1}(v+w) L_{2N}(v-w),
\end{split}
\end{align}
assuming that there are $2N$ physical lattice sites.
Following similar routes as in the main text and substituting (\ref{Ltol}) into $t$, putting $w= v$, $t(v,v)$ becomes a tensor network with boundaries. Explicitly, suppose $N=2$ we have
\be
t(v,v) = \textrm{tr}_f[K_+^f l_{4,f}]  l_{23} K_-^1 l_{12} l_{34}.
\ee

In the case of a semi-infinite lattice, the right boundary charecterized by $K_+$ would be taken off to infinity.

\begin{figure}[!h]
		\centering
		\includegraphics[width=8cm]{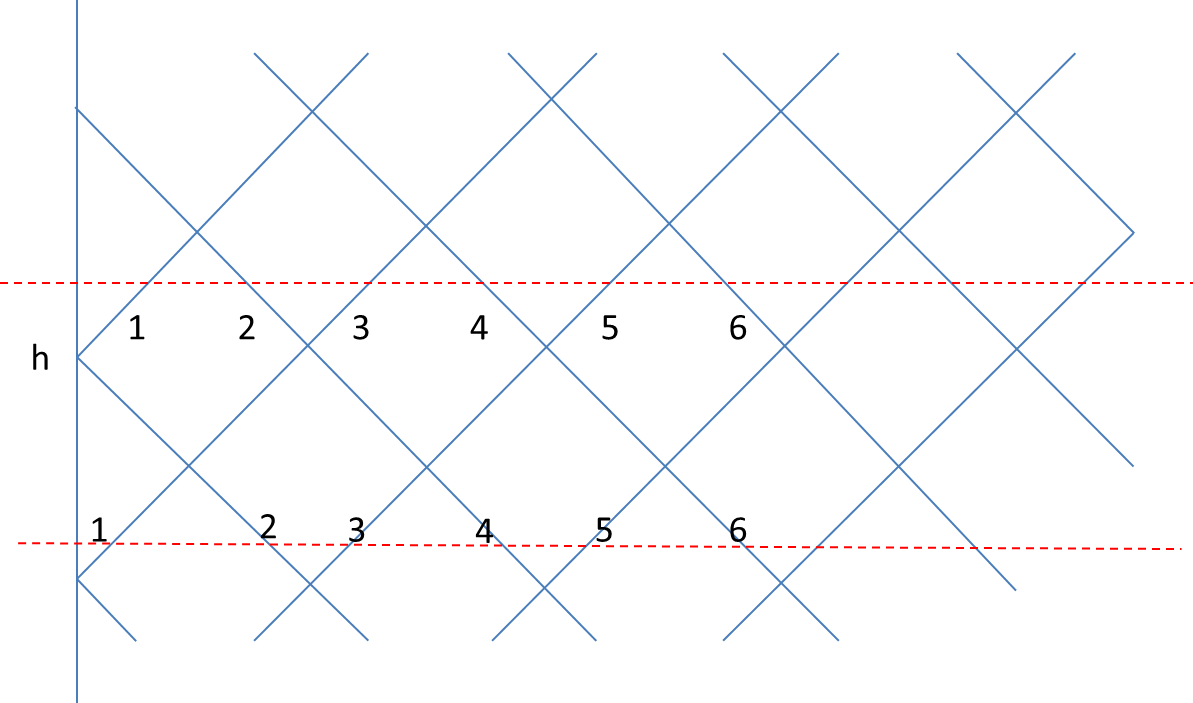}
		\label{halftensor}
		\caption{Tensor network that evolves only half of the space.}
\end{figure}

Choose a half-space Hamiltonian which is illustrated in figure \ref{halftensor}. This is a special case, in which we are simply taking $K_- $ to be proportional to the identity, which is a well known case satisfying the algebraic constraint described above.

In the case of the XXZ or 6-vertex model eigenmodes can be solved in the same manner as discussed in the previous section. The recursion relations (\ref{eigen1}) continue to apply, except that we have to include new relations that apply only at the boundary.

Suppose the boundary link is located at $I= 1$. Let the boundary be $K_ = h \mathcal{I}$.
Then the extra boundary recursion relation is given by
\be \label{boundaryrec}
\Lambda \psi_L(1) =  h( b \,\psi_L(1) + c \, \psi_L(2) ),
\ee
where $\Lambda$ has to be the same eigenvalue as in the bulk of the network determined in (\ref{eigen1}).
This can be solved by taking the modified ansatz
\be
\psi(I) = m \, e^{i p I } \bigg( \begin{tabular}{c}
 $\alpha_+(p)$  \\ $\beta_+(p) $\end{tabular}\bigg)  +
 n \, e^{-i p I } \bigg(\begin{tabular}{c} $ \alpha_-(p)$  \\$ \beta_-(p) $\end{tabular}\bigg) ,
\ee
where $\alpha_{\pm}, \beta_{\pm}$ have been determined in (\ref{solr1}) and (\ref{solr2}), and
$m,n$ are constants that can be readily solved using (\ref{boundaryrec}) and overall normalization of the wavefunction analogous to (\ref{norm}). These are simply typical solutions where the plane-waves are reflected at the boundary.
Multiple spinon solutions can be solved similarly. For semi-infinite lattice, there is no extra constraints that follow from the boundary.


\bibliographystyle{utphys}
\bibliography{bib}

\providecommand{\href}[2]{#2}\begingroup\raggedright\begin{thebibliography}{10}

\bibitem{Swingle:2009bg}
B.~Swingle, ``{Entanglement Renormalization and Holography},''
  \href{http://dx.doi.org/10.1103/PhysRevD.86.065007}{{\em Phys. Rev.}
  {\bfseries D86} (2012) 065007},
\href{http://arxiv.org/abs/0905.1317}{{\ttfamily arXiv:0905.1317
  [cond-mat.str-el]}}.

\bibitem{Pastawski:2015qua}
F.~Pastawski, B.~Yoshida, D.~Harlow, and J.~Preskill, ``{Holographic quantum
  error-correcting codes: Toy models for the bulk/boundary correspondence},''
  \href{http://dx.doi.org/10.1007/JHEP06(2015)149}{{\em JHEP} {\bfseries 06}
  (2015) 149},
\href{http://arxiv.org/abs/1503.06237}{{\ttfamily arXiv:1503.06237 [hep-th]}}.

\bibitem{Hayden:2016cfa}
P.~Hayden, S.~Nezami, X.-L. Qi, N.~Thomas, M.~Walter, and Z.~Yang,
  ``{Holographic duality from random tensor networks},''
  \href{http://dx.doi.org/10.1007/JHEP11(2016)009}{{\em JHEP} {\bfseries 11}
  (2016) 009},
\href{http://arxiv.org/abs/1601.01694}{{\ttfamily arXiv:1601.01694 [hep-th]}}.

\bibitem{Qi:2018shh}
X.-L. Qi and Z.~Yang, ``{Space-time random tensor networks and holographic
  duality},''
\href{http://arxiv.org/abs/1801.05289}{{\ttfamily arXiv:1801.05289 [hep-th]}}.

\bibitem{lattice_book}
  C.~Gattringer and C.~B.~Lang,
  ``Quantum chromodynamics on the lattice,''
  Lect.\ Notes Phys.\  {\bf 788}, 1 (2010).
  doi:10.1007/978-3-642-01850-3

\bibitem{Fredenhagen:2014lda}
K.~Fredenhagen and K.~Rejzner, ``{Quantum field theory on curved spacetimes:
  Axiomatic framework and examples},''
  \href{http://dx.doi.org/10.1063/1.4939955}{{\em J. Math. Phys.} {\bfseries
  57} no.~3, (2016) 031101},
\href{http://arxiv.org/abs/1412.5125}{{\ttfamily arXiv:1412.5125 [math-ph]}}.

\bibitem{Wald:1995yp}
R.~M. Wald, {\em{Quantum Field Theory in Curved Space-Time and Black Hole
  Thermodynamics}}.
\newblock Chicago Lectures in Physics. University of Chicago Press, Chicago,
  IL,
1995.
\newblock

\bibitem{Bombelli:1987aa}
L.~Bombelli, J.~Lee, D.~Meyer, and R.~Sorkin, ``{Space-Time as a Causal Set},''
\href{http://dx.doi.org/10.1103/PhysRevLett.59.521}{{\em Phys. Rev. Lett.}
  {\bfseries 59} (1987) 521--524}.

\bibitem{Sorkin:2003bx}
R.~D. Sorkin, \href{http://dx.doi.org/10.1007/0-387-24992-3_7}{``{Causal sets:
  Discrete gravity},''} in {\em {Lectures on quantum gravity. Proceedings,
  School of Quantum Gravity, Valdivia, Chile, January 4-14, 2002}},
  pp.~305--327.
\newblock 2003.
\newblock
\href{http://arxiv.org/abs/gr-qc/0309009}{{\ttfamily arXiv:gr-qc/0309009
  [gr-qc]}}.
\newblock

\bibitem{beny}
Cedric Beny, ``{Causal structure of the entanglement renormalization ansatz
,}" \href{http://dx.doi.org/10.1088/1367-2630/15/2/023020}{{\em {New J. Phys.},{\bfseries
  15} (2013) 023020}}.
\newblock
\href{https://arxiv.org/abs/1110.4872}{{\ttfamily arXiv:1110.4872}}.
\newblock

\bibitem{Peng:2005yg}
X.~Peng, J.~Du, and D.~Suter, ``{Quantum phase transition of ground-state
  entanglement in a Heisenberg spin chain simulated in an NMR quantum
  computer},''
\href{http://dx.doi.org/10.1103/PhysRevA.71.012307}{{\em Phys. Rev.} {\bfseries
  A71} (2005) 012307}.

\bibitem{Peng:2014kda}
X.~Peng, Z.~Luo, S.~Kou, D.~Suter, and J.~Du, ``{Experimental implementation of
  adiabatic passage between different topological orders},''
  \href{http://dx.doi.org/10.1103/PhysRevLett.113.080404}{{\em Phys. Rev.
  Lett.} {\bfseries 113} (2014) 080404},
\href{http://arxiv.org/abs/1408.3787}{{\ttfamily arXiv:1408.3787 [quant-ph]}}.

\bibitem{Lieb:1972wy}
E.~H. Lieb and D.~W. Robinson, ``{The finite group velocity of quantum spin
  systems},''
\href{http://dx.doi.org/10.1007/BF01645779}{{\em Commun. Math. Phys.}
  {\bfseries 28} (1972) 251--257}.

\bibitem{Vidaltalk}
G.~Vidal, {\em {Geometric tensor networks for critical quantum spin chains }}.
\newblock 2017.
\newblock
\url{KITP Conference: Frontiers of Quantum Information Physics}.
\newblock

\bibitem{socolovsky}
M.~Socolovsky, ``{Rindler space, Unruh effect and Hawking temperature},'' {\em
  Annales de la Foundation Louis de Broglie} {\bfseries 39} (2014) .

\bibitem{Bisognano:1976za}
J.~J. Bisognano and E.~H. Wichmann, ``{On the Duality Condition for Quantum
  Fields},''
\href{http://dx.doi.org/10.1063/1.522898}{{\em J. Math. Phys.} {\bfseries 17}
  (1976) 303--321}.

\bibitem{Haag:1992hx}
R.~Haag, {\em {Local quantum physics: Fields, particles, algebras}}.
\newblock
1992.
\newblock

\bibitem{Casini:2009sr}
H.~Casini and M.~Huerta, ``{Entanglement entropy in free quantum field
  theory},'' \href{http://dx.doi.org/10.1088/1751-8113/42/50/504007}{{\em J.
  Phys.} {\bfseries A42} (2009) 504007},
\href{http://arxiv.org/abs/0905.2562}{{\ttfamily arXiv:0905.2562 [hep-th]}}.

\bibitem{2004JSMTE..06..004P}
I.~{Peschel}, ``{On the reduced density matrix for a chain of free
  electrons},'' \href{http://dx.doi.org/10.1088/1742-5468/2004/06/P06004}{{\em
  Journal of Statistical Mechanics: Theory and Experiment} {\bfseries 6} (June,
  2004) 06004}, \href{http://arxiv.org/abs/cond-mat/0403048}{{\ttfamily
  cond-mat/0403048}}.

\bibitem{2009JPhA...42X4003P}
I.~{Peschel} and V.~{Eisler}, ``{Reduced density matrices and entanglement
  entropy in free lattice models},''
  \href{http://dx.doi.org/10.1088/1751-8113/42/50/504003}{{\em Journal of
  Physics A Mathematical General} {\bfseries 42} (Dec., 2009) 504003},
  \href{http://arxiv.org/abs/0906.1663}{{\ttfamily arXiv:0906.1663
  [cond-mat.stat-mech]}}.

\bibitem{Papadodimas:2012aq}
K.~Papadodimas and S.~Raju, ``{An Infalling Observer in AdS/CFT},''
  \href{http://dx.doi.org/10.1007/JHEP10(2013)212}{{\em JHEP} {\bfseries 10}
  (2013) 212},
\href{http://arxiv.org/abs/1211.6767}{{\ttfamily arXiv:1211.6767 [hep-th]}}.

\bibitem{Baxter:1982zz}
R.~J. Baxter, {\em {Exactly solved models in statistical mechanics}}.
\newblock 1982.
\newblock
\url{http://www.amazon.com/dp/0486462714}.
\newblock

\bibitem{Faddeev:1996iy}
L.~D. Faddeev, ``{How algebraic Bethe ansatz works for integrable model},''
  {\em {Proceedings, School of Physics, Les Houches, France, September
  26-October 6, 1995}} (1996) pp. 149--219,
\href{http://arxiv.org/abs/hep-th/9605187}{{\ttfamily arXiv:hep-th/9605187
  [hep-th]}}.

\bibitem{Davies:1989zz}
B.~Davies, ``{On the spectrum of six-vertex corner transfer matrices},''
\href{http://dx.doi.org/10.1016/0378-4371(89)90565-7}{{\em Physica} {\bfseries
  A159} (1989) 171--187}.

\bibitem{Thacker:1985gz}
H.~B. Thacker, ``{Corner Transfer Matrices and Lorentz Invariance on a
  Lattice},''
{\em Physica} {\bfseries 18D} (1986) 348--359.

\bibitem{Faddeev:1992xa}
L.~D. Faddeev and A.~{\relax Yu}. Volkov, ``{Quantum inverse scattering method
  on a space-time lattice},'' \href{http://dx.doi.org/10.1007/BF01015552}{{\em
  Theor. Math. Phys.} {\bfseries 92} (1992) 837--842}.
[Teor. Mat. Fiz.92,207(1992)].

\bibitem{Tetelman}
M.~G. Tetel'man, ``{Lorentz group for two-dimensional integrable lattice
  systems},'' {\em Zh. Eksp. Teor. Fiz.} {\bfseries 82} (1981) 528--535.

\bibitem{1999AnP...511..153P}
I.~{Peschel}, M.~{Kaulke}, and {\"O}.~{Legeza}, ``{Density-matrix spectra for
  integrable models},''
  \href{http://dx.doi.org/10.1002/(SICI)1521-3889(199902)8:2<153::AID-ANDP153>3.0.CO;2-N}{{\em
  Annalen der Physik} {\bfseries 511} (Feb., 1999) 153--164},
  \href{http://arxiv.org/abs/cond-mat/9810174}{{\ttfamily cond-mat/9810174}}.

\bibitem{1997LNP...478..167N}
T.~{Nishino} and K.~{Okunishi},
  \href{http://dx.doi.org/10.1007/BFb0104638}{``{Density matrix and
  renormalization for classical lattice models},''} in {\em Lecture Notes in
  Physics, Berlin Springer Verlag}, G.~{Sierra} and M.~A.
  {Mart{\'{\i}}n-Delgado}, eds., vol.~478 of {\em Lecture Notes in Physics,
  Berlin Springer Verlag}, p.~167.
\newblock 1997.
\newblock \href{http://arxiv.org/abs/cond-mat/9610107}{{\ttfamily
  cond-mat/9610107}}.

\bibitem{Ercolessi:2009kc}
E.~Ercolessi, S.~Evangelisti, and F.~Ravanini, ``{Exact entanglement entropy of
  the XYZ model and its sine-Gordon limit},''
  \href{http://dx.doi.org/10.1016/j.physleta.2010.03.014}{{\em Phys. Lett.}
  {\bfseries A374} (2010) 2101--2105},
\href{http://arxiv.org/abs/0905.4000}{{\ttfamily arXiv:0905.4000 [hep-th]}}.

\bibitem{Calabrese:2010rx}
P.~Calabrese, J.~Cardy, and I.~Peschel, ``{Corrections to scaling for block
  entanglement in massive spin-chains},''
  \href{http://dx.doi.org/10.1088/1742-5468/2010/09/P09003}{{\em J. Stat.
  Mech.} {\bfseries 1009} (2010) P09003},
\href{http://arxiv.org/abs/1007.0881}{{\ttfamily arXiv:1007.0881
  [cond-mat.stat-mech]}}.

\bibitem{Frahm:1991hr}
H.~Frahm and H.~B. Thacker, ``{Corner transfer matrix eigenstates for the six
  vertex model},''
{\em J. Phys.} {\bfseries A24} (1991) 5587--5604.

\bibitem{Truong:1983kz}
T.~t. Truong and K.~d. Schotte, ``{Quantum inverse scattering method and the diagonal-to-diagonal transfer matrix of vertex models},''
\href{http://dx.doi.org/10.1016/0550-3213(83)90135-9}{{\em Nucl. Phys.}
  {\bfseries B220} (1983) 77--101}.

\bibitem{Sklyanin:1988yz}
E.~K. Sklyanin, ``{Boundary Conditions for Integrable Quantum Systems},''
\href{http://dx.doi.org/10.1088/0305-4470/21/10/015}{{\em J. Phys.} {\bfseries
  A21} (1988) 2375--289}.

\bibitem{Vidal2}
  A.~Milsted and G.~Vidal,
  ``Tensor networks as path integral geometry,''
  arXiv:1807.02501 [cond-mat.str-el].

\bibitem{Qi2}
  J.~Cotler, X.~Han, X.~L.~Qi and Z.~Yang,
  ``Quantum Causal Influence,''
  arXiv:1811.05485 [hep-th].

\end{thebibliography}\endgroup


\providecommand{\href}[2]{#2}\begingroup\raggedright\endgroup

\end{document}